\documentclass[10pt]{article}

\usepackage[T1]{fontenc}
\usepackage[utf8]{inputenc} % (si pdfLaTeX)
\usepackage[english]{babel}
\usepackage{geometry}
\usepackage{microtype}
\usepackage{hyperref}

\usepackage{amsmath,amssymb,amsfonts,amsthm,mathtools}
\usepackage{mathrsfs}
\usepackage{dsfont}

\usepackage{graphicx}
\usepackage{rotating}
\usepackage{subfig} % tu peux passer à subcaption si tu préfères

\usepackage{xcolor}
\usepackage{textcomp}
\usepackage{booktabs}
\usepackage{array}
\usepackage{multirow}
\usepackage{enumitem}
\usepackage{adjustbox}
\usepackage{diagbox}
\usepackage[title]{appendix}
\usepackage[normalem]{ulem}
\usepackage{algorithm}
\usepackage{algorithmicx}
\usepackage{algpseudocode}
\usepackage{listings}

\usepackage[authoryear]{natbib}   % author–year citations
\theoremstyle{plain}
\newtheorem{theorem}{Theorem}        % Theorem 1, 2, ...

\newtheorem{lm}{Lemma}

\theoremstyle{definition}
\newtheorem{definition}{Definition}   % Definition 1, 2, ...
\newtheorem{assumption}{Assumption}   % Assumption 1, 2, ...

\newtheorem{remark}{Remark}

\newcommand\norm[1]{\left\lVert#1\right\rVert}

\newcommand{\proofeq}{%
  \setcounter{equation}{0}%
  \renewcommand{\theequation}{A\arabic{equation}}%
}

\title{\bf AN EFFICIENT HYBRID APPROACH OF QUANTILE AND EXPECTILE REGRESSION}

\usepackage{authblk}
\author[1,2]{ABDELLAH ATANANE\thanks{\texttt{Corresponding author:atanane.abdellah@courrier.uqam.ca}}}
\author[2]{ABDALLAH MKHADRI}
\author[1]{KARIM OUALKACHA}
\affil[1]{Department of Mathematics, Universit\'e du Qu\'ebec \`a Montr\'eal (UQAM), Montreal, Quebec H2X 3Y7, Canada}
\affil[2]{Department of Mathematics, Cadi Ayyad University, Marrakesh-Safi 40000, Morocco}

\date{} % vide pour ne pas afficher de date

\begin{document}
\maketitle

\begin{abstract}
Quantiles and expectiles are determined by different loss functions: asymmetric least absolute
deviation for quantiles and asymmetric squared loss for expectiles. This distinction ensures that
quantile regression methods are robust to outliers but somewhat less effective than expectile
regression, especially for normally distributed data. However, expectile regression is vulnerable
to lack of robustness, especially for heavy-tailed distributions. To address this trade-off between
robustness and effectiveness, we propose a novel approach. By introducing a parameter $\gamma$
that ranges between 0 and 1, we combine the aforementioned loss functions, resulting in a hybrid
approach of quantiles and expectiles. This fusion leads to the estimation of a new type of location
parameter family within the linear regression framework, termed Hybrid of Quantile and Expectile
Regression (HQER). The asymptotic properties of the resulting estimator are then established.
Through simulation studies, we compare the asymptotic relative efficiency of the HQER estimator
with its competitors, namely the quantile, expectile, and $k$th power expectile regression estimators.
Our results show that HQER outperforms its competitors in several simulation scenarios. In addition,
we apply HQER to a real dataset to illustrate its practical utility.
\end{abstract}


\section{Introduction}\label{sec1}
\par Quantile regression (QR) by \citet{koenker_regression_1978}  and Expectile regression (ER) by \cite{newey_asymmetric_1987} are  important methods to estimate 
the quantiles  and expectiles, respectively, of the conditional distribution of a response variable given a set of predictors. They provide a
much more accurate representation of the relationship between the response variable and the predictors than existing methods such as least squares (LS) or least absolute deviation (LAD) regression, which only measure the central tendency of the response
variable.  The QR and ER generalize the LAD and LS approaches, respectively. They provide a description of  the tails of the distribution.  
These two common regression methods and their derivatives have been widely used by researchers in the fields of social sciences, genomics, and economics due to their great advantages.

In terms of interpretability, the $\tau$-th quantile represents the location below which 100 $\tau \% $ of the distributional mass of a random variable $Y$ lies, while the $\tau$-th expectile specifies the location $\mu_{\tau}$ such that the average distance from the data below $\mu_\tau$ to $\mu_\tau$ itself is 100$\tau$\% of the average distance between $\mu_\tau$ and all the data:

\[
\tau = \frac{\mathbb{E} (|Y-\mu_\tau| \mathds{1}\{ Y\leq \mu_\tau \} )}{\mathbb{E}(|Y-\mu_\tau|) }, \quad 
\]
and $\mathds{1}\{ A \} $ denotes the indicator function, which equals the value \( 1 \) if the event \( A \) is true, and \( 0 \) otherwise
\[
\mathds{1}\{ A\} =
\begin{cases} 
1 & \text{if } A \text{ is true,} \\
0 & \text{otherwise.}
\end{cases}
\]
Thus, the $\tau$-th expectile has an intuitive interpretation similar to that of the $\tau$-th quantile, replacing the number of observations with the distance (\cite{daouia2018estimation}). \cite{bellini2017risk} provide a transparent financial meaning of expectiles in terms of the gain-loss ratio, a popular performance measure in portfolio management and well known in the literature on no-good-deal valuation in incomplete markets (see, e.g., \cite{bellini2017risk} and the references therein).

Both expectiles and quantiles have their advantages and disadvantages. The asymmetric least-squares approach (that is, expectiles) offers advantages over quantiles, such as computational efficiency and smooth estimates through sample expectiles, which make better use of the data (\cite{newey_asymmetric_1987, sobotka2012geoadditive}). This has also been pointed out in \cite{waltrup2015expectile}. Expectile regression provides easier inference compared to quantile regression, as demonstrated in several previous works, such as \cite{newey_asymmetric_1987}, \cite{abdous1995relating}, and \cite{daouia2018estimation}. However, the method has limitations, including a lack of interpretability, as it does not provide clear percentile-based thresholds. Moreover, it may not handle outliers as effectively as quantiles. In fact, quantile regression is more robust to outliers and provides easier interpretation of specific percentiles, but it may not capture tail behavior as smoothly as expectile regression. In summary, each method has its advantages and drawbacks, and the key challenge lies in finding a combination that balances these strengths and weaknesses for a given problem (\cite{615bac47-25ad-34ae-aa2c-9c666c4d612b}).

For an insight into both methods  and their
extensions, see \cite{efron1991regression}; \cite{yao_asymmetric_1996}; \cite{arcones_bahadur-kiefer_1996}; \cite{engle_caviar_2004}; \cite{koenker_quantile_2005}; \cite{kim_quantile_2007}; \cite{cai_nonparametric_2008}; \cite{taylor_estimating_2008}; \cite{kuan_assessing_2009}; \cite{cai_semiparametric_2012}; \cite{ehm_quantiles_2016}; \cite{gu_high-dimensional_2016}; \cite{koenker_quantile_2017}; \cite{farooq2017svm} and \cite{jiang_kth_2021}, among others. In particular, \cite{arcones_bahadur-kiefer_1996} examined the $L_p$ regression, which is only related
to a symmetric check function, and \cite{10.1214/16-EJS1173} dealt in detail with the asymptotic distribution of sample expectiles.

The basic idea of this paper is inspired by  \cite{efron1991regression}, who showed that the check function 
\begin{equation} \label{kth power loss}
Q_{\tau,k}(s)= \Psi_{\tau}(s) \lvert s\rvert^{k},\; 1\leq k\leq 2, \; \tau\in(0,1),\; s\in \mathbb{R},
\end{equation}
where 
\begin{equation}\label{psi}
\Psi_{\tau}(s)=\left\lvert \tau - \mathds{1}\{ s<0 \} \right\rvert,\; s\in \mathbb{R}, 
\end{equation}
with $k=1.50$ is attractive as
a compromise between the robustness of QR ($k = 1$) and the high-normal theory efficiency
of ER ($k=2$). In recent studies, \cite{jiang_kth_2021}  have studied in detail a method based on the loss function defined by  equation (\ref{kth power loss}), called the $k$th power expectile regression, for different values of \(k\). More recently, \cite{hu2021penalized} applied this method in high-dimensional settings and for variable selection purposes. Furthermore, \cite{lin2022k} used the $k$th power approach for both estimation and testing. Unlike QR, which does not impose any specific condition on the underlying distribution, common ER typically requires the existence of the mean of the true distribution underlying the data. In practice, this condition can be stringent, as it is the case with  certain financial data sets, such as high-frequency trading data or extreme events (e.g., stock market crashes). 
The $k$th power expectile regression relaxes this assumption. Instead, it focuses on the existence of the $(k - 1)$th order moment instead ($0<k<2$).

QR operates without imposing any moment conditions; however, its computational complexity, especially for tasks such as variance calculation, can be challenging because  it depends on the unknown density of the error terms.
On the other hand, the $k$th power expectile regression offers computational ease, especially with respect to variance computation (\cite{jiang_kth_2021}). This makes the $k$th power expectile regression, where $1 < k< 2$, a promising trade-off between computational efficiency and flexibility. Additionally, in certain numerical investigations, it has been observed that the asymptotic variance associated with the $k$th power expectile regression is smaller compared to both quantile and expectile regressions  (\cite{jiang_kth_2021}). 
Under specific regularity conditions, the asymptotic normality properties of the $k$th power expectile regression have been investigated in an earlier work by \citet{jiang_kth_2021} for $k \in (1.5, 2]$. More recently, \cite{lin2022k} extended this result for the entire range, $k \in (1, 2]$, under more general assumptions. This broader theoretical justification strengthens the applicability of the $k$th power expectile regression.

As an alternative to power loss functions, hybrid loss functions have gained increasing attention in recent years as flexible and principled tools for statistical modeling. Instead of relying on a single loss criterion—which may be either robust but inefficient, or efficient but sensitive to outliers—hybrid approaches combine multiple convex losses to leverage the strengths of each. This blending allows for more adaptable estimation procedures that can remain stable across a wide range of data conditions, including asymmetry, heteroskedasticity, or structural complexity. As demonstrated in \citet{bradic2011penalized}, penalized composite quasi-likelihood methods based on convex combinations of loss functions yield substantial improvements in both estimation accuracy and model selection. Their practical utility is illustrated through empirical applications on real biological data, notably in single nucleotide polymorphisms (SNPs) selection for Down syndrome studies. Similarly, \citet{fan2018single} explored hybrid loss-based models in the context of systemic risk assessment, applying them to U.S. financial market data to effectively capture interdependencies and distributional shifts that traditional models may miss. These contributions underscore that hybrid methods are not only theoretically appealing but also highly effective in diverse real-world scenarios. They provide a solid foundation for developing new loss-based estimation strategies that remain robust and computationally tractable in the face of modern data challenges.

\begin{figure}[h]
    %\centering
  \includegraphics[width=1.1 \linewidth]{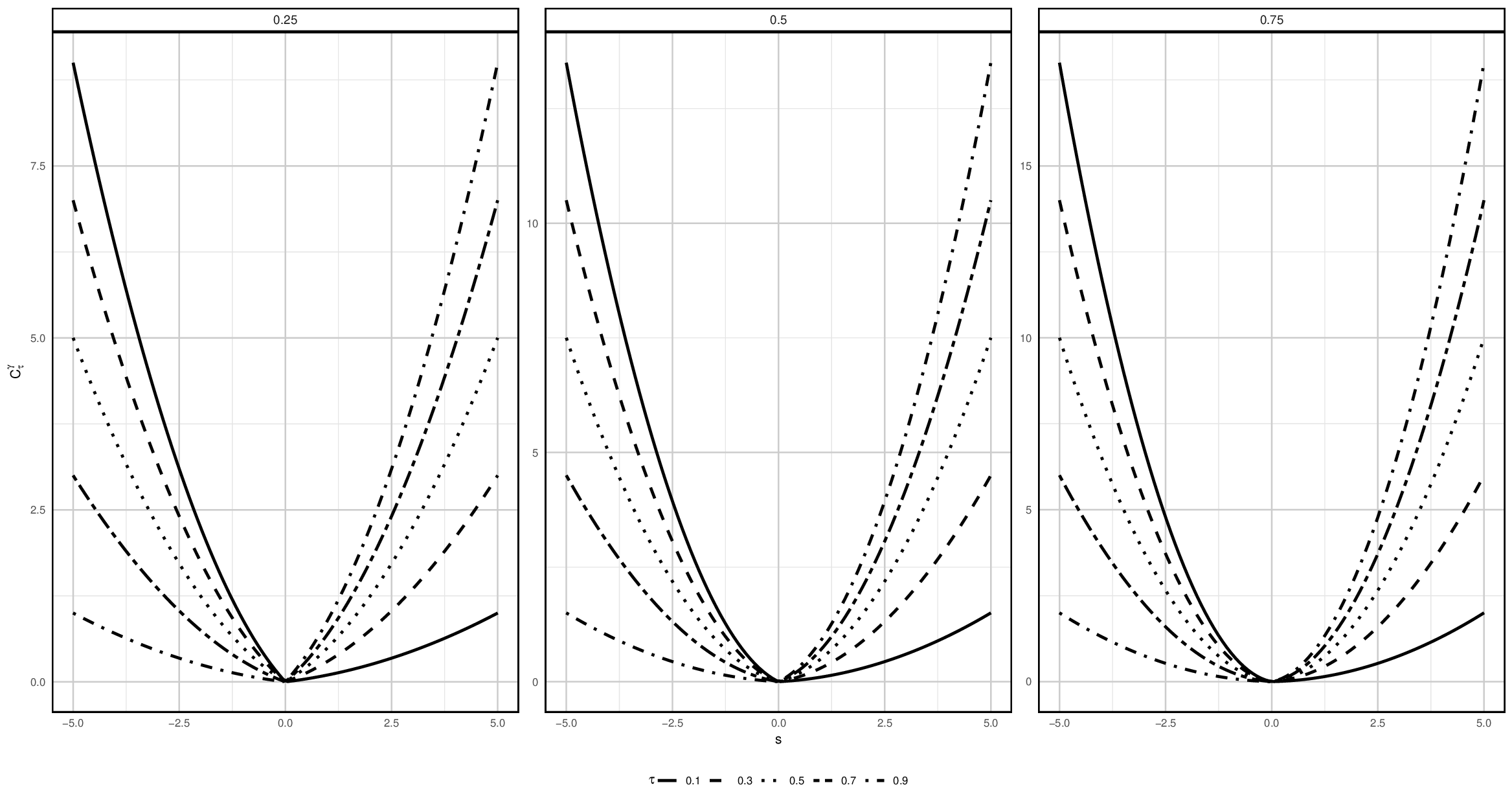} 
    \caption{The graphs depict the loss functions of the $\tau$-$\gamma$th HQER expectile regression  for different values of the tuning parameter $\gamma \in \{ 0.25,0.5,0.75 \}$ (from left to right) at levels $\tau \in \{ 0.10,0.30,0.50,0.70,0.90 \}$}
    \label{hqer loss only}

\end{figure}

This paper  proposes an efficient \underline{H}ybrid of \underline{Q}uantile and \underline{E}xpectile \underline{R}egression approach, called $\tau$-$\gamma$th HQER or HQER when there is no confusion. %\underline{H}\underline{Q}\underline{E}\underline{R}. 
It is based on a convex combination of the $k$th power loss (\ref{kth power loss}) in $k=1$ (Quantile) and $k=2$ (Expectile)   using a tuning parameter $\gamma \in [0,1]$, as follows
\begin{equation}\label{loss function here}
C_{\tau}^{\,\gamma}(s)= (1-\gamma)\cdot Q_{\tau,1}(s)+ \gamma\cdot Q_{\tau,2}(s)     ,\;  0  \leq \gamma \leq 1,  \; \tau\in(0,1), s\in \mathbb{R}.
\end{equation}
By adjusting  the $\gamma$ parameter, one can regulate the effect of each measure within the composite metric. Fig. \ref{hqer loss only} illustrates the loss function in equation (\ref{loss function here}) for various values of the tuning parameter $\gamma$, specifically considering $\gamma$ values of $0.25, 0.50,$ and $0.75$. The latter gives rise to a new family of parameter estimators. Using insights from this analysis, we give an explicit definition
of the $\tau$-$\gamma$th HQER  expectile based on this function and prove its existence and uniqueness under the assumption of the existence of the first absolute order moment. Some fundamental properties of the $\tau$-$\gamma$th HQER  expectile
are  studied. In addition, we turn our attention to the problems associated with HQER  expectile regression, such as the asymptotic properties of the proposed  estimator $\boldsymbol{ \hat{\beta }} (\tau,\gamma)$ of the unknown true regression coefficient ${\boldsymbol\beta}_0$. The proofs differ from those of the expectile regression considered in \cite{newey_asymmetric_1987}. \cite{newey_asymmetric_1987} used the theory of  \cite{huber1973robust} to prove their asymptotic
normality results, while we use arguments in \cite{pollard1991asymptotics}, \cite{knight1998limiting}, \cite{koenker_quantile_2005}, and \cite{hjort2011asymptotics} under some regularity conditions.  Our theoretical results include, to some extent, those in \cite{koenker_quantile_2005} and \cite{newey_asymmetric_1987} as special cases. A  Newton-Raphson algorithm is proposed for computing the  $\tau$-$\gamma$th HQER estimators. To illustrate the performance of the $\tau$-$\gamma$th HQER on some  simulated  data sets, a procedure for determining an appropriate value of $\gamma$ is also provided.  To highlight the advantage of the general $\tau$-$\gamma$th HQER 
 expectile regression, some comparisons with the quantile, the expectile, and the $k$th power expectile in terms of asymptotic variance
regression have been considered in detail. An application of the \(\tau\)-\(\gamma\)th HQER to the analysis of real data is presented, demonstrating the merits of the proposed method. It is worth noting that the empirical results indicate that the \(\tau\)-\(\gamma\)th HQER expectile regression (especially for \(\gamma\) close to \(1\) and at high levels of \(\tau\)) outperforms standard expectile regression and quantile regression and, in many cases, is competitive with the \(k\)th power expectile regression in terms of variance. Although the $k$th power expectile regression requires only the satisfaction of the absolute $(k-1)$th order moment condition, the $\tau$-$\gamma$th HQER emerges as an attractive alternative. This approach enhances the validation procedure, under which the theoretical asymptotic normality and consistency are established. Given the complexity associated with $k$th power expectiles, the $\tau$-$\gamma$th HQER stands out as a promising option. It offers straightforward verification conditions based solely on a simple convex combination, making it easier to confirm. 

It is essential to point out that the main difference between HQER and the methods of  \cite{bradic2011penalized} and \cite{fan2018single}, in terms of composite losses, lies in both methodology and theoretical results. HQER innovates by combining quantile and expectile regression through a hybrid loss function that adapts to different data distributions, balancing robustness (for heavy-tailed data) and efficiency (for normally distributed data). The theoretical results establish asymptotic properties for this hybrid estimator. In contrast, \cite{bradic2011penalized} focuse on ultrahigh-dimensional settings and use a composite quasi-likelihood approach that combines different convex loss functions to handle high-dimensional data and correct for bias in L1-penalization. Their theoretical contribution is to establish oracle properties for the estimator, focusing primarily on variable selection. \cite{fan2018single} use combined losses in the form of quantile and expectile regression, but their methodology focuses on dimension reduction and systemic risk analysis, with less emphasis on combined losses. The theoretical results here focus on asymptotic efficiency to capture tail dependencies. Thus, the main difference is that HQER combines losses to improve location estimation, while the other references either focus on variable selection in ultra-high dimensions (\cite{bradic2011penalized}) or tail event modeling in very high dimensional single index models (\cite{fan2018single}).

%===============================================

This paper is organized as follows. In Section $\ref{section 2}$, we explain and introduce the $\tau$-$\gamma$th HQER expectile methodology: in Section $\ref{subsection 2.1}$, we present the hybrid loss function and define the target HQER expectiles; in Section $\ref{section class of hqer}$, we describe the class of HQER expectiles; in Section $\ref{subsection 2.3}$, we detail the estimation procedure. A Newton--Raphson algorithm is provided in Section \ref{subsection 2.2}. Basic properties of the $\tau$-$\gamma$th HQER  expectile are given in Section $\ref{Sec 3}$. Numerical experiments for the $\tau$-$\gamma$th HQER  expectile regression estimators 
are presented in Section $\ref{section 4}$. Some comparisons based on simulated experiments with  quantile regression,  expectile regression, and the $k$th power expectile regression  are  given in Section $\ref{subsection 4.1}$. Section $\ref{subsection 4.2}$  includes an application to real data. A method for choosing an appropriate $\gamma$ is given in Section $\ref{section 5}$. Section $\ref{Sec 5}$ concludes the paper. All theoretical proofs are postponed to the Appendix.

%=======================================================
\section{HQER  methodology}\label{section 2}

%=======================================================

We introduce our HQER procedure and the estimation algorithm in this section.

%-----------------------------------------------------
\subsection{HQER loss function}
%{\color{cyan}We introduce our HQER procedure and the estimation algorithm in this section. }
\label{subsection 2.1}
%-----------------------------------------------------
%\subssection{HQER Loss function}

Define the $\tau$-$\gamma$th HQER loss function as follows 
%-----------------------
\begin{equation}\label{l creq loss}
C_{\tau}^{\,\gamma}(s)= (1-\gamma)\cdot \rho_{\tau}(s)+\gamma\cdot \ell_{\tau}(s),  \;\tau \in (0,1),  0  \leq \gamma \leq 1,
\end{equation}
%------------------------
where $\rho_{\tau}(\cdot)$  and $\ell_{\tau}(\cdot)$ are the check functions corresponding to the common quantile and expectile regression respectively, defined as $\rho_{\tau}(s)=\Psi_\tau(s)\cdot \lvert s \rvert,$ and $\ell_{\tau}(s)=\Psi_\tau(s) \cdot s^{2}, $ with $\Psi_\tau(\cdot)$ is defined in \eqref{psi}. The      loss functions corresponding to \( \gamma = 0 \) and \( \gamma = 1 \) are those used in quantile regression and expectile regression, as introduced by \cite{koenker_regression_1978} and \cite{newey_asymmetric_1987}, respectively.\\
In Fig. \ref{hqer loss only}, each curve represents the loss function \( C^{\gamma}_{\tau}(\cdot) \) in (\ref{l creq loss}) for three different values of \( \gamma \in \{0.25, 0.50, 0.75\} \). The loss functions are shown at different quantile levels \( \tau \), ranging from \( \{0.10, 0.30, 0.50, 0.70, 0.90\} \). This visualization illustrates how the choice of the tuning parameter \( \gamma \) affects the shape of the loss function at different quantile levels. For \( \tau = 0.50 \), the loss function assigns equal weights to positive and negative values of \( s \), regardless of the values of \( s \) and \( \gamma \). In contrast, for \( \tau = 0.70 \) and \( \tau = 0.90 \), \( C^{\gamma}_{\tau}(\cdot) \) assigns higher weights to positive values of \( s \) and lower weight to negative values of \( s \), for all the given values of \( \gamma \). Conversely, when \( \tau = 0.10 \) or \( \tau = 0.30 \), \( C^{\gamma}_{\tau}(\cdot) \) assigns lower weight to positive values of \( s \) and higher weight to negative values of \( s \).

\begin{figure}[h]
  \centering
  \includegraphics[width=1.1 \linewidth]{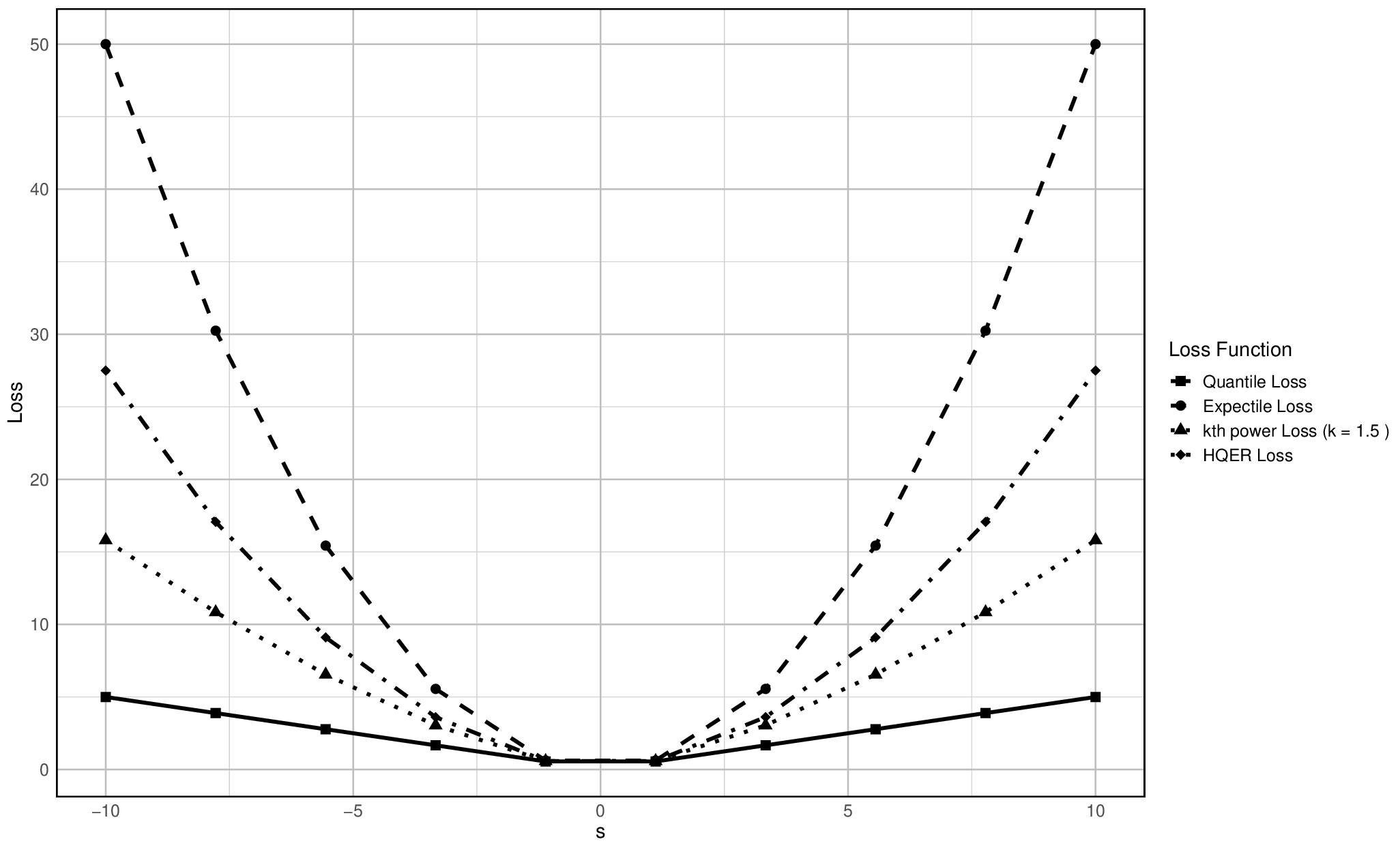} 
  
  \caption{The check  functions for QR, $\rho_{\tau}(\cdot)$, ER, $\ell_{\tau}(\cdot)$, $k$th power expectile regression, $Q_{\tau,k}(\cdot)$
  $(k=1.50, \tau=0.50),$ and the $\tau$-$\gamma$th HQER loss function, $C^{\,\gamma}_{\tau}(\cdot) (\gamma=0.50, \tau=0.50)$} 
  
  \label{checkff}
\end{figure}
\noindent
For the sake of comparison, the \(\tau\)-\(\gamma\)th HQER loss function (for \(\gamma = 0.50, \tau = 0.50\)) defined in equation (\ref{l creq loss}), along with the QR, ER, and \(k\)th power loss function (\(k = 1.50, \tau = 0.50\)) defined in equation ($\ref{kth power loss}$), are shown in Fig. $\ref{checkff}$. It can be clearly observed that both the \(k\)th power and \(\tau\)-\(\gamma\)th HQER loss functions fall between those of QR and ER, which motivates the idea of exploring them further. In the  next Section $\ref{section class of hqer}$, we provide the definition of the  $\tau$-$\gamma$th HQER expectile as used in this paper and its comparison with the classical quantile, expectile, and the $k$th power expectile.

\subsection{The class of HQER expectiles}\label{section class of hqer}

\begin{definition}\label{definition 1}
Let \( Y \) be a random variable with a cumulative distribution function (cdf) \( F(\cdot) \) that is absolutely continuous, assumed to have a finite absolute first-order moment and a continuous, positive density \( f(\cdot) \). The \(\tau\)-\(\gamma\)th HQER expectile of \( Y \) is defined as the scalar parameter \( \xi_{\gamma}(\tau, Y) \) satisfying:
\begin{equation}
\label{exp}
\xi_{\gamma}(\tau, Y) = \arg\min_{\theta \in \mathbb{R}} \mathbb{E} \left[ C_{\tau}^{\gamma}(Y - \theta) - C_{\tau}^{\gamma}(Y) \right],
\end{equation}
where \( C_{\tau}^{\gamma}(\cdot) \) is the corresponding loss function in $(\ref{l creq loss})$.
\end{definition}
\noindent
For convenience, we write ${\xi}_{\gamma}(\tau,Y)$ as ${\xi}_{\gamma}(\tau)$ or ${\xi}(\tau)$ without confusion. The problem \eqref{exp}  remains  well-defined due to the finiteness of the absolute first-order moment, i.e., $\mathbb{E}(|Y|)<+\infty$, mainly due to the subtraction of the term $C_{\tau}^{\,\gamma}(Y)$. By taking the first order condition from \eqref{exp},  it is easy to show that the parameter ${\xi}(\tau)$ is the solution to 
\begin{equation*}
(1-\gamma)(1-\tau) \cdot \int_{-\infty}^{{\xi}(\tau)}\,\mathrm{d F}(s)-(1-\gamma)\tau \cdot \int_{{\xi}(\tau)}^{+\infty}\,\mathrm{dF}(s)-2 \gamma(1-\tau)\cdot\mathbb{E}(Y-{\xi}(\tau))_{-}+2\gamma\tau\cdot\mathbb{E}(Y-{\xi}(\tau))_{+}=0,
\end{equation*}
where $(s)_{-}=\min(s,0)$ and $(s)_{+}=\max(s,0), \, s \in \mathbb{R}$. So we have,
\begin{equation}\label{first order condition}
(1-\gamma)\cdot(F({\xi}(\tau))-\tau)+2\gamma (1-\tau)\cdot \int_{-\infty}^{{\xi}(\tau)} ({\xi}(\tau) - s)\cdot \mathrm{d F}(s)-2\gamma\tau\cdot \int_{{\xi}(\tau)}^{+\infty}(s-\xi(\tau))\cdot \mathrm{d F}(s)=0.
\end{equation}
The \(\tau\)-\(\gamma\)th HQER expectile of a distribution cannot be determined explicitly. Instead, it is implicily defined by an equation. From \eqref{first order condition}, we can derive the inverse of the \(\tau\)-\(\gamma\)th HQER expectile, i.e.,  $\xi_{\gamma}^{-1}(s)=\tau$
\begin{equation}\label{InvExp}
\xi_{\gamma}^{-1}(s)=\frac{2\gamma\cdot \left( G(s)-s \cdot F(s) \right)- (1-\gamma)\cdot F(s) }{4\cdot \left( G(s)-s\,F(s) \right)+2\cdot( s-m)-(1-\gamma) },\quad s\in \mathbb{R},
\end{equation}
where $G$ is the partial moment function $G(s)=\int_{-\infty}^{s} t\cdot \mathrm{d}F(t)$ and $m$ is the mean of $Y$, i.e., $m=\mathbb{E}(Y).$ This definition shows that the $\tau$-$\gamma$th HQER expectile function is determined by the tail expectations of the distribution of $Y$. Interestingly, the $\tau$-$\gamma$th HQER expectile function  also satisfies 
 
\begin{equation}\label{exp two for xi}
  \xi(\tau)=\frac{\mathbb{E}\left[ \Psi_\tau(Y-{\xi}(\tau))\cdot Y \right] }{ \mathbb{E}\left[ \Psi_\tau(Y-{\xi}(\tau)) \right] } +\frac{(1-\gamma)}{2\gamma}\cdot \frac{\tau - F({\xi}(\tau))}{ \mathbb{E}\left[ \Psi_\tau(Y-{\xi}(\tau)) \right] }. 
\end{equation}
In fact, using the first-order condition in \eqref{exp}, we derive the following equation
\[
(1-\gamma) \cdot (F(\xi(\tau)) - \tau) - 2\gamma\cdot \mathbb{E}[\Psi_\tau(Y - \xi(\tau))(Y - \xi(\tau))] = 0,
\]
which simplifies to
\[
(1-\gamma) \cdot (F(\xi(\tau)) - \tau) - 2\gamma \cdot \mathbb{E}[\Psi_\tau(Y - \xi(\tau))(Y - \xi(\tau))] = 0.
\]
 So we have
\[
(1-\gamma) \cdot (F(\xi(\tau)) - \tau) - 2\gamma \cdot \mathbb{E}[\Psi_\tau(Y - \xi(\tau))Y] + 2\gamma \cdot \xi(\tau) \cdot \mathbb{E}[\Psi_\tau(Y - \xi(\tau))] = 0.
\]
Dividing through by \(\mathbb{E}[\Psi_\tau(Y - \xi(\tau))]\), gives \eqref{exp two for xi}. This last definition, which is 
more meaningful in the context of regression, shows that the HQER expectile can be viewed as a combination of weighted averages and the tail of the corresponding distribution. Given a random sample, $\{y_i\}_{i=1}^{n}$ from $Y$, the $\tau$-$\gamma$th HQER empirical expectile function,

\[{\hat{\xi}}(\tau)=\sum_{i=1}^{n}       \frac{  \Psi_{\tau}(y_{i}-{\hat{\xi}}(\tau))\cdot y_{i} }{ \sum_{i=1}^{n}  \Psi_{\tau}(y_{i}-{\hat{\xi}}(\tau))  } +\frac{(1-\gamma)}{2\gamma}\cdot \frac{ n \tau-  \, \sum_{i=1}^{n} \mathds{1}\{ y_{i} \leqslant {\hat{\xi}}(\tau) \}   }{ \sum_{i=1}^{n} \Psi_{\tau}(y_{i}-{\hat{\xi}}(\tau)) },\]
is also a solution that minimizes the empirical loss function
$\frac{1}{n}\, \sum_{i=1}^{n}\, \left [ C^\gamma_\tau(y_i - \theta)-C^\gamma_\tau(y_i) \right ],$ over $\theta.$

\begin{remark}
\label{definition 1-conditional}
Let $X$ be a random variable. Similarly to Definition $\ref{definition 1}$, the conditional \(\tau\)-\(\gamma\)th HQER expectile of \( Y \) given \( X = x\) is defined as
\begin{equation}\label{exp-conditional}
\xi_{\gamma}(\tau, Y \mid X = x) = \arg\min_{\theta \in \mathbb{R}} \mathbb{E} \left[ C_{\tau}^{\gamma}(Y - \theta) - C_{\tau}^{\gamma}(Y) \mid X = x \right],
\end{equation} 
where the conditional cdf of \( Y \) given \( X = x \), \( F(\cdot \mid x) \), is assumed to be absolutely continuous, to have a finite absolute first-order moment, and to have a continuous, positive conditional density \( f(\cdot \mid x) \).
\end{remark}

\begin{figure}[h]
    \includegraphics[width=1.1 \linewidth]{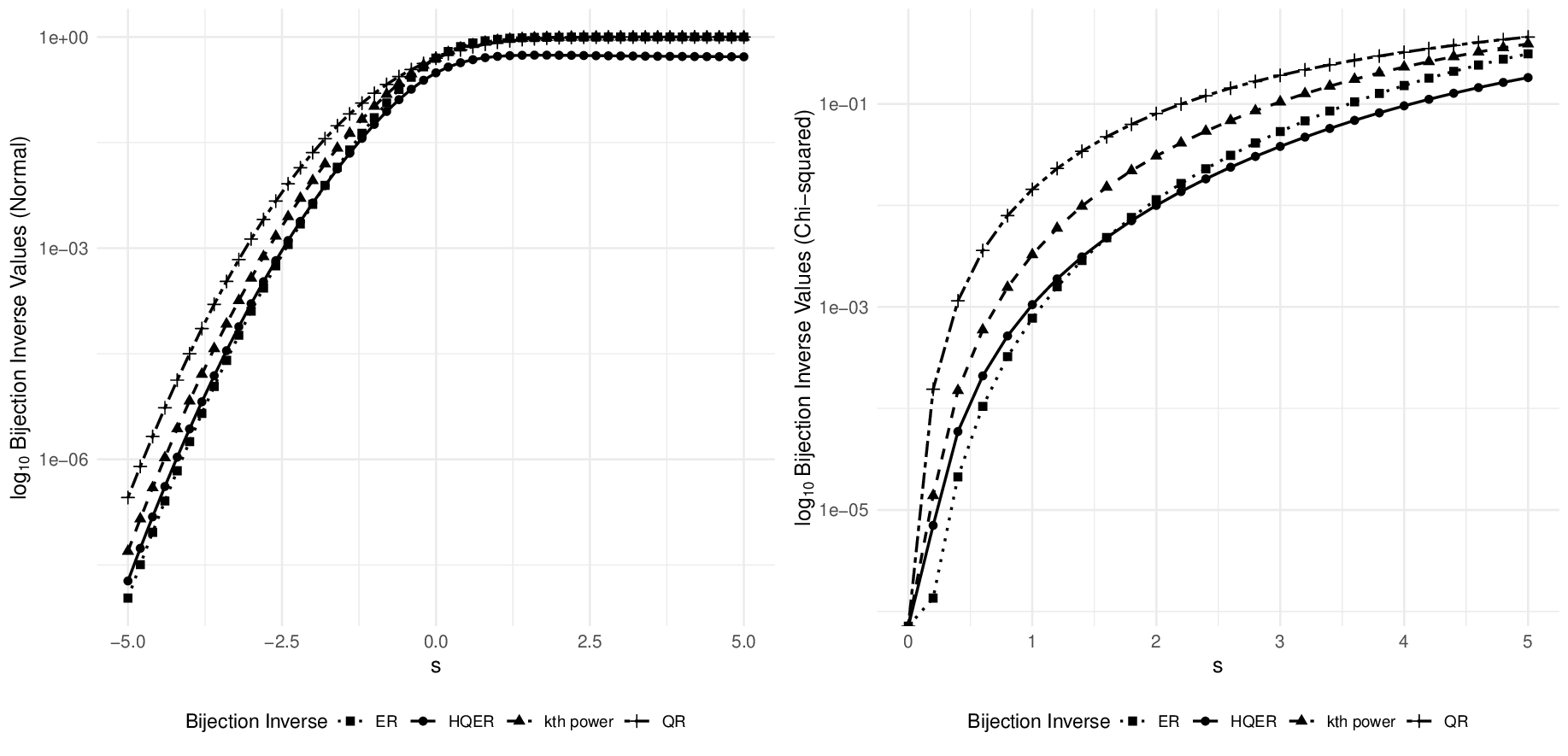}
     \caption{The plots show the functions of the cumulative distribution of the inverse of 
      the 
     quantile (cdf), the inverse  of the expectile, the inverse of  the $k$th
power expectile ($k=1.5$) and  the $\tau$-$\gamma$th HQER expectile ($\gamma=0.5$) for the normal and the Chi-squared distributions  $\chi^{2}(\nu)$  with degree of freedom $\nu=6$ (from left to right). To display the curves well, we set the ordinate
scale to $\log_{10}(s)$} 
    \label{ExpInvUni and normal}%
\end{figure}
\noindent
The  graphs in Fig. \ref{ExpInvUni and normal}  show the functions of the cdf, the
inverse of the expectile, the inverse of the $k$th power expectile $(k=1.5)$, and the inverse of the $\tau$-$\gamma$th HQER expectile $(\gamma=0.5)$ for the standard normal distribution and the chi-squared distribution $\chi^{2}(\nu)$  with degree of freedom $\nu=6.$ The values of the inverse of the $k$th  power expectile function and the inverse of the $\tau$-$\gamma$th HQER expectile function 
are approximately between those of the cumulative distribution function and the inverse of the expectile function.

\begin{table}[ht]
\centering
\caption{ $\tau_{\alpha}$ values for a given $\alpha$, such that $\xi_{\gamma}(\tau_{\alpha})=\delta(\alpha)$, under different distributions $\mathcal{N}(0,1)$, $\mathcal{E}(1)$, $\mathcal{U}([0,1])$, and $t(\nu)$, with degrees of freedom $\nu \in \{5,10,50,500\}$. These correspond to the standard normal distribution, the exponential distribution with parameter 1, the uniform distribution with support $(0,1)$, and the Student distribution with $\nu$ degrees of freedom, respectively}
\label{table 1}
%\centering
\begin{tabular}{@{}lccccccc@{}}
\toprule
$\alpha$ & $\mathcal{N}(0,1)$ & $\mathcal{E}(1)$ & $\mathcal{U}([0,1])$ & $t(500)$ & $t(50)$ & $t(10)$ & $t(5)$ \\ 
\midrule
0.01 & 0.002 & 0.002 & 0.003 & 0.002 & 0.002 & 0.002 & 0.002 \\
0.02 & 0.004 & 0.004 & 0.007 & 0.004 & 0.004 & 0.004 & 0.005 \\
0.03 & 0.006 & 0.006 & 0.011 & 0.006 & 0.006 & 0.007 & 0.008 \\
0.07 & 0.018 & 0.014 & 0.027 & 0.018 & 0.019 & 0.020 & 0.022 \\
0.12 & 0.039 & 0.024 & 0.052 & 0.039 & 0.039 & 0.041 & 0.044 \\
0.20 & 0.080 & 0.040 & 0.102 & 0.081 & 0.081 & 0.084 & 0.088 \\
0.25 & 0.112 & 0.049 & 0.139 & 0.112 & 0.113 & 0.115 & 0.119 \\
\bottomrule
\end{tabular}
\end{table}

\noindent
When $\gamma$ is set to $0$, representing the usual quantiles, we denote the corresponding probability as $\alpha$ instead of $\tau$. Table \ref{table 1} shows various pairs of $\tau$ and $\alpha$ values where $\tau$-$\gamma$th HQER expectiles $(\xi_{\gamma}(\tau))$ equal $\alpha$-th quantiles $(\delta(\alpha))$ for selected common distributions, i.e., $\xi_{ \gamma }(  \tau_{\alpha}   )=\delta(\alpha)$. Note that for the Student distribution with degrees of freedom \(\nu\) tending to toward \(\infty\), it gives the same values as the standard normal distribution, which is consistent with the fact that $t(\nu) \xrightarrow{\mathcal{D}} \mathcal{N}(0,1) \quad \text{as} \quad \nu \to \infty$ in theory, where $\xrightarrow{\mathcal{D}}$ denotes the convergence in the distribution.

\begin{figure}[H]
    \centering
     \includegraphics[width=1.1 \linewidth]{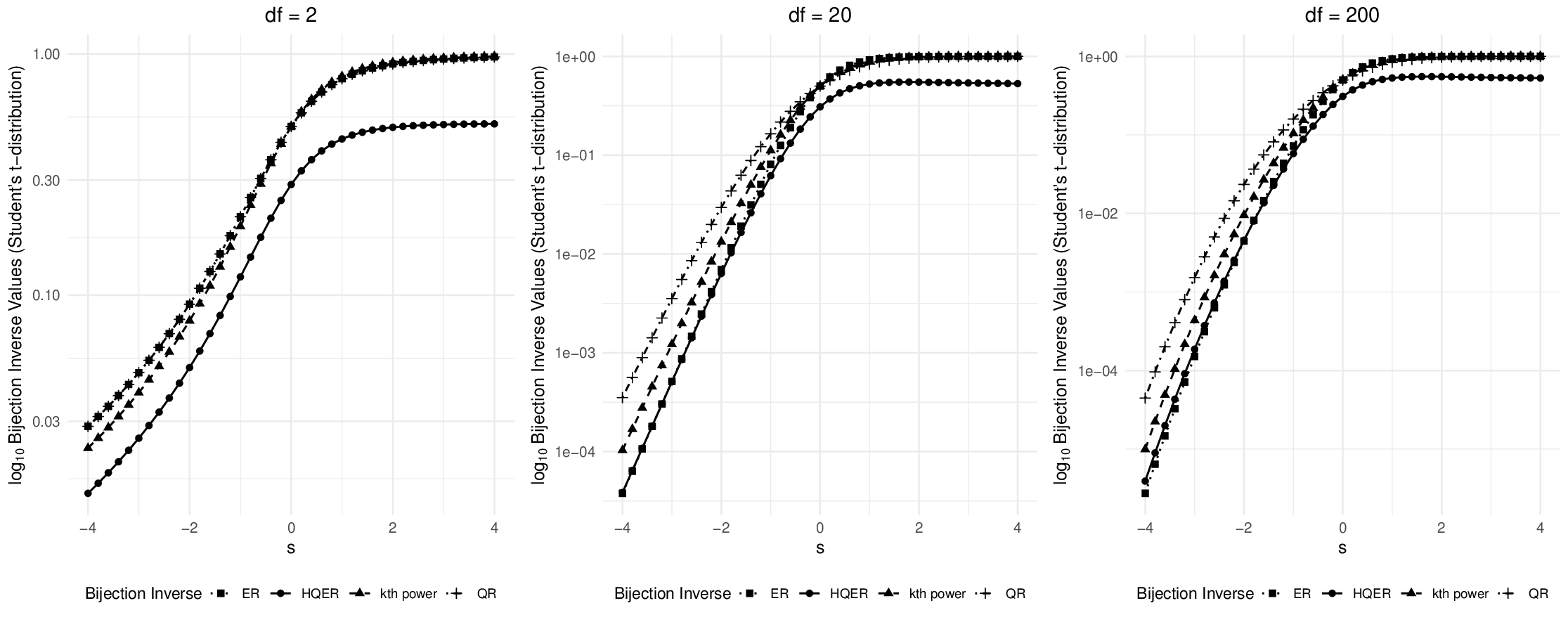}

    \caption{The plots show the functions of the cumulative distribution function, the inverse  of the expectile and the inverse of $k$th
power expectile  $(k=1.5) $ and the inverse of the $\tau$-$\gamma$th HQER expectile $ (\gamma=0.5,\tau=0.5) $ for the Student distribution with different degrees of freedom (df=$\nu$) $\nu\in\{2,20,200\}$ (from left to right). To display the curves well, we set the ordinate
scale to $\log_{10}(s)$}
    \label{ExpInvStt}%

\end{figure}
\noindent
The graphs in Fig. \ref{ExpInvStt}, which give the same result as in Fig. \ref{ExpInvUni and normal} but for the $t(\nu)$ distribution, for different degrees of freedom $\nu \in\{ 2,20,200\}$, gives a negative answer. Graphically, the cumulative distribution function is exactly the same as the inverse of the expectile function, as expected, but they
are very different from the inverse of the $k$th power expectile function ($k=1.5$) and the invere of the $\tau$-$\gamma$th HQER expectile function ($\gamma=0.5$).\\
Based on the observations and analyses presented in this section, we conclude that, like the \(k\)th power expectiles in \cite{jiang_kth_2021}, the \(\tau\)-\(\gamma\)th HQER expectiles are significantly different from both the quantiles, expectiles, and $k$th power expectiles. However, they can be used to define a distribution in a manner analogous to the quantile function, the expectile function, and $k$th power expectiles. The one-to-one correspondence between these four functions shown in Figs. $\ref{ExpInvUni and normal}$ and $\ref{ExpInvStt}$ highlights an intrinsic property of a distribution. In particular, a distribution can also be fully characterized by its \(\tau\)-\(\gamma\)th HQER expectile function $\xi(\tau)$.

\subsection{HQER estimation}\label{subsection 2.3}
We assume that the observed data $\left\{ (y_{i},\boldsymbol{x}_{i}) \right\}_{i=1}^{n}$ come from the linear model:
\begin{equation}\label{model}
y_{i} = \boldsymbol{x}_{i}^\top \, \boldsymbol{\beta}_{0} + \varepsilon_{i}\, , \; i=1,2,\ldots, n,
\end{equation}
where $y_{i}$ is the $i$th value of the response variable, $\{\boldsymbol{x}_{i}\}_{i=1}^{n}$ is a sequence of regression vectors of dimension $p$ with the first component $x_{i,1} = 1$. The vector $\boldsymbol{\beta}_{0} \in \mathbb{R}^{p}$ is an unknown parameter vector, and $\{ \varepsilon_{i} \}_{i=1}^{n}$ is a sequence of scalar error terms.\\
We define the HQER's population objective function as  
\begin{equation}\label{population}
   T(\boldsymbol{\beta},\gamma,\tau)=\mathbb{E} [C_{\tau}^{\,\gamma}(Y-\boldsymbol{X}^{\top}\!\boldsymbol{\beta})-C_{\tau}^{\,\gamma}(Y)].
\end{equation}

\begin{definition}
The HQER estimator   $\boldsymbol{\hat{\beta}}^{\,\gamma}(\tau)$ minimizes the sample analog based on the empirical distribution. Specifically, it solves for:
\begin{equation}\label{total asymmetric creq error}
T_{n}(\boldsymbol{\beta},\tau,\gamma)=\frac{1}{n}\sum_{i=1}^{n} [C_{\tau}^{\,\gamma}(y_{i}-\boldsymbol{x}_{i}^{\top} \boldsymbol{\beta})-C_{\tau}^{\,\gamma}(y_{i})].
\end{equation}   
\end{definition}
\noindent
For simplicity, we denote \(\boldsymbol{\hat{  \beta}}^{  \gamma}(\tau) \) as \( \boldsymbol{\hat{\beta}}(\tau) \) or \(\boldsymbol{ \hat{\beta}} \) without ambiguity. Subsequently, in alternative notations related to \( \gamma \), we often omit the symbol \( \gamma \). For both QR and ER, when $\varepsilon_i$ and $\boldsymbol{x}_i$  are independent (homoscedasticity), we have (\cite{koenker_regression_1978} and \cite{newey_asymmetric_1987})

\begin{equation}
\label{cvv}
\boldsymbol{\hat{\beta}} (\tau) \;  \xrightarrow[]{\mathcal{P}}            \; \boldsymbol{\beta}_{0}+\tilde{\delta}_{\varepsilon}(\tau)\, \boldsymbol{e}_{1}, \; \forall \tau \in (0,1),
\end{equation}
where $\xrightarrow[]{\mathcal{P}}$ denotes the convergence in probability and the vector \( \boldsymbol{e}_{1} \) refers to the first element of the canonical basis of \( \mathbb{R}^{p} \), which has $1$ at the first position and $0$ elsewhere and $\tilde \delta_{\varepsilon}(\tau):= F^{\,\!-}(\tau),$ with $F^{-}(\tau) = \inf \{ x \mid F(x) \geq \tau \}
$ for the quantile regression and  $\tilde \delta_{\varepsilon}(\tau):= \xi_{1}(\tau,\varepsilon)$
for the expectile regression. This result means that the probability limits of \( \boldsymbol{\hat{\beta}}(\tau) \) differ from \( \boldsymbol{\beta}_{0} \) only in their intercept terms. Similarly, the general regression HQER for $0 \leq \gamma \leq 1$ shares this aforementioned property as it will be  discussed  in Remark $\ref{remark 5}$. 

\begin{remark}
If the error terms $\varepsilon_i$ and $\boldsymbol{x}_i$ are not necessarily independent (heteroscedasticity), the convergence in probability for the
slope coefficients will depend on $f(\varepsilon_i,\boldsymbol{x}_i)$, i.e., the joint distribution of $\varepsilon_i$ and
$\boldsymbol{x}_i,$ which means that the convergence in $(\ref{cvv})$ doesn't hold (\cite{koenker_regression_1978} and \cite{newey_asymmetric_1987}). This result is  consistent with a similar one in \cite{jiang_kth_2021}.
\end{remark}

%----------------------------------------------------------------
 \subsection{Algorithm}\label{subsection 2.2}
%----------------------------------------------------------------
 To apply the Newton-Raphson algorithm, we need differentiability conditions. However, the function related to quantiles is not differentiable at $0$, and the function related to expectiles is not twice differentiable at $0$. Write the residual 
$e_{i}({\boldsymbol{\beta}})=y_{i}-\boldsymbol{x}_{i}^{\top}{\boldsymbol{\beta}}, \, i=1,\ldots,n.$ Assuming that the residuals \( e_i(\boldsymbol{\beta}) \) satisfy \( e_i(\boldsymbol{\beta}) < 0 \) or \( e_i(\boldsymbol{\beta}) > 0 \), we can calculate the first and second derivatives of \( T_n(\boldsymbol{\beta}, \tau, \gamma) \) defined in $(\ref{total asymmetric creq error})$ as follows. The first derivative of \( T_n(\boldsymbol{\beta}, \tau, \gamma) \) with respect to \( \boldsymbol{\beta} \) is given by:

$$
\nabla T_{n}(\boldsymbol{\beta},\tau,\gamma)=\frac{1}{n} \sum_{i=1}^{n} \boldsymbol{x}_{i} \varphi_{\tau}(y_{i}-\boldsymbol{x}_{i}^{\top}\,\boldsymbol{\beta}),
$$  
where

\begin{equation}\label{varphi}
\varphi_{\tau}(s)=\{ (-1)^{(1-\gamma)\mathds{1}\{ \,s>0\,\} } -2\gamma s\} \Psi_{\tau}(s), 
\end{equation}
with $\Psi_{\tau}(\cdot)$ is defined in \eqref{psi}.\\
The second derivative of $T_{n}(\boldsymbol{\beta},\tau,\gamma)$ with respect to $\boldsymbol{\beta}$ can be expressed as

$$\nabla^2 T_{n}(\boldsymbol{\beta},\tau,\gamma)=\frac{1}{n}\sum_{i=1}^{n}\boldsymbol{ x}_{i}\, \boldsymbol{x}_{i}^{\top}\Delta_{\tau}(y_{i}-\boldsymbol{x}^{\top}_{i}\, \boldsymbol{\beta}),
$$
where $\Delta_\tau(s)=2\gamma \Psi_\tau(s).$ 
So we get a Newton–Raphson updating formula
\begin{equation*}
\label{raphson*}
\boldsymbol{\hat{\beta}}^{(t)}=\, \boldsymbol{\hat{\beta}}^{(t-1)}-\left \{\nabla^2 T_{n}(\,  \boldsymbol{\hat{\beta}}^{(t-1)},\tau,\gamma)  \right \}^{-1}  \nabla T_{n}(\boldsymbol{\hat{\beta}}^{(t-1)},\tau,\gamma).
\end{equation*}
%--------------------------
For any $\tau\in(0,1)$, we choose the estimate of the corresponding least squares 
regression or least absolute regression as the iterative initial value, $\boldsymbol{\hat{\beta}}^{(0)}$.\\ 
Because we
use the condition that none of the residuals is such that $e_{i}(\boldsymbol{\beta})=0,$ there may be some
problems in the algorithm in practice, but this happens with almost zero probability if
the  data is from a continuous distribution and the sample size $n$ is larger than the dimension $p$ (\cite{jiang_kth_2021}). For our empirical study in Section \ref{section 4}, we use the \texttt{R} function \texttt{optim} to estimate the parameter vector of HQER. More stable and efficient algorithms based on the majorize-minimize principle  (see, e.g., \cite{hunter_quantile_2000} and \cite{mkhadri2017coordinate}) or the smoothing of the quantile function  could be considered (see, e.g., \cite{horowitz1998bootstrap}, \cite{fernandes2021smoothing}, and \cite{he2023smoothed}). These are left for future work for variable selection in high-dimensional settings where the Newton-Raphson algorithm is not computationally efficient.

\section{Theoretical analysis for HQER} \label{Sec 3}
Fundamental and standard asymptotic properties of HQER are prensetd in this section.
The theoretical development focuses more on HQER for $\gamma \in (0,1)$. For $\gamma = 0$ and  $\gamma = 1$, HQER reduces to quantile and expectile regression, respectively. The theoretical results for these two methods are well established in the literature.

\noindent
In the regression setting, let the vector \( \boldsymbol{\tilde{\beta}}_{0}(\tau) \) be the minimizer of the HQER population-level loss function
$ 
\mathbb{E}[ C_{\tau}^{\,\gamma}(y_{i} - \boldsymbol{x}_{i}^{\top} \boldsymbol{\beta}) - C_{\tau}^{\,\gamma}(y_{i})],
$
with respect to \( \boldsymbol{\beta} \). Its value depends on the distribution of \( y_{i} \) given \( \boldsymbol{x}_{i} \).  We denote 
%---------------------------------------
   $
\boldsymbol{\hat{\beta}}(\tau,\boldsymbol{y},\boldsymbol{x})=\arg\min\limits_{\boldsymbol{\beta}\, \in \mathbb{R}^{p}} T_{n}(\boldsymbol{\beta},\tau,\gamma)
   $
%-----------------------------------------
 as an estimator of $\boldsymbol{\tilde{\beta}}_{0}(\tau)$ given data $\boldsymbol{y}=(y_{1},\dots,y_{n})$ and $\boldsymbol{x}^{\top}=(\boldsymbol x_{1},\dots,\boldsymbol x_{n}).$ 
 
%------------------------------------------------------------------------------
 \subsection{Some basic properties for HQER expectile} 
%------------------------------------------------------------------------------
Some basic properties of HQER expectiles and their regression estimators are presented below.
%In the following, we presentsome basic properties of HQER expectiles and their regression estimators.

\begin{theorem}\label{thorem 1}
Let \( Y \) be a random variable as  in Definition $\ref{definition 1}$, and let \( \xi(\tau) \) denotes its corresponding \(\tau\)-\(\gamma\)th HQER expectile. Then, for each \( (\tau, \gamma) \in (0,1) \times (0,1), \) one has

\begin{enumerate}[label=(\roman*)]
\item \label{1-theorem 1}
${\xi}(\cdot )$ exists and is unique.
\item \label{2-theorem 1}
The function ${\xi}(\cdot),$ is strictly monotonically increasing.
\item \label{3-theorem 1}
For $\tilde{Y} = Y+a$, where $a\in \mathbb{R}$, $\xi(\tau)$ is the $\tau$-$\gamma$th HQER  expectile of $Y,$ and the $\tau$-$\gamma$th HQER expectile ${\tilde{\xi}}(\tau)$ of $\tilde{Y}$ satisfies
 $\tilde{\xi}(\tau)= \xi(\tau)+a $.
\end{enumerate}
\end{theorem}

\noindent
In the regression framework, the HQER estimator possesses the following properties.
\begin{theorem} \label{theorem 2}

Denote 
%---------------------------------------
   $
\boldsymbol{\hat{\beta}}(\tau,\boldsymbol{y},\boldsymbol{x})=\arg\min\limits_{\boldsymbol{\beta}\, \in \mathbb{R}^{p}} T_{n}(\boldsymbol{\beta},\tau,\gamma)
   $
%-----------------------------------------
 as an estimator of $\boldsymbol{\tilde{\beta}}_{0}(\tau)$ given data $\boldsymbol{y}=(y_{1},\dots,y_{n})$ and $\boldsymbol{x}^{\top}=(\boldsymbol x_{1},\dots,\boldsymbol x_{n}).$ Then, for any vector $b\in \mathbb{R}^p$ and any nonsingular matrix $\boldsymbol{A}\in \mathbb{R}^{p\times p},$ one has
\begin{enumerate}[label=(\roman*)]
\item $\boldsymbol{\hat{\beta}}(\tau,\boldsymbol{y}+\boldsymbol{x}^{\top} \boldsymbol{b},\boldsymbol{x})=\,\boldsymbol{\hat{\beta}}(\tau,\boldsymbol{y},\boldsymbol{x})+\boldsymbol{b},$\\
\item \label{theorem 2-2}$\boldsymbol{\hat{\beta}}(\tau,\boldsymbol{y},\boldsymbol{x}^{\top} \boldsymbol{A})=\,\boldsymbol{A^{-1}} \boldsymbol{\hat{\beta}}(\tau,\boldsymbol{y},\boldsymbol{x})$.\\
\end{enumerate}
\end{theorem} 
\noindent
The proof of Theorems \ref{thorem 1} and \ref{theorem 2} is postponed to the Appendix.

\begin{remark}\label{remark scale inv}
\noindent
Notice that HQER inherits the location equivariance property from its particular cases, QR and ER, as stated in $\ref{3-theorem 1}$ of Theorem \ref{thorem 1}. Although it lacks the scale equivariance property (that is, for $\tilde{Y} = sY$, with $s>0$, in general $\tilde{\xi}(\tau) \neq s\xi(\tau)$), in the regression setting, HQER is invariant to linear changes of scale on the covariates. In fact, property $\ref{theorem 2-2}$ of Theorem $\ref{theorem 2}$ states that invariance holds for any full-rank linear transformation of the $\boldsymbol{x}$ variables.

\end{remark}

%-----------------------------------------------------------------------------------------
\subsection{Large sample properties of $\tau$-$\gamma$th HQER estimator} \label{subsection 3.2}
%------------------------------------------------------------------------------------------
Under the following assumptions, the asymptotic theory for the HQER estimators is considered.
For a matrix \( \boldsymbol{A} = \{ a_{i,j} \} \), where \( i = 1, \ldots, n \) and \( j = 1, \ldots, p \), define the norm of \( \boldsymbol{A} \) as
\[
\| \boldsymbol{A} \| = \max_{(i,j) \in \{1, \ldots, n\} \times \{1, \ldots, p\}} \lvert a_{i,j} \rvert.
\]

\begin{assumption} \label{assumption 1}
The sample $\boldsymbol{z}_i=(y_i,\boldsymbol x_i)\in\mathbb{R}\times\mathbb{R}^p,$ $(i=1,\ldots ,n)$ are i.i.d. with the same law as $\boldsymbol{Z}=(Y,\boldsymbol{X})$ and have a probability density function $f(y|\boldsymbol{x})g(\boldsymbol{x})$ with respect to a measure $\boldsymbol{d}_{z}=d \times \boldsymbol{d}_{x}$, with $\boldsymbol{d}_{x}$ being the measure related to the continuous  $g(\boldsymbol{x})$ and $d$ the Lebesgue measure on $\mathbb{R}$. 
\end{assumption}

\begin{assumption} \label{assumption 2} 
%\begin{enumerate}[label=(\roman*)]
%\item 
The cumulative distribution function \(F(y \lvert \boldsymbol{x})\) of \(Y\) given \(\boldsymbol{x}\) is absolutely continuous, with a continuous and strictly positive bounded density \(f(y \lvert \boldsymbol{x})\) in \(y\) for almost all \(\boldsymbol{x}\), and is well-defined in \(\boldsymbol{X}^\top \boldsymbol{\tilde{\beta}}_0(\tau)\), i.e.,
\[
F(y \mid \boldsymbol{x}) = \int_{-\infty}^y f(t \mid \boldsymbol{x}) \, \mathrm{d}t, \quad f(y \mid \boldsymbol{x}) > 0, \quad \int_{-\infty}^\infty f(y \mid \boldsymbol{x}) \, \mathrm{d}y = 1,\quad \text{and} \quad f(y \mid \boldsymbol{x}) \leq c_0,
\]
for some constant \( c_0 > 0 \) and for almost all \( \boldsymbol{x} \). Furthermore, \( F(y \mid \boldsymbol{x}) \) is well-defined for all \( y \in \mathbb{R} \), particularly for  \( y=\boldsymbol{X}^\top \boldsymbol{\tilde{\beta}}_0(\tau) \).

%\end{enumerate}

\end{assumption}

\begin{assumption}\label{assumption 3}
There is a constant $c_1>0$  such  that $\mathbb {E} (\norm{  \boldsymbol{Z}} ^{4}) <c_1,$ i.e., \( \int_{\mathbb{R}^p} \int_{-\infty}^{+\infty} \max(|y|,\norm{\boldsymbol x})^{4}f(y|\boldsymbol{x})g(\boldsymbol{x}) \mathrm{d}\boldsymbol{x}\mathrm{d} y<c_1.\)
\end{assumption}

\begin{assumption}\label{assumption 5}
$\mathbb{E}(\boldsymbol{x}_{i}\boldsymbol{x}_{i}^{\top})$ is a nonsingular matrix.
\end{assumption}

\begin{remark}

Assumption \(\ref{assumption 1}\) is similar to assumption $1$ in \cite{newey_asymmetric_1987}. Assumption $\ref{assumption 3}$ is less stringent than assumption $3$ in \cite{newey_asymmetric_1987} and is easier to be satisfied than assumption $3$ in  \cite{jiang_kth_2021}. Assumption $\ref{assumption 5}$ represents a common constraint in dealing with regression problems. 
\end{remark}
%--------------------------------------------------------------------------
\begin{theorem} \label{theorem 3}

Set  $\boldsymbol{\tilde{\beta}}_{0}(\tau)=\arg\min\limits_{\boldsymbol{\beta}\in \mathbb{R}^p}\mathbb{E}\left [ C_{\tau}^{\,\gamma}(Y-\boldsymbol{X}^{\top} \boldsymbol{\beta})-C_{\tau}^{\,\gamma}(Y) \right ]
.$ \\
Under Assumptions $\ref{assumption 1}$, $\ref{assumption 2}$, $\ref{assumption 3}$, and $\ref{assumption 5}$, \(\boldsymbol{\tilde{\beta}}_{0}(\tau)\) exists and is unique. Furthermore, \(\boldsymbol{\hat{\beta}}(\tau) \xrightarrow[]{\mathcal{P}} \boldsymbol{\tilde{\beta}}_{0}(\tau)\).

\end{theorem}
%---------------------------------------------------------------------------
\begin{remark}\label{remark 5}
In the classical linear model \eqref{model} and Theorem $\ref{theorem 3}$, the homoscedasticity means that $\varepsilon_{i}$ is independent
of $\boldsymbol{x}_{i}$ and only the location of $y_{i}$ depends on $\boldsymbol{x}_{i}$. The  property $\ref{3-theorem 1}$ in Theorem $\ref{thorem 1}$
implies that $\xi ( \tau, y_{i} )=\boldsymbol{x}_{i}^{\top} \boldsymbol{\beta}_{0} + \xi(\tau,\varepsilon),$  
where $\xi(\tau,y_{i})$ is the $\tau$-$\gamma$th HQER expectile of $y_{i}$. The linearity of the $\tau$-$\gamma$th HQER expectile yields 
$\xi(\tau, y_{i}) = \boldsymbol{x}_{i}^{\top}\, \boldsymbol{\tilde{\beta}}_{0}(\tau),$ 
where $\boldsymbol{\tilde{\beta}}_{0} (\tau) =\boldsymbol{\beta}_{0} +\xi(\tau,\varepsilon )\, \boldsymbol{e}_{1}.$ Only the intercept coefficient in the expression of $\xi(\tau, y_{i} )$ varies with $\tau$.
\end{remark}
%---------------------------------------------------------------------
\noindent
To establish the asymptotic normality of the HQER estimator in the following theorem, we introduce Assumption \ref{assumption 6}, which is commonly used in quantile and expectile regression. Specifically, Assumption \ref{assumption 6}-\ref{assumption-d}) is employed to satisfy the Lindeberg-Feller condition for the HQER.

\begin{assumption} \label{assumption 6}  
\begin{enumerate}[label=(\roman*)]
\item  \label{assumption 6-i}
The errors \(\{\varepsilon_i\}_{i=1}^n\) are i.i.d with a common cumulative distribution function \(F_{\varepsilon}(\cdot)\).  

\item \label{assumption-ii}
There exist positive definite matrices \(\boldsymbol{K}\) and \(\boldsymbol{J} \) such that
\begin{enumerate}
    \item  \label{assumption-a}
    \(\frac{1}{n} \sum_{i=1}^n \boldsymbol{x}_i \boldsymbol{x}_i^\top \to \boldsymbol{K},\)
    \item \label{assumption-b}
    \( \frac{1}{n} \sum_{i=1}^n f(\boldsymbol {x}^\top_i\boldsymbol{\tilde {\beta}}_0(\tau)\mid \boldsymbol{x}_i) \boldsymbol{x}_i \boldsymbol{x}_i^\top \to \boldsymbol J, \)
    \item \label{assumption-c}
    \(\max_{i=1,\ldots,n} \frac{\norm{\boldsymbol{x}_i}}{\sqrt{n}} \to 0.\)
    \item \label{assumption-d}
    \(  \frac{1}{n}\sum_{i=1}^{n} \mathbb{E}\{[(1-\gamma)\cdot a_\tau(\varepsilon_i) -\varepsilon_i\Delta_\tau(\varepsilon_i)]^2 \boldsymbol{x}_i \boldsymbol{x}_i^\top  \} \to 0\), with $a_\tau(\varepsilon_i)=\tau-\mathds{1}\{\varepsilon_i<0\},$ and $\Delta_\tau(\varepsilon_i)=2\gamma\Psi_\tau(\varepsilon_i).$
\end{enumerate}

\end{enumerate}

\end{assumption}

\begin{theorem}\label{theorem 4}
 If Assumptions $\ref{assumption 1}$, $\ref{assumption 2}$, $\ref{assumption 3}$, and $\ref{assumption 6}$ are satisfied, then

$$\sqrt{n} \boldsymbol{\Sigma}^{-1/2} \left(\boldsymbol{\hat{\beta}}(\tau)-\boldsymbol{\tilde{\beta}}_{0}(\tau)\right)\xrightarrow[]{\mathcal D    } \mathcal{N} \left( \boldsymbol{0}_{p},\mathbb{I}_p \right),$$
where
  $
\boldsymbol{\Sigma}=:\boldsymbol{\Sigma}(\gamma,\tau)=\boldsymbol{\tilde J^{-1}}\,\boldsymbol{\tilde K}\,(\boldsymbol{\tilde J^{-1})}^{\top}, \quad
\boldsymbol{\tilde{J}}:=\boldsymbol{\tilde{J}}(\gamma,\tau)= (1-\gamma)\cdot \boldsymbol{J} +2 \gamma\, d(\tau)\cdot \boldsymbol{K}, $  and  $\quad \boldsymbol{\tilde K}:=\boldsymbol{ K}(\gamma,\tau)=\varsigma{(\gamma,\tau)}\cdot\boldsymbol{K},$ 
such that $d(\tau) = (1-\tau)F_\varepsilon(0) + \tau\big(1-F_\varepsilon(0)\big),$ $\varsigma(\gamma,\tau)=\mathbb{V}\mathrm{ar}((1-\gamma)\cdot a_\tau(\varepsilon) -\varepsilon\Delta_\tau(\varepsilon)),$   $a_\tau(\varepsilon)=\tau-\mathds{1}\{ \varepsilon<0 \},$ $\Delta_\tau(\varepsilon)=2\gamma\Psi_\tau(\varepsilon)$ and $\mathbb{I}_p$ represents the identity matrix of size \( p \times p \), which has 1's on the diagonal and 0's elsewhere.

\end{theorem}

\begin{remark}
 Obviously, for \(\gamma = 0\) and \(\gamma = 1\), the results in the two theorems correspond to those in \cite{koenker_quantile_2005} and \cite{newey_asymmetric_1987}, respectively. In Theorem $\ref{theorem 4}$ we see that the asymptotic covariance matrix depends on the error density, which highlights the importance of the estimation of  the error distribution accurately for robust statistical inference. Methods such as kernel density estimation (see, e.g., \cite{efron1993introduction}) are commonly used. To avoid estimating the density, bootstrap techniques are often employed, depending on the error structure and model assumptions (see, e.g., \cite{silverman1998density}). In our numerical examples, we relied on bootstrapping to calculate the covariance of the proposed estimator.

\end{remark}

%=====================================================
\section{Numerical Experiments}\label{section 4}
%======================================================
 We conduct a simulation study to compare the asymptotic efficiency of  the HQER estimator  with the maximum likelihood estimator (MLE) for scale-location and location-shift models. We also investigate the HQER method for analyzing data on child malnutrition in India consisting of $n=4000$ observations described by six covariates.

 \subsection{Asymptotic Variance Comparisons} \label{subsection 4.1}
Location-shift models and scale-location models are fundamental to statistical modeling, especially in regression analysis and robust estimation. This section examines the variations in the asymptotic efficiency of estimates in HQER expectile regression relative to MLE for different values of $\gamma$, and compares these efficiencies to those of QR, ER, and $k$th power expectile regression.

%------------------------------------------
\subsubsection{Scale-Location Models}\label{section4.1.1}
%------------------------------------------
We consider a simple model in which there are no covariates. The data of $n$ observations
are generated by a scale-location family
\begin{equation}\label{zY MODEL}
  y_{i}=\mu +\sigma \varepsilon_{i}, \;\;  i=1,2,\ldots,n,  
\end{equation}
where $\varepsilon_{1}$, $\varepsilon_{2},\ldots, \varepsilon_{n}$ are i.i.d. random variables with mean $0$ and variance  $1$  following a known probability density function $f(\cdot)$.\\
Let  
\begin{align*}
   \beta^{\gamma}(\tau, y) = \arg\min\limits_{\theta \in \mathbb{R}} \mathbb{E} \left[ C_{\tau}^{\,\gamma}(y - \theta) - C_{\tau}^{\,\gamma}(y) \right]\; \textit{and}\; \beta^{\gamma}(\tau, \varepsilon) = \arg\min\limits_{\theta \in \mathbb{R}} \mathbb{E} \left[ C_{\tau}^{\,\gamma}(\varepsilon - \theta) - C_{\tau}^{\,\gamma}(\varepsilon) \right] 
\end{align*}
denote the true $\tau$-$\gamma$th HQER expectiles of $y$ and $\varepsilon$, respectively.\\
For $\gamma = 0$, which corresponds to standard quantiles, we use $\alpha$ instead of $\tau$ as the notation for the associated probability. Comparing HQER with QR, ER, and the $k$th power expectile, we establish a relationship between $\tau$ and $\alpha$. Specifically, for every $\alpha$, we define suitable values $\tau_{\gamma}, \tau_k, \tau_1$, and $\tau_0$ such that
\begin{equation*}
    \beta^{\gamma}(\tau_{\gamma}, \varepsilon) = \delta_\varepsilon(\alpha), \quad
    \beta^{k}(\tau_k, \varepsilon) = \delta_\varepsilon(\alpha), \quad
    \beta^{1}(\tau_1, \varepsilon) = \delta_\varepsilon(\alpha), \quad
    \tau_0 := \alpha.
\end{equation*}
where
\begin{itemize}
    \item $\beta^k(\tau_k, \varepsilon)$ denotes the $k$th power expectile of $\varepsilon$, i.e., \begin{align}
       \beta^{k}(\tau_{k}, \varepsilon) := \arg\min\limits_{\theta \in \mathbb{R}} \mathbb{E} \left[ Q_{\tau,k}(\varepsilon - \theta) - Q_{\tau,k}(\varepsilon) \right],\; k\in(1,2],\notag
    \end{align} where $Q_{\tau,k}(\cdot)$ is given in equation \eqref{kth power loss}.
    \item $\beta^1(\tau_1, \varepsilon)$ is the standard expectile of $\varepsilon$,
    \item $\delta_\varepsilon(\alpha)$ represents the $\alpha$-th quantile of $\varepsilon$.
\end{itemize}
For notational convenience, we denote \( \beta^\gamma(\tau_{\alpha,\gamma}, \varepsilon) \) simply as \( \beta^\gamma(\tau_{\gamma}, \varepsilon) \) when there is no ambiguity.\\
Set $\sigma=1,$ then, the $\tau$-$\gamma$th HQER expectile $\beta^{\gamma}(\tau_{\,\gamma},y)$ of $y,$
can be calculated as
$\beta^{\,\gamma}(\tau_{\,\gamma},y)=\mu+\sigma\, \beta^{\gamma}(\tau_{\gamma},\varepsilon)=\mu+ \beta^{\gamma}(\tau_{\gamma},\varepsilon).$
The estimator of $\beta^{\gamma}(\tau_{\,\gamma},y),$ based on the empirical HQER loss function, is denoted by $\hat{\beta}^{\,\gamma}(\tau_{\,\gamma},y)$. We compare the asymptotic variance of $\hat{\beta}^{\gamma}(\tau_{\gamma},y)$ with that one of the MLE estimator of $\beta^{\gamma}(\tau_{\,\gamma},y)$ which we denote by  $\hat{\beta}_{\textit{MLE}}^{\gamma}(\tau_{\gamma},y)$. It is useful to have some notations about the MLE. The Fisher information matrix for estimating $(\mu, \sigma$) in $(\ref{zY MODEL})$ is
$$
\mathbb{F}_{(\mu, \sigma)}=\begin{pmatrix}
F_{11}&F_{12}\\
F_{12}&F_{22}
\end{pmatrix},
$$
where $F_{11} = \sigma^{-2}\cdot\mathbb{E} [  g(\varepsilon)^2 ]$, $ F_{12}=F_{21}=\sigma^{-2} \cdot \mathbb{E} [ g(\varepsilon )\cdot \varepsilon ],$ $F_{22}=\sigma^{-2}\cdot \mathbb{E}[ g(\varepsilon)\cdot \varepsilon ]^{2}$ and  $g(s)=\nabla_{s}\log(f(s)),\; s\in \mathbb{R}$. Let $\hat{\mu}_{\textit{MLE}}$ and $\hat{\sigma}_{\textit{MLE}}$ be the MLEs of $\mu$ and $\sigma$, respectively. Then, the asymptotic variance ($\mathbb{V}\mathrm{ar}_{a}$)
$$
\mathbb{V}\mathrm{ar}_{a} (\hat{\beta}_{\textit{MLE}}^{\,\gamma}(\tau_{\,\gamma},y))= \lim_{n \to +\infty} n\cdot \mathbb{V}\mathrm{ar}(\hat{\beta}_{\textit{MLE}}^{\gamma}(\tau_{\gamma},y))
$$ 
of the MLE, 
$\hat{\beta}_{\textit{MLE}}^{\,\gamma} (\tau_\gamma,y) = \hat{\mu}_{\textit{MLE}}+\hat{\sigma}_{\textit{MLE}}\cdot {\beta}^{\gamma}(\tau_{\,\gamma},\varepsilon)$ is defined by 

\begin{equation}\label{eq mle}
   \mathbb{V}\mathrm{ar}_{a} (\hat{\beta}_{\textit{MLE}}^{\,\gamma} (\tau_\gamma,y) )=\sigma^{2}\cdot  \left[ F_{22}-2\ F_{12}\cdot {\beta}^{\gamma}(\, \tau_{\gamma},\varepsilon) +F_{11}\cdot ({\beta}^{\gamma}(\tau_{\,\gamma},\varepsilon))^{2}   \right] /( F_{11}\cdot F_{22}-F^{2}_{12} )\cdot
\end{equation}
Referring to \cite{newey_asymmetric_1987}, the asymptotic variance of the expectile
regression estimator ${\hat{\beta}}^{\,1}(\tau_{\,1},y)$ is
%\cite{newey_asymmetric_1987} is
$$
 \mathbb{V}\mathrm{ar}_{a}  ({\hat{\beta}}^{\,1}(\tau_{\,1},y)) =\sigma^{2}\cdot\mathbb{E}\left[ \Psi_{\tau_{\, 1}}(\,\varepsilon-{\beta}^{1}(\tau_{1},\varepsilon))\right]^{2} / \left[1-\tau_{\, 1}+(2\tau_{\, 1}-1)\cdot \mathbb{P}( \varepsilon > {\beta}^{\,1}(\tau_{1},\varepsilon)) \right]^{2}.
$$
In light of Theorem \ref{theorem 4}, the asymptotic variance of the $\tau$-$\gamma$th HQER expectile regression estimator  is defined by $\mathbb{V}\mathrm{ar}_{a}({\hat{\beta}}^{\,\gamma}(\tau_{\,\gamma},y))=\frac{\mathbb{V}\mathrm{ar}((1-\gamma)\cdot a_\tau(\varepsilon) -\varepsilon\Delta_\tau(\varepsilon))}{[(1-\gamma)f(\delta_\varepsilon(\alpha))+2\gamma d(\tau)]^2}$, where $d_\tau(\cdot)$, $a_\tau(\cdot)$, and $\Delta_\tau(\cdot)$ are defined in Theorem $\ref{theorem 4}$.\\
From  Theorem $4$  of \cite{jiang_kth_2021}, the asymptotic variance of the $k$th power expectile regression estimator $\hat{\beta}^k(\tau_k,y)$ is  
$$
\mathbb{V}\mathrm{ar}_{a} (\hat{\beta}^{\,k}(\tau_{\,k},y))=\sigma^{2}\cdot  \mathbb{E}\left [  \Psi_{\tau_k}^{2} (\varepsilon-\beta^{k}(\tau_k,\varepsilon))\, \lvert   \varepsilon-\beta^{k}(\tau_k,\varepsilon) \rvert^{2(k-1)} \right ]               /  (\mathbb{E}[ \Upsilon (\varepsilon,\beta^{k}(\tau_k,\varepsilon)) ] )^{2},
$$
with 
  $\Upsilon(s,y)=  Q''_{\tau_k,k}(s-y)/k, \; (s,y)\in \mathbb{R}^2, \; k \in (1, 2],$ 
where $Q''_{\tau_k,k}(s)=k(k-1)|s|^{k-2}\Psi_{\tau_k}(s),\; s\in \mathbb{R}.$\\
According to \cite{koenker_robust_1982}, the asymptotic variance of the quantile regression estimator $\hat{\beta}^{0}(\alpha,y)$ is 
 $\mathbb{V}\mathrm{ar}_{a}(\hat{\beta}^{0}(\alpha,y)\,)= \sigma^{2}\cdot \alpha(1-\alpha)/f^{2}(\delta_{\varepsilon}(\alpha)),   $
where $\delta_\varepsilon(\alpha)$ is the $\alpha$-th quantile of $\varepsilon$.  \\ 
We assume that $\varepsilon_{i}$ in \eqref{zY MODEL} comes from one of three types of distributions:
\begin{enumerate}[label=(\roman*)]
    \item $\textbf{Case I:}$ $\mathcal{N}(0,1),$ the standard normal distribution;
    \item $\textbf{Case II:}$ $t(\nu)$, the Student distribution with  degree of freedom $\nu=3$; 
    \item $\textbf{Case III:}$ $\chi^{2}(\nu)$, the Chi-square distribution with  degree of
freedom $\nu=6$.
\end{enumerate}
The rationale behind the selection of these distributions is rooted in their widespread utility: the normal distribution stands out as one of the most commonly used distributions, the Student distribution is known for its heavy-tail properties and is used extensively in various domains such as finance, while the Chi-square distribution is emblematic of skewed distributions. In each scenario, we  elucidate the efficiency behavior of our method alongside these distributions.  For each case, we will give the efficiency change of our method
 with $\gamma$.\\
For the normal, the Student with degree of freedom $\nu=3$ and the Chi-squared distributions with degree of freedom $\nu=6$, equation \eqref{eq mle} can be
further written as
$\mathbb{V}\mathrm{ar}_{a}(\hat{\beta}_{\textit{MLE}}^{\gamma}(\tau_{\,\gamma},y))=\sigma^{2}\cdot(1+\delta_\varepsilon(\alpha)^2) / 3, \quad \mathbb{V}\mathrm{ar}_{a}(\hat{\beta}_{\textit{MLE}}^{\gamma}(\tau_{\,\gamma},y))=\sigma^{2} \cdot \frac{\nu + 3}{\nu + 1} \left( 1 + \frac{\left(\delta_\varepsilon(\alpha)\right) ^2}{3} \right), \nu>0$, 
and  
$\mathbb{V}\mathrm{ar}_{a}  (\hat{\beta}_{\textit{MLE}}^{\gamma}(\tau_{\,\gamma},y))=\sigma^{2}\cdot ( 1 + \frac{\nu}{2} - \delta_\varepsilon(\alpha) + \frac{\nu^2 -16\nu +52}{16(\nu -4)} \cdot \delta_\varepsilon(\alpha)^2 )/ [\frac{(\nu^2 -16\nu +52)\cdot (1 + \frac{\nu}{2})}{16(\nu -4)} - \frac{1}{4}] , \nu>4,$ respectively. \\
We compute the ratio between the asymptotic variances of the MLE estimates and those of the expectile, the $\tau$-$\gamma$th HQER expectile, quantile, and $k$th power expectile regression defined as follows
\[
\mathbb{A}\mathrm{RE}({\hat{\beta}}^{\,1}(\tau_{\,1}, y)) =
\cfrac{\mathbb{V}\mathrm{ar}_{a}({\hat{\beta}}_{\textit{MLE}}^{\,1}(\tau_{\,1}, y))}{
\mathbb{V}\mathrm{ar}_{a}({\hat{\beta}}^{\,1}(\tau_{\,1}, y))},
\quad
\mathbb{A}\mathrm{RE}({\hat{\beta}}^{\,\gamma}(\tau_{\,\gamma}, y)) =
\cfrac{\mathbb{V}\mathrm{ar}_{a}({\hat{\beta}}_{\textit{MLE}}^{\,\gamma}(\tau_{\,\gamma}, y))}{
\mathbb{V}\mathrm{ar}_{a}({\hat{\beta}}^{\,\gamma}(\tau_{\,\gamma}, y))},
\]
\[
\mathbb{A}\mathrm{RE}({\hat{\beta}}^{\,0}(\tau_0,y)) =
\frac{\mathbb{V}\mathrm{ar}_{a}({\hat{\beta}}_{\textit{MLE}}^{\,0}(\tau_0,y))}{
\mathbb{V}\mathrm{ar}_{a}({\hat{\beta}}^{\,0}(\tau_0,y))},
\quad
\mathbb{A}\mathrm{RE}({\hat{\beta}}^{\,k}(\tau_{\,k}, y)) =
\cfrac{\mathbb{V}\mathrm{ar}_{a}({\hat{\beta}}_{\textit{MLE}}^{\,k}(\tau_{\,k}, y))}{
\mathbb{V}\mathrm{ar}_{a}({\hat{\beta}}^{\,k}(\tau_{\,k}, y))}.
\]
This ratio represents the asymptotic relative efficiency ($\mathbb{A}\mathrm{RE}$) of the estimator with respect to the MLE.

\begin{table}[h]
\centering
\caption{Asymptotic relative efficiency ($\mathbb{A}\mathrm{RE}$) of Expectile Regression (ER), $\hat{\beta}^{\,1.00}(\tau_{\,1.00})$; Quantile Regression (QR), $\hat{\beta}^{\,0.00}(\tau_{\,0.00})$; $k$th Power Expectile Regression  ($k$thER ) for $k=1.50$, $\hat{\beta}^{\,1.50}(\tau_{\,1.50})$; and the HQER Expectile Regression (HQER), $\hat{\beta}^{\,\gamma}(\tau_{\,\gamma}),\, \gamma\in\{\frac{i}{10}\mid i=1,\ldots,9  \}$, in \textbf{Case I} ($\varepsilon \sim \mathcal{N}(0,1)$)}
\label{table 2}

\begin{tabular}{@{}lcccccc@{}}
\toprule
& &  & $\alpha\footnotemark[1]$&  &  &  \\
\cmidrule(lr){2-7}
$\hspace{1cm}  \mathbb{A}\mathrm{RE}(\cdot) $ & 0.55 or 0.45 & 0.63 or 0.37 & 0.7 or 0.3 & 0.83 or 0.17 & 0.92 or 0.08 & 0.97 or 0.03 \\
\midrule
$\mathbb{A}\mathrm{RE}\left(\hat{\beta}^{1.00}(\tau_{\,1.00})\right) \textit{(ER)}$ & \textbf{0.996} & 0.974 & 0.935 & 0.787 & 0.571 & 0.339 \\
$\mathbb{A}\mathrm{RE}\left(\hat{\beta}^{1.50}(\tau_{\,1.50})\right) \textit{(kthER)}$ & 0.925 & 0.912 & 0.889 & 0.788 & 0.610 & 0.388 \\
$\mathbb{A}\mathrm{RE}\left(\hat{\beta}^{0.90}(\tau_{\,0.90})\right) \textit{(HQER)}$ & \textbf{0.995} & \textbf{0.996} & \textbf{0.988} & \textbf{0.946} & \textbf{0.839} & \textbf{0.640} \\
$\mathbb{A}\mathrm{RE}\left(\hat{\beta}^{0.80}(\tau_{\,0.80})\right) \textit{(HQER)}$ & 0.972 & 0.975 & 0.970 & 0.904 & 0.732 & 0.540 \\
$\mathbb{A}\mathrm{RE}\left(\hat{\beta}^{0.70}(\tau_{\,0.70})\right) \textit{(HQER)}$ & 0.933 & 0.931 & 0.916 & 0.816 & 0.635 & 0.479 \\
$\mathbb{A}\mathrm{RE}\left(\hat{\beta}^{0.60}(\tau_{\,0.60})\right) \textit{(HQER)}$ & 0.885 & 0.873 & 0.848 & 0.727 & 0.558 & 0.429 \\
$\mathbb{A}\mathrm{RE}\left(\hat{\beta}^{0.50}(\tau_{\,0.50})\right) \textit{(HQER)}$ & 0.829 & 0.809 & 0.775 & 0.644 & 0.489 & 0.381 \\
$\mathbb{A}\mathrm{RE}\left(\hat{\beta}^{0.40}(\tau_{\,0.40})\right) \textit{(HQER)}$ & 0.768 & 0.740 & 0.698 & 0.564 & 0.422 & 0.328 \\
$\mathbb{A}\mathrm{RE}\left(\hat{\beta}^{0.30}(\tau_{\,0.30})\right) \textit{(HQER)}$ & 0.699 & 0.664 & 0.618 & 0.482 & 0.350 & 0.267 \\
$\mathbb{A}\mathrm{RE}\left(\hat{\beta}^{0.20}(\tau_{\,0.20})\right) \textit{(HQER)}$ & 0.622 & 0.580 & 0.529 & 0.392 & 0.267 & 0.192 \\
$\mathbb{A}\mathrm{RE}\left(\hat{\beta}^{0.10}(\tau_{\,0.10})\right) \textit{(HQER)}$ & 0.534 & 0.486 & 0.431 & 0.290 & 0.171 & 0.103 \\
$\mathbb{A}\mathrm{RE}\left(\hat{\beta}^{0.00}(\tau_{0.00})\right) \textit{(QR)}$ & 0.636 & 0.634 & 0.628 & 0.592 & 0.498 & 0.347 \\
\bottomrule
\end{tabular}
\footnotetext[1]{$\alpha$ denotes the $\alpha$-th quantile level such that $\delta_{\varepsilon}(\alpha)=\beta^\gamma(\tau_\gamma,\varepsilon)$.}
\end{table}

\begin{figure}%
    \centering
    \includegraphics[width=14cm]{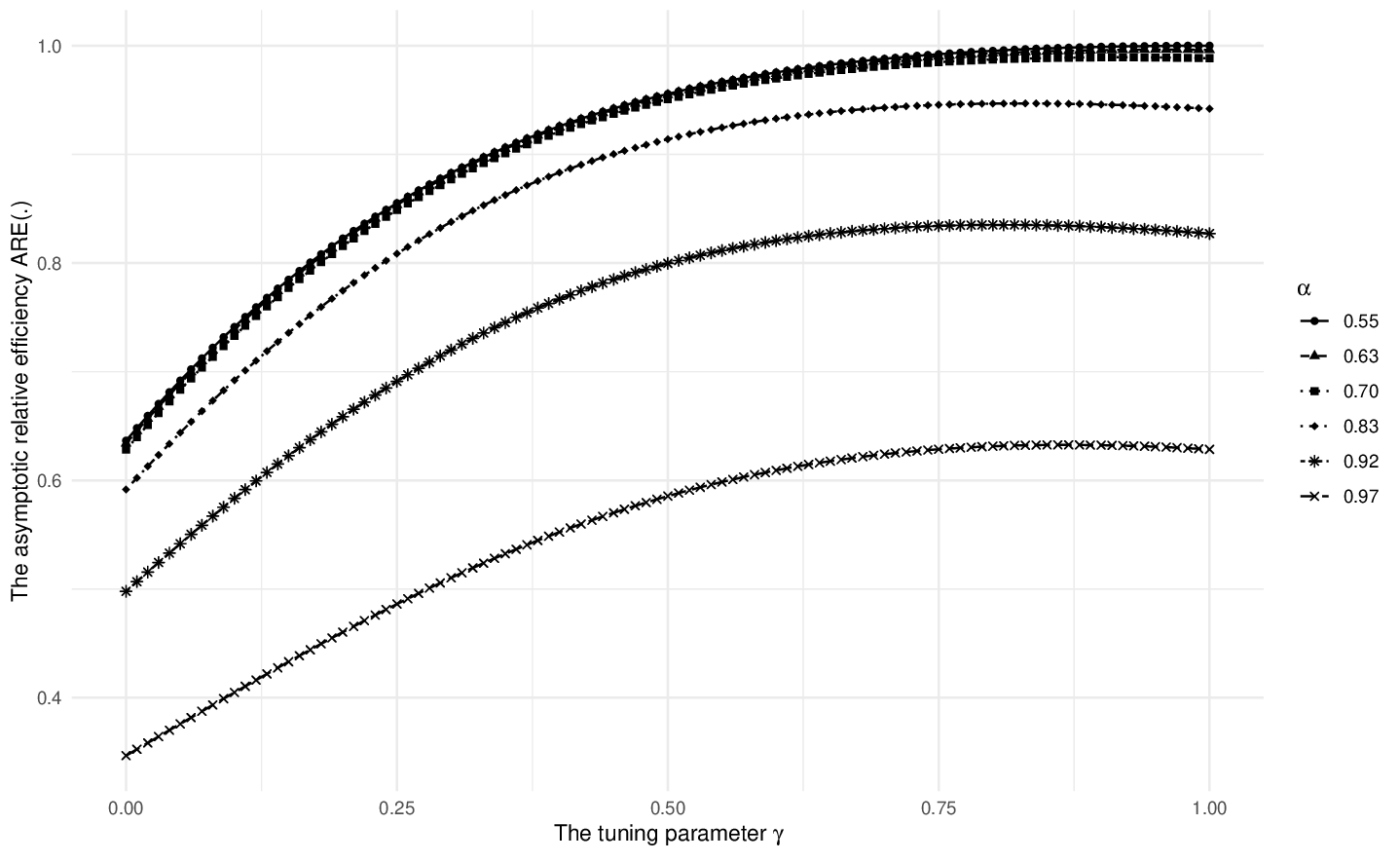}  %
   \caption{Asymptotic relative efficiency of the HQER expectile regression, $\hat{\beta}^{\,\gamma}(\tau_{\,\gamma})$, for the standard normal distribution (\textbf{Case I}), as the tuning parameter $\gamma$ varies in $(0,1)$}
\label{Figure normal asymptotic}
\end{figure}
\begin{figure}%
    \centering
       \includegraphics[width=14cm]{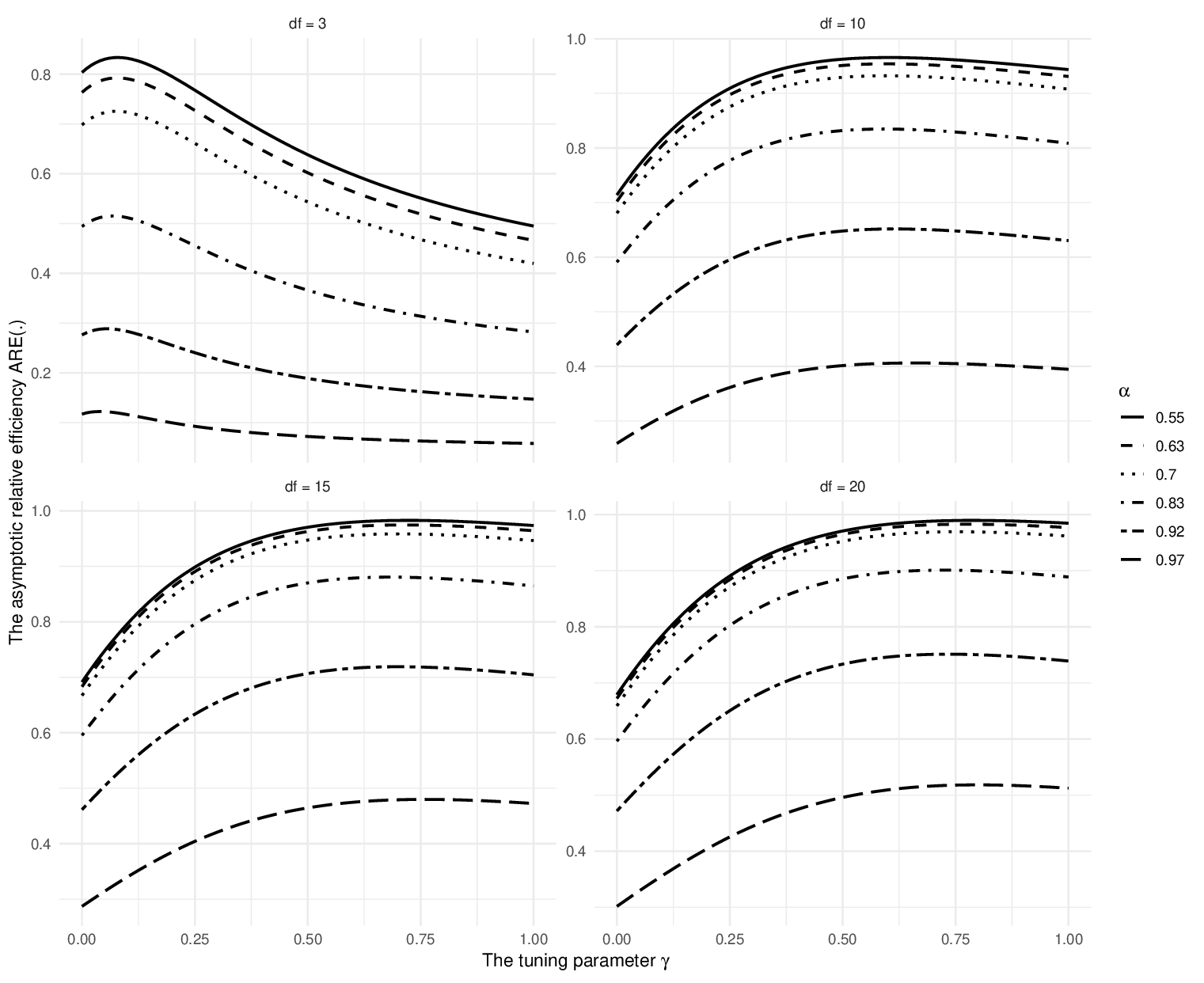}  %
    \caption{Asymptotic relative efficiency of  $\hat{\beta}^{\,\gamma}(\tau_{\,\gamma})$ for the Student  distribution ($\textbf{Case II}$), with degree of freedom (df=$\nu$) $\nu\in \{3,10,15,20\}$  as the tuning parameter $\gamma$ varies in $(0,1)$}%
    \label{Figure student}

\end{figure}
\begin{figure}
    \centering
           \includegraphics[width=15cm]{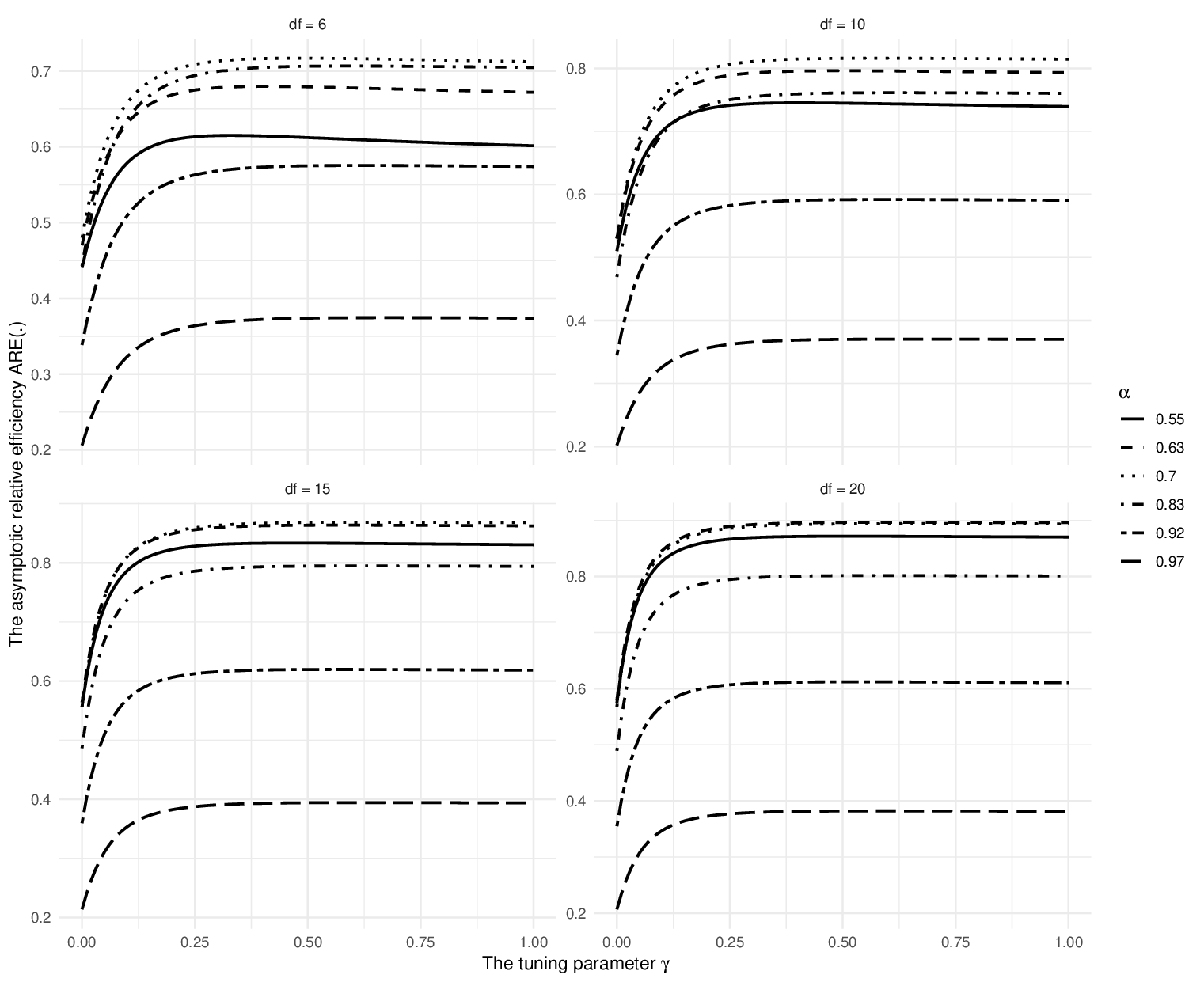}  %
    \caption{Asymptotic relative efficiency of  $\hat{\beta}^{\,\gamma}(\tau_{\,\gamma} )$ for the Chi-square distribution ($\textbf{Case III}$), with degree of freedom (df=$\nu$) $\nu\in \{6,10,15,20\}$  as the tuning parameter  $\gamma$ varies in (0,1)}%
    \label{Figure chi square}

\end{figure}

\noindent
For convenience, we suppress \(y\) in the terms \(\mathbb{A}\mathrm{RE}\) to ease notation.\\
The results are shown in Tables \ref{table 2}, \ref{ARE student 3}, and \ref{ARE chi squared 6} for the three distributions under different quantile levels of $\alpha$. The best performances are emphasized in solid  for clarity. HQER consistently outperforms its competitors across different quantile levels of \(\alpha\) for specific values of \(\gamma\), demonstrating superior efficiency in various cases. The results, summarized in Tables \ref{table 2}, \ref{ARE student 3}, and $\ref{ARE chi squared 6}$, highlight these trends across different distributions.  
For the normal distribution (Table \(\ref{table 2}\)), the efficiency of HQER for \(\gamma = 0.90\) surpasses that of QR, \(k\)th power expectile regression, and ER across different \(\alpha\) values (except possibly for \(\alpha = 0.55\), where HQER achieves approximately the same efficiency as ER), reaching the highest efficiency levels. Moreover, for \(\alpha = 0.97\) or \(0.03\), \(\mathbb{A}\mathrm{RE}\left(\hat{\boldsymbol{\beta}}^{1.50}(\tau_{\,1.50})\right) = 0.388\) is surpassed by \(\mathbb{A}\mathrm{RE} \left(\hat{\boldsymbol{\beta}}^{0.90}(\tau_{\,0.90})\right) = 0.640\). The efficiency trend accelerates as \(\alpha\) approaches \(0.55\) but may decline below that of QR as \(\gamma\) approaches \(0\), particularly when \(\alpha\) approaches \(1\). Furthermore, the efficiency of \(\hat{\beta}^{\,\gamma}(\tau_{\,\gamma})\) improves as \(\gamma\) increases from \(0\) to \(1\). These findings are further illustrated in Fig. \(\ref{Figure normal asymptotic}\).
 For the Student distribution (Table \(\ref{ARE student 3}\)), HQER with \(\gamma = 0.60\) achieves superior efficiency over all its competitors and is particularly effective for extreme quantiles (\(\alpha \in \{0.83\,\textit{or}\; 0.17,\ldots,0.97\,\textit{or}\; 0.03\}\)), which are crucial in estimation. As shown in Fig. \(\ref{Figure student}\), HQER maintains strong performance across different values of \(\gamma\) and degrees of freedom \(\nu \in \{3,10,15,20\}\), further demonstrating its robustness in capturing tail behaviors. For quantiles \(\alpha \in \{0.55\,\textit{or}\; 0.45,\ldots,0.70\,\textit{or}\; 0.30\}\), \( k \)th power expectile regression shows slightly better efficiency but remains overall less competitive than HQER for extreme quantiles.
 For the Chi-square distribution (Table \(\ref{ARE chi squared 6}\)), HQER demonstrates strong efficiency for all \(\alpha \in \{0.75, 0.83, 0.90, 0.92\}\), particularly for \(\gamma\) values around \(0.60\) and \(0.50\), where it significantly outperforms its competitors. This effect is especially pronounced for extreme quantiles, where precise estimation is crucial. As illustrated in Fig. \(\ref{Figure chi square}\), the efficiency of HQER stabilizes quickly as \(\nu\) increases, reinforcing its effectiveness in heavy-tailed distributions. For quantiles in $\{ 0.03,0.08,0.17,0.25,0.33,0.37,0.45,0.55,0.63 \}$, \( k \)th power expectile regression shows slightly better performance than the alternative methods.
Across all cases, we observe that selecting an appropriate value of \(\gamma\) is crucial for maximizing efficiency. A detailed discussion on this choice is provided in Section $\ref{section 5}$.

\begin{table}[h]
\centering
\caption{Asymptotic relative efficiency ($\mathbb{A}\mathrm{RE}$) of Expectile Regression (ER), $\hat{\beta}^{\,1.00}(\tau_{\,1.00})$; Quantile Regression (QR), $\hat{\beta}^{\,0.00}(\tau_{\,0.00})$; $k$th Power Expectile Regression  ($k$thER ) for $k=1.50$, $\hat{\beta}^{\,1.50}(\tau_{\,1.50})$; and the HQER Expectile Regression (HQER), $\hat{\beta}^{\,\gamma}(\tau_{\,\gamma}),\, \gamma\in\{\frac{i}{10}\mid i=1,\ldots,9  \}$, in \textbf{Case II} ($\varepsilon \sim t(\nu),\, \nu=3$)}
\label{ARE student 3}

\begin{tabular}{@{}lcccccc@{}}
\toprule
 & &  & $\alpha\footnotemark[1]$ &  &  &  \\
\cmidrule(lr){2-7}
$\hspace{1cm}  \mathbb{A}\mathrm{RE}(\cdot) $ & 0.55 or 0.45 & 0.63 or 0.37 & 0.7 or 0.3 & 0.83 or 0.17 & 0.92 or 0.08 & 0.97 or 0.03 \\
\midrule
$\mathbb{A}\mathrm{RE}\left(\hat{\beta}^{1.00}(\tau_{\,1.00})\right) \textit{(ER)}$ & 0.492 & 0.450 & 0.387 & 0.226 & 0.103 & 0.037 \\
$\mathbb{A}\mathrm{RE}\left(\hat{\beta}^{1.50}(\tau_{\,1.50})\right) \textit{(kthER)}$ & \textbf{0.874} & \textbf{0.818} & \textbf{0.730} & 0.476 & 0.241 & 0.093 \\
$\mathbb{A}\mathrm{RE}\left(\hat{\beta}^{0.90}(\tau_{\,0.90})\right) \textit{(HQER)}$ & 0.519 & 0.501 & 0.461 & 0.348 & 0.266 & 0.260 \\
$\mathbb{A}\mathrm{RE}\left(\hat{\beta}^{0.80}(\tau_{\,0.80})\right) \textit{(HQER)}$ & 0.538 & 0.535 & 0.514 & 0.441 & 0.394 & 0.406 \\
$\mathbb{A}\mathrm{RE}\left(\hat{\beta}^{0.70}(\tau_{\,0.70})\right) \textit{(HQER)}$ & 0.550 & 0.556 & 0.544 & 0.491 & 0.449 & 0.444 \\
$\mathbb{A}\mathrm{RE}\left(\hat{\beta}^{0.60}(\tau_{\,0.60})\right) \textit{(HQER)}$ & 0.559 & 0.567 & 0.557 & \textbf{0.504} & \textbf{0.454} & \textbf{0.437} \\
$\mathbb{A}\mathrm{RE}\left(\hat{\beta}^{0.50}(\tau_{\,0.50})\right) \textit{(HQER)}$ & 0.567 & 0.571 & 0.556 & 0.491 & 0.431 & 0.410 \\
$\mathbb{A}\mathrm{RE}\left(\hat{\beta}^{0.40}(\tau_{\,0.40})\right) \textit{(HQER)}$ & 0.575 & 0.569 & 0.545 & 0.460 & 0.390 & 0.370 \\
$\mathbb{A}\mathrm{RE}\left(\hat{\beta}^{0.30}(\tau_{\,0.30})\right) \textit{(HQER)}$ & 0.583 & 0.562 & 0.523 & 0.413 & \t0.333 & 0.314 \\
$\mathbb{A}\mathrm{RE}\left(\hat{\beta}^{0.20}(\tau_{\,0.20})\right) \textit{(HQER)}$ & 0.585 & 0.544 & 0.487 & 0.347 & 0.256 & 0.236 \\
$\mathbb{A}\mathrm{RE}\left(\hat{\beta}^{0.10}(\tau_{\,0.10})\right) \textit{(HQER)}$ & 0.570 & 0.503 & 0.425 & 0.255 & 0.152 & 0.125 \\
$\mathbb{A}\mathrm{RE}\left(\hat{\beta}^{0.00}(\tau_{0.00})\right) \textit{(QR)}$ & 0.804 & 0.764 & 0.698 & 0.494 & 0.276 & 0.117 \\
\bottomrule
\end{tabular}
\footnotetext[1]{$\alpha$ denotes the $\alpha$-th quantile level such that $\delta_{\varepsilon}(\alpha)=\beta^\gamma(\tau_\gamma,\varepsilon)$.}
\end{table}

\begin{sidewaystable}
\centering
\caption{Asymptotic relative efficiency ($\mathbb{A}\mathrm{RE}$) of Expectile Regression (ER), $\hat{\beta}^{\,1.00}(\tau_{\,1.00})$; Quantile Regression (QR), $\hat{\beta}^{\,0.00}(\tau_{\,0.00})$; $k$th Power Expectile Regression  ($k$thER ) for $k=1.50$, $\hat{\beta}^{\,1.50}(\tau_{\,1.50})$; and the HQER Expectile Regression (HQER), $\hat{\beta}^{\,\gamma}(\tau_{\,\gamma}),\, \gamma\in\{\frac{i}{10}\mid i=1,\ldots,9  \}$, in \textbf{Case III} ($\varepsilon \sim \chi^2(\nu),\, \nu=6$)}
\label{ARE chi squared 6}

\begin{tabular}{@{}lccccccccccccccc@{}}
\toprule
& &&&& $\alpha\footnotemark[1]$ \\
\cmidrule(lr){2-16}
$\mathbb{A}\mathrm{RE}(\cdot)$ & 0.03 & 0.08 & 0.17 & 0.25 & 0.33 & 0.37 & 0.45  & 0.55 & 0.63 & 0.70 & 0.75 & 0.83 & 0.90 & 0.92 & 0.97 \\
\midrule
$\mathbb{A}\mathrm{RE}\left(\hat{\beta}^{1.00}(\tau_{\,1.00})\right) \textit{(ER)}$ & 0.387 & 0.406 & 0.440 & 0.479 & 0.522 & 0.545 & 0.593 &   0.651 & 0.691 & 0.717 & 0.725 & 0.710 & 0.640 & 0.600 & \textbf{0.412} \\
$\mathbb{A}\mathrm{RE}\left(\hat{\beta}^{1.50}(\tau_{\,1.50})\right) \textit{(kthER)}$ & \textbf{0.620} & \textbf{0.657} & \textbf{0.667} & \textbf{0.676} & \textbf{0.690} & \textbf{0.697} & \textbf{0.709} &   \textbf{0.715} & \textbf{0.708} & 0.686 & 0.659 & 0.586 & 0.471 & 0.423 & 0.247 \\
$\mathbb{A}\mathrm{RE}\left(\hat{\beta}^{0.90}(\tau_{\,0.90})\right) \textit{(HQER)}$ & 0.411 & 0.420 & 0.449 & 0.484 & 0.526 & 0.549 & 0.596 &   0.653 & 0.693 & 0.718 & 0.726 & 0.711 & 0.640 & 0.601 & \textbf{0.412} \\
$\mathbb{A}\mathrm{RE}\left(\hat{\beta}^{0.80}(\tau_{\,0.80})\right) \textit{(HQER)}$ & 0.438 & 0.436 & 0.458 & 0.491 & 0.531 & 0.553 & 0.599 &   0.655 & 0.694 & 0.719 & 0.727 & 0.711 & \textbf{0.641} & 0.601 & \textbf{0.412} \\
$\mathbb{A}\mathrm{RE}\left(\hat{\beta}^{0.70}(\tau_{\,0.70})\right) \textit{(HQER)}$ & 0.468 & 0.455 & 0.469 & 0.499 & 0.537 & 0.558 & 0.603 &   0.658 & 0.696 & 0.720 & \textbf{0.728} & \textbf{0.712} & \textbf{0.641} & \textbf{0.602} & \textbf{0.412} \\
$\mathbb{A}\mathrm{RE}\left(\hat{\beta}^{0.60}(\tau_{\,0.60})\right) \textit{(HQER)}$ & 0.501 & 0.477 & 0.483 & 0.508 & 0.544 & 0.564 & 0.607 &   0.660 & 0.697 & \textbf{0.721} & \textbf{0.728} & \textbf{0.712} & \textbf{0.641} & \textbf{0.602} & \textbf{0.412} \\
$\mathbb{A}\mathrm{RE}\left(\hat{\beta}^{0.50}(\tau_{\,0.50})\right) \textit{(HQER)}$ & 0.535 & 0.502 & 0.498 & 0.520 & 0.552 & 0.571 & 0.612 &   0.663 & 0.699 & \textbf{0.721} & \textbf{0.728} & \textbf{0.712} & \textbf{0.641} & \textbf{0.601} & \textbf{0.412} \\
$\mathbb{A}\mathrm{RE}\left(\hat{\beta}^{0.40}(\tau_{\,0.40})\right) \textit{(HQER)}$ & 0.566 & 0.529 & 0.516 & 0.533 & 0.561 & 0.579 & 0.617 &   0.665 & 0.700 & \textbf{0.721} & 0.727 & \textbf{0.710} & 0.639 & \textbf{0.600} & 0.410 \\
$\mathbb{A}\mathrm{RE}\left(\hat{\beta}^{0.30}(\tau_{\,0.30})\right) \textit{(HQER)}$ & 0.585 & 0.555 & 0.536 & 0.546 & 0.571 & 0.586 & 0.621 &   0.666 & 0.698 & 0.717 & 0.723 & 0.705 & 0.634 & 0.594 & 0.405 \\
$\mathbb{A}\mathrm{RE}\left(\hat{\beta}^{0.20}(\tau_{\,0.20})\right) \textit{(HQER)}$ & 0.581 & 0.569 & 0.550 & 0.556 & 0.576 & 0.589 & 0.619 &  0.659 & 0.688 & 0.705 & 0.709 & 0.690 & 0.619 & 0.580 & 0.393 \\
$\mathbb{A}\mathrm{RE}\left(\hat{\beta}^{0.10}(\tau_{\,0.10})\right) \textit{(HQER)}$ & 0.540 & 0.548 & 0.538 & 0.543 & 0.559 & 0.569 & 0.594 &  0.627 & 0.649 & 0.662 & 0.663 & 0.641 & 0.570 & 0.533 & 0.356 \\
$\mathbb{A}\mathrm{RE}\left(\hat{\beta}^{0.00}(\tau_{\,0.00})\right) \textit{(QR)}$ & 0.461 & 0.458 & 0.443 & 0.441 & 0.447 & 0.452 & 0.463 &   0.477 & 0.484 & 0.483 & 0.476 & 0.446 & 0.384 & 0.354 & 0.227 \\
\bottomrule
\end{tabular}
\footnotetext[1]{$\alpha$ denotes the $\alpha$-th quantile level such that $\delta_{\varepsilon}(\alpha)=\beta^\gamma(\tau_\gamma,\varepsilon)$.}
\end{sidewaystable}

\newpage

%====================================
\subsubsection{Location-Shift Models} \label{section location shift}
%====================================
In this section, we analyze a simple stochastic linear model
\begin{equation}\label{loca shift model}
y_{i}=\beta_{0}+\beta_{1}  x_{i}+\varepsilon_{i},  \quad i=1,\ldots, n,  
\end{equation}
where the vectors \((x_i, \varepsilon_i),\, i = 1, \ldots, n\), are i.i.d. with the same distribution as \( (x, \varepsilon)\); \(y_i\) is the response variable, \(x_i\) is a scalar explanatory variable, \(\varepsilon_i\) is the error term, and \(\beta_{0}\) and \(\beta_{1}\) are constant parameters. We assume that \(x\) follows a uniform distribution $x\sim\mathcal{U}(0,1)$  and \(\varepsilon\) follows one of the two distributions 
\begin{enumerate}[label=(\roman*)]
    \item $\varepsilon \sim \mathcal{N}(0,1),$ the normal distribution;
    \item $\varepsilon \sim t(\nu),$ the Student distribution with degree of freedom $\nu=3.$
\end{enumerate}
We set \(\beta_0 = 15\) and \(\beta_1 = 90\). Although these values are chosen only for the sake of simulation convenience, they have no effect on the asymptotic variances if \(\varepsilon\) is symmetrically distributed.
The model is designed to explore how the parameter $\gamma$ influences the asymptotic variances (A.variance) of the HQER estimators $\hat{\beta}_0$ and $\hat{\beta}_1$ for $\beta_0$ and $\beta_1$, respectively.\\
For a given \(\alpha\) value in the range of \(0\) to \(1\), we use the following procedure to compute the asymptotic variances. In the context of quantile regression, we use the following formula
$
\alpha (1-\alpha) \cdot \mathbb{E} \left((1,x)\, (1,x)^{\top} \right)^{-1}/ f_\varepsilon^{2}(\delta_\varepsilon(\alpha)) ,
$ 
with $f_\varepsilon(\cdot)$ being the density function of the error $\varepsilon$. For the HQER regression, we use the asymptotic variance expression in Theorem $\ref{theorem 4}$ to complete the calculation after deriving the $\tau$-$\gamma$th HQER expectile of \(\varepsilon\), denoted by \(\xi_{\gamma}( \alpha, \varepsilon)\).

\begin{table}[h]
\centering
\caption{Asymptotic variances of the estimators in the linear model when the error term follows $\mathcal{N}(0,1)$}
\label{table normal variance}
\begin{tabular}{@{}lccccccccccc@{}}
\toprule
&&&&&&$\alpha\footnotemark[1]$\\
\cmidrule(lr){2-12}
A. variance  & 0.04 & 0.09 & 0.16 & 0.2 & 0.33 & 0.55 & 0.6 & 0.75 & 0.8 & 0.91 & 0.96 \\
\midrule
$\beta_{0}^{1.00}\textit{(ER)}$ & 2.27 & 1.58 & 1.27 & 1.19 & 1.05 & \textbf{1.00} & \textbf{1.01} & \textbf{1.11} & \textbf{1.19} & \textbf{1.58} & \textbf{2.27} \\
%\rowcolor{lightgray}%
$\beta_{1}^{1.00}$ & 6.81 & 4.73 & 3.82 & 3.56 & 3.14 & \textbf{3.01} & \textbf{3.04} & \textbf{3.34} & \textbf{3.56} & \textbf{4.73} & \textbf{6.81} \\
$\beta_{0}^{1.50}\textit{(kthER)}$ & 2.80 & 1.84 & 1.43 & 1.32 & 1.14 & 1.08 & 1.10 & 1.23 & 1.32 & 1.84 & 2.80 \\
$\beta_{1}^{1.50}$ & 8.41 & 5.53 & 4.30 & 3.96 & 3.41 & 3.25 & 3.29 & 3.68 & 3.96 & 5.53 & 8.41 \\
$\beta_{0}^{0.90}\textit{(HQER)}$ & 2.22 & 1.56 & 1.27 & \textbf{1.18} & \textbf{1.05} & 1.01 & 1.02 & 1.12 & 1.19 & \textbf{1.58} & 2.28 \\
$\beta_{1}^{0.90}$ & 6.65 & 4.67 & 3.80 & \textbf{3.55}& \textbf{3.14} & 3.02 & 3.05 & 3.35 & 3.56 & 4.74 & 6.85 \\
$\beta_{0}^{0.80}\textit{(HQER)}$ & 2.18 & 1.54 & \textbf{1.26} & \textbf{1.18} & 1.05 & 1.01 & 1.03 & 1.13 & 1.20 & 1.60 & 2.32 \\
$\beta_{1}^{0.80}$ & 6.53 & 4.64 & \textbf{3.79} & \textbf{3.55} & 3.16 & 3.04 & 3.08 & 3.38 & 3.60 & 4.79 & 6.95 \\
$\beta_{0}^{0.70}\textit{(HQER)}$ & 2.15 & 1.54 & 1.27 & 1.19 & 1.07 & 1.03 & 1.04 & 1.15 & 1.22 & 1.63 & 2.38 \\
$\beta_{1}^{0.70}$ & 6.44 & 4.62 & 3.81 & 3.58 & 3.20 & 3.10 & 3.13 & 3.44 & 3.66 & 4.90 & 7.14 \\
$\beta_{0}^{0.60}\textit{(HQER)}$ & \textbf{2.13} & \textbf{1.54} & 1.28 & 1.21 & 1.09 & 1.06 & 1.07 & 1.18 & 1.26 & 1.69 & 2.48 \\
$\beta_{1}^{0.60}$ & \textbf{6.40} & \textbf{4.64} & 3.85 & 3.62 & 3.27 & 3.18 & 3.22 & 3.54 & 3.78 & 5.08 & 7.46 \\
$\beta_{0}^{0.50}\textit{(HQER)}$ & 2.14 & 1.56 & 1.31 & 1.24 & 1.12 & 1.10 & 1.12 & 1.23 & 1.32 & 1.79 & 2.65 \\
$\beta_{1}^{0.50}$ & 6.43 & 4.69 & 3.93 & 3.71 & 3.37 & 3.30 & 3.35 & 3.70 & 3.95 & 5.36 & 7.96 \\
$\beta_{0}^{0.40}\textit{(HQER)}$ & 2.17 & 1.60 & 1.35 & 1.28 & 1.18 & 1.16 & 1.18 & 1.31 & 1.41 & 1.93 & 2.91 \\
$\beta_{1}^{0.40}$ & 6.52 & 4.80 & 4.06 & 3.85 & 3.53 & 3.49 & 3.55 & 3.94 & 4.22 & 5.80 & 8.73 \\
$\beta_{0}^{0.30}\textit{(HQER)}$ &2.23 & 1.66 & 1.42 & 1.35 & 1.25 & 1.26 & 1.28 & 1.44 & 1.55 & 2.16 & 3.31 \\
$\beta_{1}^{0.30}$ & 6.70 & 4.99 & 4.26 & 4.06 & 3.76 & 3.77 & 3.84 & 4.32 & 4.64 & 6.48 & 9.94 \\
$\beta_{0}^{0.20}\textit{(HQER)}$ & 2.33 & 1.76 & 1.52 & 1.46 & 1.37 & 1.40 & 1.44 & 1.64 & 1.77 & 2.53 & 3.97 \\
$\beta_{1}^{0.20}$ & 6.99 & 5.27 & 4.56 & 4.37 & 4.12 & 4.21 & 4.31 & 4.91 & 5.32 & 7.59 & 11.90 \\
$\beta_{0}^{0.10}\textit{(HQER)}$ & 2.47 & 1.90 & 1.68 & 1.62 & 1.56 & 1.65 & 1.70 & 1.98 & 2.16 & 3.18 & 5.12 \\
$\beta_{1}^{0.10}$ & 7.42 & 5.71 & 5.04 & 4.87 & 4.70 & 4.94 & 5.09 & 5.93 & 6.48 & 9.55 & 15.37 \\
$\beta_{0}^{0.00}\textit{(QR)}$ & 5.17 & 3.11 & 2.27 & 2.04 & 1.69 & 1.58 & 1.61 & 1.86 & 2.04 & 3.11 & 5.17 \\
$\beta_{1}^{0.00}$ & 15.51 & 9.32 & 6.81 & 6.12 & 5.06 & 4.74 & 4.82 & 5.57 & 6.12 & 9.32 & 15.51 \\
\bottomrule
\end{tabular}
\footnotetext[1]{$\alpha$ denotes the $\alpha$-th quantile level such that $\delta_{\varepsilon}(\alpha)=\beta^\gamma(\tau_\gamma,\varepsilon)$.}
\end{table}

\newpage

\begin{table}[h]
\centering
\caption{Asymptotic variances of the estimators in the linear model for QR, ER, $k$thER, and HQER expectile regression when the error term follows the Student distribution $t(\nu)$ with degree of freedom $\nu=3$}
\label{table student variance}
%\begin{center}
%\begin{adjustbox}{width=1.1\textwidth}
\begin{tabular}{@{}lccccccccccc@{}}
\toprule
&&&&&& $\alpha\footnotemark[1] $\\
\cmidrule(lr){2-12}
A. variance & 0.04 & 0.09 & 0.16 & 0.20 & 0.33 & 0.55 & 0.60 & 0.75 & 0.80 & 0.91 & 0.96 \\
\midrule
$\beta_{0}^{1.00}\textit{(ER)}$ & 33.12 & 14.78 & 7.22 & 5.63 & 3.57 & 3.04 & 3.18 & 4.50 & 5.63 & 14.78 & 45.62 \\
$\beta_{1}^{1.00}$ & 99.34 & 44.34 & 21.67 & 16.90 & 10.70 & 9.13 & 9.53 & 13.50 & 16.90 & 44.34 & 136.87 \\
$\beta_{0}^{1.50}\textit{(kthER)}$ & 16.26 & 7.44 & 3.78 & 3.00 & \textbf{1.98} & \textbf{1.72} & \textbf{1.79} & \textbf{2.44} & \textbf{3.00} & \textbf{7.44} & \textbf{22.25} \\
$\beta_{1}^{1.50}$ & 48.77 & 22.33 & 11.33 & 9.00 & \textbf{5.94} & \textbf{5.16} & \textbf{5.36} & \textbf{7.32} & \textbf{9.00} & \textbf{22.33} & \textbf{66.74} \\
$\beta_{0}^{0.90}\textit{(HQER)}$ & 29.79 & 13.44 & 6.67 & 5.24 & 3.38 & 2.94 & 3.09 & 4.43 & 5.59 & 15.00 & 47.29 \\
$\beta_{1}^{0.90}$ & 89.36 & 40.32 & 20.00 & 15.71 & 10.13 & 8.83 & 9.26 & 13.30 & 16.76 & 45.01 & 141.86 \\
$\beta_{0}^{0.80}\textit{(HQER)}$ & 26.35 & 12.05 & 6.09 & 4.82 & 3.18 & 2.84 & 3.00 & 4.37 & 5.55 & 15.24 & 49.06 \\
$\beta_{1}^{0.80}$ & 79.05 & 36.15 & 18.26 & 14.47 & 9.54 & 8.53 & 8.98 & 13.12 & 16.64 & 45.72 & 147.17 \\
$\beta_{0}^{0.70}\textit{(HQER)}$ & 22.84 & 10.63 & 5.49 & 4.40 & 2.98 & 2.75 & 2.91 & 4.32 & 5.52 & 15.51 & 50.99 \\
$\beta_{1}^{0.70}$ & 68.53 & 31.88 & 16.48 & 13.20 & 8.93 & 8.24 & 8.73 & 12.97 & 16.56 & 46.54 & 152.98 \\
$\beta_{0}^{0.60}\textit{(HQER)}$ & 19.34 & 9.20 & 4.89 & 3.97 & 2.78 & 2.66 & 2.84 & 4.29 & 5.52 & 15.86 & 53.20 \\
$\beta_{1}^{0.60}$ & 58.02 & 27.59 & 14.68 & 11.91 & 8.33 & 7.98 & 8.51 & 12.87 & 16.56 & 47.57 & 159.60 \\
$\beta_{0}^{0.50}\textit{(HQER)}$ & 15.92 & 7.79 & 4.30 & 3.54 & 2.58 & 2.59 & 2.78 & 4.30 & 5.57 & 16.34 & 55.88 \\
$\beta_{1}^{0.50}$ & 47.77 & 23.38 & 12.90 & 10.63 & 7.75 & 7.76 & 8.35 & 12.90 & 16.70 & 49.01 & 167.62 \\
$\beta_{0}^{0.40}\textit{(HQER)}$ & 12.70 & 6.46 & 3.73 & 3.14 & 2.40 & 2.55 & 2.77 & 4.38 & 5.71 & 17.09 & 59.41 \\
$\beta_{1}^{0.40}$ & 38.09 & 19.37 & 11.19 & 9.41 & 7.22 & 7.64 & 8.30 & 13.14 & 17.14 & 51.26 & 178.22 \\
$\beta_{0}^{0.30}\textit{(HQER)}$ & 9.79 & 5.24 & 3.21 & 2.77 & 2.26 & 2.57 & 2.82 & 4.61 & 6.06 & 18.42 & 64.75 \\
$\beta_{1}^{0.30}$ & 29.36 & 15.71 & 9.62 & 8.30 & 6.77& 7.71 & 8.47 & 13.82 & 18.18 & 55.26 & 194.25 \\
$\beta_{0}^{0.20}\textit{(HQER)}$ & 7.34 & 4.20 & 2.77 & 2.46 & 2.17 & 2.72 & 3.04 & 5.18 & 6.89 & 21.28 & 74.88 \\
$\beta_{1}^{0.20}$ & 22.02 & 12.60 & 8.30 & 7.39 & 6.52 & 8.15 & 9.11 & 15.54 & 20.67 & 63.84 & 224.64 \\
$\beta_{0}^{0.10}\textit{(HQER)}$ & \textbf{5.53} & \textbf{3.44} & \textbf{2.47} & \textbf{2.28} & 2.22 & 3.19 & 3.67 & 6.82 & 9.32 & 30.06 & 105.02 \\
$\beta_{1}^{0.10}$ & \textbf{16.58} & \textbf{10.31} & \textbf{7.41} & \textbf{6.85} & 6.67 & 9.57 & 11.02 & 20.47 & 27.96 & 90.18 & 315.07 \\
$\beta_{0}^{0.00} \textit{(QR)}$ & 23.07 & 9.91 & 4.66 & 3.59 & 2.22 & 1.88 & 1.96 & 2.83 & 3.59 & 9.91 & 32.21 \\
$\beta_{1}^{0.00}$ & 69.21 & 29.73 & 13.98 & 10.76 & 6.65 & 5.63 & 5.89 & 8.49 & 10.76 & 29.73 & 96.64 \\
\bottomrule
\end{tabular}
\footnotetext[1]{$\alpha$ denotes the $\alpha$-th quantile level such that $\delta_{\varepsilon}(\alpha)=\beta^\gamma(\tau_\gamma,\varepsilon)$.}
\end{table}
\noindent
Tables \(\ref{table normal variance}\) and \(\ref{table student variance}\) summarize the asymptotic variances for the normal and Student distributions, respectively. The highest-performing results are highlighted in solid to draw attention. The first column of both tables presents the asymptotic variance of the HQER regression coefficients, denoted as \(\beta_0^\gamma\) (the intercept) and \(\beta_1^\gamma\) (the slope), for different values of \(\gamma \in \{\frac{i}{10} \mid i=1,2, \ldots, 9 \}\). Additionally, \(\beta_{0}^{0}\) and \(\beta_{1}^{0}\) correspond to the coefficients of quantile regression, while \(\beta_{0}^{1.50}\) and \(\beta_{1}^{1.50}\) represent the coefficients of \(k\)th power expectile regression for \(k=1.50\).
HQER can provide  superior performance compared to its competitors, particularly for small values of $\alpha$ and specific values of $\gamma$. In Table \(\ref{table normal variance}\), for the normal distribution case, HQER delivers the smallest variances for quantiles \(\alpha \in \{0.04, 0.09, 0.16, 0.20, 0.33\}\) and \(\gamma \in \{0.60, 0.80, 0.90\}\). For example, when \(\alpha = 0.04\), HQER produces lower asymptotic variances than ER, OR, and the \(k\)th power expectile regression for its best choice of \(k=1.50\). In contrast, ER provides the best performance for all quantiles greater than \(\alpha=0.55\).  For the Student distribution case in Table \(\ref{table student variance}\), a different phenomenon occurs. Specifically, HQER excels when \(\gamma \to 0\) and \(\alpha \to 0\), yielding significantly lower variances than ER, QR, and the \(k\)th power expectile regression for \(k = 1.50\). For instance, when \(\gamma = 0.10\), HQER achieves its lowest variance at \(\alpha = 0.04\) for the intercept, with a value of \(5.53\), outperforming all other methods. For quantiles greater than \(\alpha=0.55\), ER yields the lowest variances compared to the other approaches. Finally, one can remark that for both scale-shift and location-shift models, the performance of HQER depends on selecting an optimal \(\gamma\), a topic that will be discussed later in Section~\ref{section 5}.

%=========================================================
 \subsection{Empirical Analysis} \label{subsection 4.2}
\begin{figure}[h]
  \centering
  \includegraphics[width=1\linewidth]{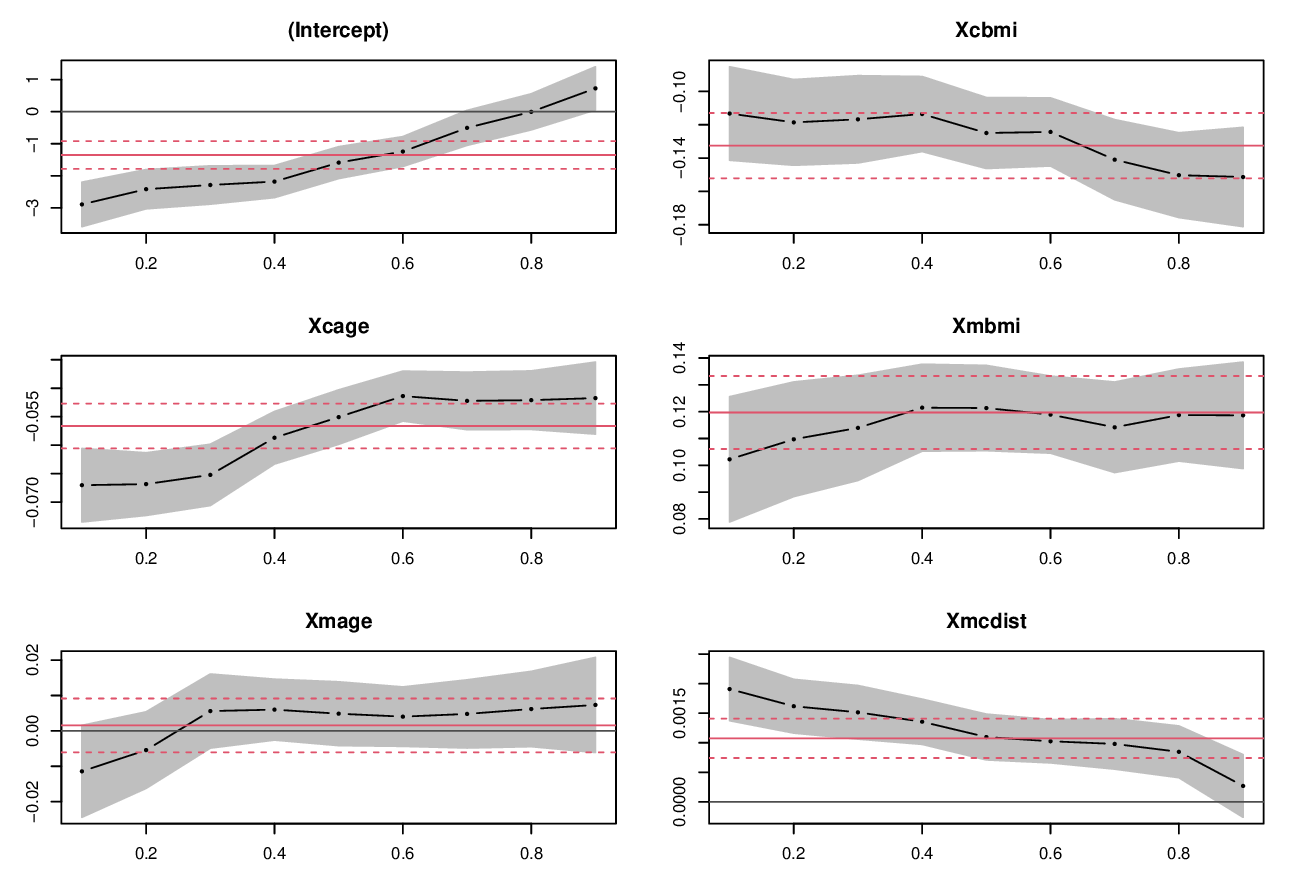} 
  \caption{QR results for the India dataset at quantile levels \(\alpha \in \{\frac{i}{10} \mid i=1, 2, \ldots, 9\}\)} 
  \label{quantile regression}
\end{figure}

%=========================================================
We used the dataset \texttt{india} from the \texttt{R} package \(\texttt{gamboostLSS}\) (\cite{miftahuddin2016modelling}) for the real data example. The original dataset is publicly available at \url{http://www.measuredhs.com}. This dataset contains several covariates from a study conducted in India on malnutrition in children up to the age of three during 1998 and 1999. Malnutrition hinders growth, so reduced development serves as an indicator of a child's nutritional status. 
The focus is on stunted growth, which is measured by a \texttt{z-score} ranging from \(-6\) to \(6\). A negative \texttt{z-score} indicates that a child is below the expected height for their age, suggesting chronic malnutrition. Children with a \texttt{z-score} below \(-2\) are classified as stunted (height-for-age) (\cite{waldmann2018quantile}).  
The dataset includes the \texttt{z-score} of \(n = 4,000\) children, along with covariates such as the child's age in months (\texttt{cage}), the mother’s body mass index (BMI) (\texttt{mbmi}), the subregion of India (\texttt{mcdist}) where they lived during the study, and the child’s BMI (\texttt{cbmi}). For a more detailed content-related analysis of this dataset, refer to \cite{fenske2011identifying}.  
Finally, in our setting, the response variable is \texttt{z-score}, and the model includes \(p=5\) covariates: \texttt{cbmi}, \texttt{cage}, \texttt{mbmi}, \texttt{mage}, and \texttt{mcdist}. The model we investigate is  
\[
\texttt{z-score}_{i}=\beta_{0}+\texttt{cbmi}_i\beta_1+\texttt{cage}_i\beta_2+\texttt{mbmi}_i\beta_3+\texttt{mage}_i\beta_4+\texttt{mcdist}_i\beta_5 +\varepsilon_{i}, \quad i=1,\ldots, n.
\]
Recent studies by \cite{fenske2011identifying}, \cite{waldmann2018quantile}, and \cite{mokalla2022application} suggest that the effects of age, the mother's BMI, and geographic factors on stunting vary across quantiles. Age has a stronger negative impact on severely stunted children, the mother's BMI has a greater positive effect on the most malnourished, and geographic disparities are more pronounced in lower quantiles. Quantile regression (QR) captures these varying influences, offering a deeper understanding of malnutrition risks compared to median and mean regression.  
In our analysis, to gain an initial insight into the data, a quantile regression plot is shown in Fig. \ref{quantile regression}. It displays the parameter estimates (represented by the dashed black lines) and the 95\% confidence intervals (represented by the dark gray shaded regions) as a function of the quantile level. The plot shows that \texttt{cage}, \texttt{cbmi}, and \texttt{mbmi} significantly affect the lower tail of the \texttt{z-score} distribution because the lower confidence limits are greater than 0 for quantiles less than 0.5.  
Additionally, the variables \texttt{mage} and \texttt{mcdist} exhibit a decreasing trend across quantiles, suggesting that their effects are more pronounced in the upper tail of the \texttt{z-score} distribution. The \texttt{mbmi} variable shows an increasing trend, indicating a stronger effect in higher quantiles. The comparison with ordinary least squares (OLS), represented by the solid  horizontal lines and their confidence intervals (represented by  dashed horizontal lines), highlights differences between mean-based and quantile-based estimation, demonstrating that quantile regression captures heterogeneity in the effects of covariates across the distribution.  

\noindent
For our analysis, we assume that the $\alpha$-HQER levels  take values within the set \( \{\frac{i}{10} \mid i=1, 2, \ldots, 9\}\). We examine the subsequent values of \(\gamma \in \{\frac{i}{10} \mid i=1, 2, \ldots, 9\}\) within the interval \((0, 1)\). Through careful comparisons, the appropriate \(\gamma\) is determined for each \(\alpha\) value using the  methodology outlined in Section \ref{section 5}, and the corresponding results are presented in Table \ref{table real data}. 

\noindent
Of note, the expectile-based predictions of HQER do not automatically correspond to the empirical quantile of order $\alpha$ when $\gamma > 0$. To address this, and following the suggestion of \cite{efron1991regression}, we compute, for each fitted HQER model at level $\alpha$, the empirical proportion of residuals $e_i = y_i - \boldsymbol{x}_i^\top \boldsymbol{\hat{\beta}}{(\alpha)}$ that are less than or equal to zero. More precisely, for each fixed $\alpha$, we define $p(\alpha) = \frac{1}{n}\sum_{i=1}^n \mathds{1}\{e_i\leq 0\}$, which corresponds to the proportion of negative residuals. This provides a diagnostic of the effective quantile level represented by each fitted HQER hyperplane, $H_{\alpha} = \{(\boldsymbol{x},y): y=\boldsymbol{x}^\top \boldsymbol{\hat{\beta}}(\alpha)\}$, and allows us to assess how closely the model aligns with the target $\alpha$ level.
The reported results in Table~\ref{table real data} include both $\alpha$-HQER level and its corresponding $p(\alpha)$-th empirical quantile for the whole grid of $\alpha$ values. Table \ref{table real data} also presents the regression parameter estimates and their corresponding standard errors. The results indicate that the BMI of both the child and the mother affects stunting, particularly, when \(\alpha\) is set to \(0.20\), \(0.40\), \(0.50\), \(0.60\), \(0.70\), \(0.80\), or \(0.90\). However, the child's age and mother's age have minimal effects on stunting; in particular, the mother's age has no effect when \(\alpha = 0.30\). The subregion has the smallest effect, especially when \(\alpha = 0.90\). As \(\alpha\) approaches 0.90 or 0.50, stunting demonstrates a significant dependence on both the child's BMI and the mother's BMI, with a particularly strong dependence at \(\alpha = 0.90\). The HQER results highlight that both child and maternal BMI have a significant impact on stunting, with stronger effects observed at higher quantiles. The child's age shows a stable effect across all quantiles, while maternal age and distance to health care show minimal effects, especially at intermediate quantiles. The variability in the intercept and certain predictor coefficients at extreme quantiles (\(\alpha= 0.10\) and \(\alpha= 0.90\)) underscores the complexity of modeling stunting in these cases. These findings are consistent with established health research, particularly in the context of malnutrition in India, where child's BMI and maternal nutritional status play a critical role in determining stunting severity (\cite{mokalla2022application}). This emphasizes the importance of HQER in capturing the nuanced relationships between predictors and stunting severity.

\begin{table}[h]
    \centering
    \caption{HQER regression results with optimal $\gamma \in \{\frac{i}{10}\mid i = 1, \ldots, 9\}$ for various quantile levels $\alpha \in \{\frac{i}{10} \mid i=1, 2, \ldots, 9\}$, including estimates and their standard deviations (SD), and the observed proportion of residuals satisfying $e_i \leq 0$}
    \label{table:real_data}
    \renewcommand{\arraystretch}{1.5}
    \begin{tabular}{@{}lccccccccc@{}}
        \toprule
        $\alpha$-HQER level & 0.1 & 0.2 & 0.3 & 0.4 & 0.5 & 0.6 & 0.7 & 0.8 & 0.9 \\
        \midrule
        Optimal $\gamma$ & 0.10 & 0.40 & 0.50 & 0.40 & 0.20 & 0.10 & 0.10 & 0.30 & 0.30 \\
        \midrule
        $p(\alpha)$     & 0.995 & 0.356 & 0.388 & 0.456 & 0.672 & 0.995 & 0.995 & 0.505 & 0.506 \\
        \midrule
        \textbf{Variables} & \multicolumn{9}{c}{\textbf{Estimates (SD)}} \\
        \midrule
        Intercept & 3.0074 & -1.5066 & -1.5654 & -1.2191 & -0.6707 & 2.8777 & 2.4147 & -0.5991 & -0.7982 \\
                  & (0.2621) & (0.4293) & (0.3289) & (0.3338) & (0.8226) & (0.2906) & (2.7094) & (0.6872) & (0.7973) \\
        cbmi      & -0.1253 & -0.1205 & -0.1264 & -0.1338 & -0.0722 & -0.1181 & -0.1114 & -0.1811 & -0.1825 \\
                  & (0.0128) & (0.0192) & (0.0155) & (0.0126) & (0.0378) & (0.0160) & (0.0256) & (0.0286) & (0.0292) \\
        cage      & -0.0566 & -0.0591 & -0.0580 & -0.0588 & -0.0860 & -0.0561 & -0.0561 & -0.0526 & -0.0511 \\
                  & (0.0023) & (0.0035) & (0.0028) & (0.0027) & (0.0316) & (0.0024) & (0.0046) & (0.0052) & (0.0048) \\
        mbmi      & 0.1187 & 0.1050 & 0.1121 & 0.1126 & 0.0935 & 0.1146 & 0.0999 & 0.1245 & 0.1255 \\
                  & (0.0083) & (0.0108) & (0.0096) & (0.0091) & (0.0335) & (0.0081) & (0.0871) & (0.0176) & (0.0204) \\
        mage      & 0.0012 & -0.0094 & -0.0036 & -0.0025 & 0.0033 & 0.0026 & 0.0022 & 0.0037 & 0.0118 \\
                  & (0.0043) & (0.0068) & (0.0052) & (0.0055) & (0.0142) & (0.0057) & (0.0069) & (0.0102) & (0.0142) \\
        mcdist    & 0.0012 & 0.0013 & 0.0012 & 0.0011 & 0.0030 & 0.0013 & 0.0013 & 0.0001 & 0.0001 \\
                  & (0.0002) & (0.0003) & (0.0002) & (0.0002) & (0.0014) & (0.0002) & (0.0003) & (0.0005) & (0.0005) \\
        \bottomrule
    \end{tabular}\label{table real data}
\end{table}

\section{Selecting the tuning parameter $\gamma$ of HQER} \label{section 5}

%----------------------------------------------------------------
We propose a method for selecting the optimal value of \( \gamma \) to enhance the efficiency of the HQER approach when applied to real-world data. Rigorous examination of this issue presents several challenges, including conducting specification tests on the relationship between the HQER expectile and various covariates, as well as performing variable selection. For simplicity, we assume a linear model of the form \( y = \beta_0 + \beta_1 x_1 + \dots + \beta_p x_p + \varepsilon \), where the covariates \( (1, x_1, \dots, x_p) \) are pre-identified. Our primary focus is on selecting the optimal value of \( \gamma \) for the HQER expectile regression within this framework.
The proposed method for determining the appropriate value of \( \gamma \) relies on the HQER score function and is based on a cross-validation procedure. This approach can be summarized, for data $(y_i,\boldsymbol{x}_i),\, i=1,2,\ldots,n$, as follows:

\begin{enumerate}[label=(\roman*)]
    \item Let the range of values for \( \gamma \) be \( \left\{ \frac{i}{\tilde{n}} \mid i = 1, 2, \dots, \tilde{n}-1 \right\} \), for some sufficiently large \( \tilde{n} \).
    \item Split the data into two parts: the first part contains \( n_c \) observations, and the second part contains \( n_v = n - n_c \) observations.
    \item For each value of \( \gamma \), perform an HQER expectile regression on the \( n_c \) observations to obtain the estimator vector \( \boldsymbol{\hat{\beta}}(\alpha) \).
    \item Use the second part of the data to evaluate the same value of \( \gamma \) as in Step 3. Specifically, compute the term
    \[
    \boldsymbol{\hat{Z}}(\gamma, \alpha, \boldsymbol{\hat{\beta}}(\alpha)) = \frac{1}{n_v} \sum_{i=1}^{n_v} \boldsymbol{x}_i \varphi_{\alpha}(y_i - \boldsymbol{x}_i^\top \boldsymbol{\hat{\beta}}(\alpha)),
    \]
    where \( \varphi_\alpha(\cdot) \) is defined in Equation \eqref{varphi}.
    \item There will be \( \tilde{n}-1 \) error terms \( \boldsymbol{\hat{Z}}(\gamma, \alpha, \boldsymbol{\hat{\beta}}(\alpha)) \). Compare these terms and find the value of \( \gamma \) that minimizes the norm of the error term, denoted as \( \gamma_0 \), i.e.,
    \[
\gamma_0 = \operatorname*{arg\,min}_{\gamma \in \left\{ \frac{i}{\tilde{n}} \mid i = 1,\dots, \tilde{n}-1 \right\}} \| \boldsymbol{\hat{Z}}(\gamma, \alpha, \boldsymbol{\hat{\beta}}(\alpha)) \|.
\]

\end{enumerate}

We apply this method, based on a cross-validation approach with the score function as the optimality criterion, to simulated data for Model \eqref{loca shift model}. For the simulation, we set the sample size to \( n = 10,000 \), selecting values of \( \alpha \) from the set \( \{0.10, 0.20, 0.40, 0.60, 0.80, 0.90\} \), and values of \( \gamma \) from the set \( \{ \frac{i}{10} \mid i = 1, 2, \dots, 9 \} \). The results are summarized in Table \ref{suitable gamma}, where the values in parentheses represent the estimators \( (\hat{\beta}_0, \hat{\beta}_1) \) along with their corresponding standard deviations. Estimators with smaller standard deviations are highlighted in bold for clarity. Based on the results, we identify appropriate values of \( \gamma \), such as \( \gamma = 0.90 \) for \( \alpha = 0.20 \) and \( \gamma = 0.80 \) for \( \alpha = 0.60 \). The smallest deviations are observed around \( \alpha = 0.60 \) for \( \gamma = 0.90 \), suggesting that this combination offers the most stable estimation performance at this quantile level.

%==============================================================
\begin{table}[h]
\centering
\caption{The appropriate choice of $\gamma$ for simulation data where the error term is normal for different quantiles $\alpha$}
\label{suitable gamma}
\centering
\begin{tabular}{@{}lcccccc@{}}
\toprule
\;\;$\alpha$ & 0.10 & 0.20 & 0.40 & 0.60 & 0.80 & 0.90 \\
\cmidrule(lr){2-7}
\textit{optimal} $\gamma$ & 0.40 & 0.40 & 0.50 & 0.90 & 0.50 & 0.60 \\
\midrule
($\hat{\beta}_0,\hat{\beta}_1$) & (15.05, 91.01) & (15.62, 90.35) & (15.16, 90.71) & (15.70, 89.98) & (16.20, 89.34) & (16.35, 89.61) \\
(\textit{SD}) & (0.88, 1.15) & (0.92, 1.17) & (0.81, 1.06) & (\textbf{0.80}, \textbf{0.95}) & (1.00, 1.30) & (0.93, 1.25) \\
\bottomrule
\end{tabular}
\end{table}

\section{Conclusion} \label{Sec 5}

In this paper, we explore the $\tau$-$\gamma$th HQER expectile and the $\tau$-$\gamma$th HQER expectile regression method, with particular focus on $0 < \gamma < 1$. HQER is a hybrid approach of QR and ER. This study is a partial extension of the work of  \cite{newey_asymmetric_1987} and \cite{koenker_quantile_2005}, aiming to establish a connection between quantiles and expectiles. We provide proofs of the existence and uniqueness of $\tau$-$\gamma$th HQER expectiles under mild conditions. Additionally, we study the consistency and asymptotic normality of the estimators for the $\tau$-$\gamma$th HQER expectile regression. Comparative analyses between $\tau$-$\gamma$th HQER expectile estimators and common quantile, expectile, and $k$th power expectile regression estimators demonstrate the advantages of HQER expectile regression. In particular, the properties of HQER expectile regression converge to those of quantile regression as $\gamma$ approaches $0$, while they converge to those of both expectile regression and $k$th power expectile regression (when $k$ tends to $2$) as $\gamma$ tends to $1$. One can choose an appropriate $\gamma$ value to perform a satisfactory $\tau$-$\gamma$th HQER expectile regression based on specific problem requirements and preferences. Our real data analysis examines  childhood malnutrition in India. The results indicate that $\tau$-$\gamma$th HQER expectile regression yields smaller variances for the majority of $\gamma$ values. \\
We acknowledge that interpretability remains a critical aspect in evaluating regression methods. Although HQER does not inherit the direct probabilistic interpretation of quantiles, it offers a flexible modeling tool that interpolates between robust and efficient objectives. Its data-driven model construction enables the HQER loss function to approximate the underlying (log)likelihood structure, providing useful summaries of conditional behavior. This balance between quantile and expectile perspectives makes HQER particularly appealing in applications where model misspecification, noise, or asymmetry are present. We hope that this work encourages further investigation into the interpretability and practical use of hybrid loss functions in complex statistical settings. \\
\noindent
Looking ahead, we highlight several promising directions for future research that were introduced earlier. First, due to the nondifferentiability of the quantile loss component in HQER, developing smoothed loss approximations is essential for efficient estimation and inference. Recent advances---such as convolution smoothing techniques that have been developed for quantile regression (\cite{fernandes2021smoothing, tan2022high, he2023smoothed})---can be adapted to HQER to improve computational tractability and facilitate  high-dimensional inference. Second, extending HQER to accommodate dependent data structures (e.g., longitudinal or clustered data) is a natural next step. This builds on literature on the quantile regression under dependence (\cite{koenker2004quantile}) and more recent GEE-based approaches for expectile regression that account for heteroscedasticity (\cite{barry2022new}). These developments would expand the practical applicability of HQER to more complex data settings involving correlation, model misspecification, or heterogeneity.

\proofeq
\section*{Appendix} 
We list the two lemmas that we will use in the proofs. The interested reader is referred to \cite{newey_asymmetric_1987} for Lemma $\ref{lemma hjort 1}$ (Lemma A in \cite{newey_asymmetric_1987}), and to \cite{hjort2011asymptotics} for Lemma \ref{lemme hjort 2} (Lemma $1$ in \cite{hjort2011asymptotics}). 

\begin{lm}\label{lemma hjort 1}
Let $\boldsymbol{\theta}_{0}$ be a point in $\mathbb{R}^{p}$ and $\mathcal{O}$ an open set containing $\boldsymbol{ \theta }_{0} \cdot$ If
\begin{enumerate}[label=(\Alph*)]
\item
 $M_{n}(\cdot)$ converges in probability to $M(\cdot)$ uniformely on $\mathcal{O},$\label{Q A}
\item $M(\cdot)$ has a unique minimum on $\mathcal{O}$ at $\boldsymbol{\theta}_{0}$, \label{Q B}
\item  $M_{n}(\cdot)$ is convex, \label{Q C}
\end{enumerate}
Then for $\boldsymbol{\hat{\theta}}=\arg\min\limits_{\boldsymbol{\theta} } M_{n}(\boldsymbol{\theta}),$
\begin{enumerate}[label=(\roman*)]
\item  $\boldsymbol{\hat{\theta}}$ exists with a probability approaching one,
\item  $\boldsymbol{\hat{\theta}}$ converges in probability  to $\boldsymbol{\theta}_{0}.$
\end{enumerate}

\end{lm}

\begin{lm} \label{lemme hjort 2}
 Suppose $M_{n}(\boldsymbol{\theta}) $ is a sequence of convex random
functions defined on an open convex set $\mathcal{O}$ in $\mathbb{R}^{p}$, which converges in probability to some $M(\boldsymbol{\theta})$, for
every  $\boldsymbol{\theta} $. Then  $\sup_{\boldsymbol{\theta} \in \mathcal K} \norm{ M_{n}(\boldsymbol{\theta}) - M(\boldsymbol{\theta}) } $ goes to zero in probability, for any compact subset ${\mathcal{K}}$ of $\mathcal{O}$.
\end{lm}

\section*{The proof of Theorem \ref{thorem 1}}

\begin{enumerate}[label=(\roman*)]
\item We will first show that the solution of \eqref{first order condition} exists and is unique. Suppose the equation \eqref{first order condition} has a solution $\vartheta(\tau,\gamma)$, which we  refer to as $\vartheta$ for convenience and to avoid confusion.\\
After rewriting \eqref{first order condition}, we get
\begin{equation}
\label{eq}
\frac{(1-\gamma)}{(1-\tau)}\cdot (F(\vartheta)-\tau)-2\gamma \cdot \left \{m-\vartheta+\alpha(\tau) \cdot T_{F}(\vartheta) \right \}=0,
\end{equation}
with $\alpha(\tau)=\frac{2\tau-1}{1-\tau},$  $T_{F}(\vartheta)=\int_{\vartheta}^{\infty}(s-\vartheta)\cdot \mathrm{dF}(s)$, and $m=\mathbb{E}(Y)$.\\
We write 
$$h(\vartheta)=\frac{(1-\gamma)}{(1-\tau)}\cdot (F(\vartheta)-\tau)-2\gamma \cdot\left \{m-\vartheta+\alpha(\tau)\cdot T_{F}(\vartheta) \right \}.$$
Since $\int_{-\infty}^{+\infty}  \lvert y\rvert  \cdot   f(y)\mathrm{d}y<+\infty,$ then
\begin{align*}
h(\vartheta)=&\frac{(1-\gamma)}{(1-\tau)}\cdot (F(\vartheta)-\tau)-2\gamma\cdot \left \{m-\vartheta+\alpha(\tau) (\int_{\vartheta}^{\infty} s\cdot  \mathrm{dF}(s)-\vartheta\int_{\vartheta}^{+\infty} \mathrm{dF}(s) )\right \}\nonumber\\
=&\frac{(1-\gamma)}{(1-\tau)}\cdot (F(\vartheta)-\tau)-2\gamma\cdot \left \{m-\vartheta+\alpha(\tau) \left( \int_{\vartheta}^{\infty} s \cdot \mathrm{dF}(s)-\vartheta(1-F(\vartheta)) \right)\right \}\nonumber.\\
\end{align*}
Therefore, one has 
\begin{align*}
  h'(\vartheta)=&\frac{(1-\gamma)}{(1-\tau)}\cdot f(\vartheta)-2\gamma \left \{-1+\alpha(\tau)\cdot \left( -\vartheta f(\vartheta)-1+\vartheta f(\vartheta)+F(\vartheta) \right)\right \}\\
         =&\frac{(1-\gamma)}{(1-\tau)}\cdot f(\vartheta)+2\gamma \cdot \left \{1+\alpha(\tau)\cdot \left( 1-F(\vartheta) \right)\right \}. 
\end{align*}
Note that $\forall \tau \in(0,1),\, \alpha(\tau)>-1 ,$ since $\alpha(\tau)+1=\frac{\tau}{1-\tau}>0.$
This implies that 
$$
h'(\vartheta) >0,\forall \vartheta. 
$$
We will first prove that for any $\tau \in (0,1)$, there exists a $\vartheta(\tau)$ such that \eqref{eq} holds. 
Taking into account the improper integral theorem, it follows that $T_{F}(\vartheta) $ goes to $0$ as $\vartheta$ goes to infinity, then $h$ goes to  $+\infty$ and $-\infty$ as $\vartheta$ goes to  $+\infty$ and $-\infty$, respectively. Then for $\gamma\in(0,1)$ and  $\tau \in (0,1)$  fixed, we have
\begin{itemize}
    \item $h(\vartheta)<0$, for $-\vartheta$ large enough, $\lim_{\vartheta \to -\infty } h(\vartheta)=-\infty,$ 
    \item $h(\vartheta)>0$, for $\vartheta$ large enough, $\lim_{\vartheta \to +\infty } h(\vartheta)=+\infty$. 
\end{itemize}
The intermediate value theorem therefore ensures that the solution of \eqref{eq} exists. Since  the objective function is strictly convex (since $(\tau,\gamma)\in(0,1)\times (0,1)$) the solution is unique.  We denote this unique solution as $\xi(\tau)$. Therefore, the property $\ref{1-theorem 1}$ is proved.\\
\item
The strictly monotonic property of $\xi(\cdot)$ can be proved by the equation

\begin{equation}\label{h equation}
   h(\xi(\tau))=0,\, \forall \tau \in (0,1)  
\end{equation}
As $F(\cdot)$ is continuously differentiable (since $f(\cdot)$ is continuous), then $h(\cdot)$ is continuously differentiable and since $h(\vartheta)>0,\, \forall \vartheta$, then according to the implicit function theorem, $\xi(\cdot)$ is continuously differentiable.\\
Taking the derivative with respect to $\tau$ in \eqref{h equation}, we get
$$\frac{(1-\gamma)}{(1-\tau)^{2}}\cdot (\xi'(\tau)f(\xi(\tau))-1)-2\gamma \cdot\left \{-\xi'(\tau)+\frac{1}{(1-\tau)^{2}}\cdot T_{F}(\xi(\tau)) -\alpha(\tau)\cdot  [1-F(\xi(\tau))]\cdot \xi'(\tau) \right \}=0,$$
i.e.,
$$\left\{ \frac{(1-\gamma)}{(1-\tau)^{}}\cdot f(\xi(\tau)) +2\gamma[ 1 +\alpha(\tau) \cdot (1-F(\xi(\tau)))] \right \}\cdot \xi'(\tau) = \frac{1-\gamma}{(1-\tau)^{2}}\cdot  [1-F(\xi(\tau))]   +\frac{2\gamma}{(1-\tau)^{2}} \cdot T_{F}(\xi(\tau)).$$
Then, $$ \xi'(\tau) = [\frac{1-\gamma}{(1-\tau)^{2}}\cdot  [1-F(\xi(\tau))]   +\frac{2\gamma}{(1-\tau)^{2}} \cdot T_{F}(\xi(\tau))]/ \left\{ \frac{(1-\gamma)}{(1-\tau)^{}}\cdot f(\xi(\tau)) +2\gamma[ 1 +\alpha(\tau) \cdot (1-F(\xi(\tau)))] \right \}.$$
Therefore $\xi'(\tau)>0,\, \forall \tau \in(0,1),$ then $\xi(\cdot)$ is strictly monotonically increasing and the property  $\ref{2-theorem 1}$ holds. 

\item
According to the definition of the $\tau$-$\gamma$th HQER expectile ${\tilde{\xi}}(\cdot)$ of $\tilde{Y},$ we have

$$\frac{(1-\gamma)}{(1-\tau)}\cdot (\tilde{F}(\tilde{\xi}(\tau))-\tau)-2\gamma\cdot \left \{\mathbb{E}(\tilde{Y})-\tilde{\xi}(\tau)+\alpha(\tau)\cdot T_{\Tilde{F}}(\Tilde{\xi}(\tau)) \right \}=0.$$
Considering that $\tilde{Y}= Y+a$ and changing the variable, we get 

$$\frac{(1-\gamma)}{(1-\tau)}\cdot (F(  \tilde{\xi}(\tau)-a)-\tau)-2\gamma\cdot \left \{\mathbb{E}(Y)-\tilde{\xi}(\tau)-a+\alpha(\tau)\cdot T_{F}(\Tilde{\xi}(\tau)-a) \right \}=0.$$
By the uniqueness of $\xi(\cdot)$, we have
  $$ \tilde{\xi}(\tau)=\xi(\tau)+a,\; \forall \tau \in(0,1).$$
Hence, the property $\ref{3-theorem 1}$ is proved.
\end{enumerate}

\section*{The proof of Theorem \ref{theorem 2}} 
\begin{enumerate}[label=(\roman*)]
 \item For $\boldsymbol{b} \in \mathbb{R}^{p}$, we set
$$
B_{\boldsymbol{\beta}+}=\{ i,\; y_{i}+\boldsymbol{x}_{i}^{\top}\, \boldsymbol{\beta} \geq \boldsymbol{x}_{i}^{\top}\,(\boldsymbol{\beta}+\boldsymbol{b})   \}\; \textit{and} \; B_{\boldsymbol{\beta}-}=\{i,\; y_{i}+\boldsymbol{x}_{i}^{\top}\boldsymbol{\beta} < \boldsymbol{x}_{i}^{\top}\,(\boldsymbol{\beta}+\boldsymbol{b})   \}.
$$
Let 
$$
\boldsymbol{\zeta}(\tau, \boldsymbol{\beta}+\boldsymbol{b}, \boldsymbol{y}+\boldsymbol{x}^{\top}\boldsymbol{b},\boldsymbol{ x })=(1-\tau)\cdot \sum_{i \in B_{\boldsymbol{\beta}-}}\, (1-\gamma)\cdot ( \boldsymbol{x}_{i}^{\top}\,(\boldsymbol{\beta}+\boldsymbol{b})-[y_{i}+\boldsymbol{x}_{i}^{\top}\, \boldsymbol{b}])+\gamma \cdot ( y_{i}+\boldsymbol{x}_{i}^{\top}\, \boldsymbol{b} - \boldsymbol{x}_{i}^{\top}\,(\boldsymbol{\beta}+\boldsymbol{b}))^2+
$$
$$\;\;\;\;\;\;\;\;\;\; \tau\cdot \sum_{i \in B_{\boldsymbol{\beta}+}}\, (1-\gamma)\cdot ( y_{i}+\boldsymbol{x}_{i}^{\top}\boldsymbol{b} - \boldsymbol{x}_{i}^{\top}(\boldsymbol{\beta}+\boldsymbol{b}))+\gamma\cdot ( y_{i}+\boldsymbol{x}_{i}^{\top}\boldsymbol{b} - \boldsymbol{x}_{i}^{\top}\,(\boldsymbol{\beta}+\boldsymbol{b}))^2, 
\, 
$$
then 
$$
\boldsymbol{\zeta}(\tau, \boldsymbol{\beta}+\boldsymbol{b}, \boldsymbol{y}+\boldsymbol{x}^{\top}\boldsymbol{b},\boldsymbol{x})=\boldsymbol{\zeta}(\tau, \boldsymbol{\beta}, \boldsymbol{y},\boldsymbol{x}).
$$
So
$$
\boldsymbol{\hat{\beta}}(\tau,\boldsymbol{y}+\boldsymbol{x}^{\top}\boldsymbol{b},\boldsymbol{x})=\boldsymbol{\hat{\beta}}(\tau,\boldsymbol{y},\boldsymbol{x})+\boldsymbol{b}.
$$ 

\item To prove the property \ref{theorem 2-2}, we set 
 $$C_{\boldsymbol{\beta}+}=\{ i, \;y_{i} \geq \boldsymbol{x}_{i}^{\top}\boldsymbol{\beta}   \}\; \textit{and} \; C_{\boldsymbol{\beta}-}=\{ i,\; y_{i}< \boldsymbol{x}_{i}^{\top}\,\boldsymbol{\beta}  \}.
 $$
 For a nonsingular matrix $\boldsymbol{A}$, we get
$$
\boldsymbol{\zeta}(\tau,\boldsymbol{A^{-1}} \boldsymbol{\beta},\boldsymbol{y},\boldsymbol{x}^{\top}\boldsymbol{A})=(1-\tau)\cdot\sum_{i \in C_{\boldsymbol{\beta}-}}\, (1-\gamma) ( \boldsymbol{x}_{i}^{\top} \boldsymbol{A}\boldsymbol{A^{-1}}\boldsymbol{\beta}-y_i)+\gamma \cdot ( y_{i} - \boldsymbol{x}_{i}^{\top}\,\boldsymbol{ A}\boldsymbol{A^{-1}}\boldsymbol{\beta})^2+
$$ 
$$\;\;\;\;\;\;\;\;\;\; \tau \,\sum_{i \in C_{\boldsymbol{\beta}+}}\, (1-\gamma)\cdot (y_i-\boldsymbol{x}_{i}^{\top}\, \boldsymbol{A}\,\boldsymbol{A^{-1}}\,\boldsymbol{\beta}   )+\gamma\cdot( y_{i} - \boldsymbol{x}_{i}^{\top}\, \boldsymbol{A}\boldsymbol{A^{-1}} \boldsymbol{\beta})^2.
\, $$
It leads to 
$$
\boldsymbol{\zeta}(\tau,\boldsymbol{A^{-1}} \boldsymbol{\beta},\boldsymbol{y},\boldsymbol{x}^{\top}\boldsymbol{A})=\boldsymbol{\zeta}(\tau, \boldsymbol{\beta}, \boldsymbol{y},\boldsymbol{x}),
$$
therefore \ref{theorem 2-2} follows.
\end{enumerate}

\section*{The proof of Theorem \ref{theorem 3}} \label{proof theorem 3}

The proof of this result is based on Lemma $\ref{lemma hjort 1}$ from \cite{newey_asymmetric_1987}. We mainly verify the conditions \ref{Q A},   \ref{Q B}, and  \ref{Q C} of this Lemma $\ref{lemma hjort 1}$ under the Assumptions \ref{assumption 1}--\ref{assumption 3}.\\
Firstly, to verify \ref{Q A}, we suggest using Lemma \ref{lemme hjort 2}. In our work, the convexity of $T_{n}(\boldsymbol{\beta},\tau,\gamma)$ is obvious, and the convergence of $T_{n}(\boldsymbol{\beta},\tau,\gamma)$ to $T(\boldsymbol{\beta},\tau,\gamma)$ in probability can be proved under the i.i.d. setting of $\boldsymbol{z}_{i}$ and Assumption $\ref{assumption 3}$. Therefore, a direct application of the result is sufficient to verify \ref{Q A}.\\
Next, verifying \ref{Q C}, is obvious since $T(\cdot)$ is convex.\\
The focus now is on verifying \ref{Q B}, consider the function 
$T:\boldsymbol{\beta} \mapsto \mathbb{E} \left\{ C_{\tau}^{\,\gamma}(Y - \boldsymbol{X}^{\top} \boldsymbol{\beta}) - C_{\tau}^{\,\gamma}(Y) \right\},$ and denote $T=(1-\gamma)\cdot T_1+\gamma \cdot T_2$, such that, 
 $
T_1:\boldsymbol{\beta} \mapsto \mathbb{E} \left\{ \rho_{\tau}(Y - \boldsymbol{X}^{\top} \boldsymbol{\beta}) - \rho_{\tau}(Y) \right\}$ and $T_2:\boldsymbol{\beta} \mapsto \mathbb{E} \left\{ \ell_{\tau}(Y - \boldsymbol{X}^{\top} \boldsymbol{\beta}) - \ell_{\tau}(Y) \right\}.$\\
For \(T_1\) to be differentiable with respect to $\boldsymbol{\beta}$, it suffices that   $ \tilde T_1:\boldsymbol{\beta} \mapsto \mathbb{E} \left\{ \rho_{\tau}(Y - \boldsymbol{X}^{\top} \boldsymbol{\beta}) \right\}$  to be differentiable.
One can see that
\begin{align*}
\mathbb E\{\rho_\tau (Y - \boldsymbol{X}^\top \boldsymbol{\beta}) \mid \boldsymbol{X}  \} 
&= \int_{-\infty}^{+\infty} \rho_\tau (y - \boldsymbol{X}^\top \boldsymbol{\beta}) \mathrm{d}F(y\mid \boldsymbol X) \\
&= \int_{-\infty}^{+\infty} (y - \boldsymbol{X}^\top \boldsymbol{\beta})(\tau - \mathds{1}\{y - \boldsymbol{X}^\top \boldsymbol{\beta} < 0\}  ) \mathrm{d}F(y \mid \boldsymbol X) \\
&= \tau \int_{-\infty}^{+\infty} y \,\mathrm{d}F(y \mid \boldsymbol X) - \tau \boldsymbol{X}^\top \boldsymbol{\beta} - \int_{-\infty}^{\boldsymbol{X}^\top \boldsymbol{\beta}} y \, \mathrm{d}F(y \mid \boldsymbol X) + (\boldsymbol{X}^\top \boldsymbol{\beta}) \cdot F(\boldsymbol{X}^\top \boldsymbol{\beta}\mid \boldsymbol{X}).
\end{align*}
Assumption $\ref{assumption 2}$ implies that this expression is well-defined and differentiable with respect to \(\boldsymbol{\beta}\). This means that
\begin{align*}
\tilde {T}_1(\boldsymbol \beta )
&= \mathbb E \left\{ \mathbb E\{\rho_\tau (Y - \boldsymbol{X}^\top \boldsymbol{\beta})| \boldsymbol{X}\} \right\}\\
&= \mathbb E \left\{ \tau \int_{-\infty}^{+\infty} y \, \mathrm{d}F(y\mid \boldsymbol X) - \tau \boldsymbol{X}^\top \boldsymbol{\beta} - \int_{-\infty}^{\boldsymbol{X}^\top \boldsymbol{\beta}} y \, \mathrm{d}F(y|\boldsymbol X) + (\boldsymbol{X}^\top \boldsymbol{\beta}) \cdot F(\boldsymbol{X}^\top \boldsymbol{\beta} \mid \boldsymbol{X}) \right\}\\
&= \int_{\mathbb{R}^p} \left\{ \tau \int^{+\infty}_{-\infty} y \, \mathrm{d}F(y|\boldsymbol x) - \tau \boldsymbol{x}^\top \boldsymbol{\beta} - \int_{-\infty}^{\boldsymbol{x}^\top \boldsymbol{\beta}} y\, \mathrm{d}F(y\mid \boldsymbol x) + (\boldsymbol{x}^\top \boldsymbol{\beta}) \cdot F(\boldsymbol{x}^\top \boldsymbol{\beta}|\boldsymbol{x}) \right\} g(\boldsymbol x)\mathrm{d}\boldsymbol x\\
&=\int_{\mathbb{R}^p} \Gamma(\boldsymbol \beta,\boldsymbol x) \mathrm{d}\boldsymbol x,
\end{align*}
 where $\Gamma(\boldsymbol \beta,\boldsymbol x)= \left\{ \tau \int^{+\infty}_{-\infty} y \, \mathrm{d}F(y \mid \boldsymbol x) - \tau \boldsymbol{x}^\top \boldsymbol{\beta} - \int_{-\infty}^{\boldsymbol{x}^\top \boldsymbol{\beta}} y \, \mathrm{d}F(y\mid\boldsymbol x) + (\boldsymbol{x}^\top \boldsymbol{\beta}) \cdot F(\boldsymbol{x}^\top \boldsymbol{\beta}\mid \boldsymbol{x}) \right\} g(\boldsymbol x).$
The gradient of $\Gamma(\boldsymbol \beta,\boldsymbol x) $  with respect to $\boldsymbol \beta$ 
is $\nabla \Gamma(\boldsymbol \beta,\boldsymbol x)   =\boldsymbol{x} \{F(\boldsymbol{x}^\top \boldsymbol{\beta} \mid \boldsymbol{x} ) - \tau\} g(\boldsymbol x) $. From

\begin{align*}
 \int_{\mathbb{R}^p} \norm{\boldsymbol{x} \{F(\boldsymbol{x}^\top \boldsymbol{\beta} \mid \boldsymbol{x} ) - \tau\} g(\boldsymbol x)  }  \mathrm{d}\boldsymbol x 
\leq  \int_{\mathbb{R}^p} \norm{\boldsymbol{x}   } (1+\tau) g(\boldsymbol x) \mathrm{d}\boldsymbol x 
\leq ( 1+\tau ) \mathbb  E  (\norm{\boldsymbol{Z}}),  
\end{align*}
the gradient \(\nabla \Gamma(\boldsymbol{\beta}, \boldsymbol{x})\) is uniformly bounded (in norm) with respect to \(\boldsymbol{\beta}\) by the integrable variable \((1+\tau)\boldsymbol{Z}\) under Assumption~\ref{assumption 3}. In addition, the functions \(\boldsymbol{\beta} \mapsto \nabla_{\boldsymbol{\beta}} \Gamma(\boldsymbol{\beta}, \boldsymbol{x})\) and \(\boldsymbol{x} \mapsto \nabla_{\boldsymbol{x}} \Gamma(\boldsymbol{\beta}, \boldsymbol{x})\) are continuous for almost every \(\boldsymbol{X}\) and \(\boldsymbol{\beta}\), respectively.\\
Due to the derivation under the integral sign, \(\tilde{T}_1\) is differentiable, and therefore, \(T_1\) is also differentiable with $\nabla_{\boldsymbol \beta}T_1(\boldsymbol \beta )=\int_{\mathbb{R}^p}  \nabla_{\beta} \Gamma(\boldsymbol \beta,\boldsymbol x) \mathrm{d}\boldsymbol x  $. We prove that $T_1$ has a second derivative by a similar procedure.\\
The gradient of $\nabla_{\boldsymbol \beta }\Gamma(\boldsymbol \beta,\boldsymbol X),$  with respect to $\boldsymbol \beta$, is $\nabla^2_{\boldsymbol \beta} \Gamma(\boldsymbol \beta,\boldsymbol X)   =\boldsymbol{X} \boldsymbol{X}^\top  \{f(\boldsymbol{X}^\top \boldsymbol{\beta} | \boldsymbol{X} ) \} g(\boldsymbol X) $  and

\begin{align*}
 \int_{\mathbb{R}^p} \norm{\boldsymbol{x}\, \boldsymbol{x}^{\top}  \{f(\boldsymbol{x}^\top \boldsymbol{\beta} \mid \boldsymbol{x} ) \} g(\boldsymbol x)  }  \mathrm{d}\boldsymbol x 
&\leq  \int_{\mathbb{R}^p} \norm{\boldsymbol{x}   }^2 c_1 g(\boldsymbol x) d\boldsymbol x \leq c_1 \mathbb  E  (\norm{\boldsymbol Z}^2).
\end{align*}
Since the function $\boldsymbol{\beta} \mapsto \nabla_{\boldsymbol{\beta}} \mathbb E\{\rho_\tau (Y - \boldsymbol{X}^\top \boldsymbol{\beta})| \boldsymbol{X}\} $ is differentiable under Assumption \ref{assumption 2}, then

$T_1(\boldsymbol \beta)= \mathbb{E}\left\{ \mathbb{E} \left\{ \rho_{\tau}(Y - \boldsymbol{X}^\top \boldsymbol{\beta}) - \rho_{\tau}(Y) |\boldsymbol{X} \right\}    \right\}=  \mathbb{E} [ \rho_{\tau}(Y - \boldsymbol{X}^\top \boldsymbol{\beta}) - \rho_{\tau}(Y) ],$ is of class $\mathcal{C}^2$, i.e., twice continuously differentiable with $\nabla^2 T_1(\boldsymbol{\beta})=\mathbb{E}[\boldsymbol{X}\boldsymbol{X}^{\top}f(\boldsymbol{X}^\top\boldsymbol{\beta})].$\\
To prove that the function $T_2(\boldsymbol{\beta}) = \mathbb{E} \left\{ \ell_{\tau}(Y - \boldsymbol{X}^\top \boldsymbol{\beta}) - \ell_{\tau}(Y) \right\}$ is differentiable, 
we analyze the integral
\[
\mathbb{E} \left\{ \ell_{\tau}(Y - \boldsymbol{X}^\top \boldsymbol{\beta})- \ell_{\tau}(Y) |\boldsymbol{X} \right\} = \int_{-\infty}^\infty \left( \ell_{\tau}(y - \boldsymbol{X}^\top \boldsymbol{\beta})- \ell_{\tau}(y)  \right) f(y \mid \boldsymbol{X}) \, \mathrm{d}y,
\]

Conditions for differentiability of this term are

\begin{enumerate}
    \item Continuity of the Integrand:
The function \(\ell_\tau(s)\) is continuous in \(s\), and hence the difference $\ell_\tau(y - \boldsymbol{X}^\top \boldsymbol{\beta}) - \ell_\tau(y)$ is continuous in both \(y\) and \(\boldsymbol{\beta}\). Combined with the continuity of \(f(y \mid \boldsymbol{X})\), the integrand is continuous.

 \item Smoothness of the Integrand:
The derivative of the integrand, with respect to \(\boldsymbol{\beta}\), is $
\nabla_{\boldsymbol{\beta}} \Big( \ell_{\tau}(y - \boldsymbol{X}^\top \boldsymbol{\beta}) - \ell_{\tau}(y) \Big)
= -\ell_\tau'(y - \boldsymbol{X}^\top \boldsymbol{\beta}) \boldsymbol{X}, $
where $\ell'_{\tau}(s)=2\Psi_\tau(s) s$. This derivative is continuous in \(y - \boldsymbol{X}^\top \boldsymbol{\beta}\), so the integrand is smooth.

\item Absolute Integrability of the Derivative:
The derivative of the integrand with respect to $\boldsymbol{\beta}$ is $-\ell_\tau'(y - \boldsymbol{X}^\top \boldsymbol{\beta}) \boldsymbol{X}.$
Since \(\ell_\tau'(s)\) grows linearly with \(s\), the integrand is dominated by a term proportional to \(|y| f(y\mid \boldsymbol{x})\), which is integrable over \(\mathbb{R}\) given the decay of \(f(y \mid \boldsymbol{X})\) as \(y \to \pm \infty\) since $\mathbb{E}(|Y|)<+\infty.$

\item Uniform Convergence:
The boundedness of \(\boldsymbol{X}\) and the decay of \(f(y \mid \boldsymbol{X})\) ensure uniform convergence of the integral. This allows differentiation under the integral sign.

\item Differentiating \(T_2(\boldsymbol{\beta})\):\\
Using differentiation under the integral sign, we have
\begin{align*}
\nabla \mathbb{E} \left\{ \ell_{\tau}(Y - \boldsymbol{X}^\top \boldsymbol{\beta}) - \ell_{\tau}(Y) \mid \boldsymbol{X} \right\} 
&= \int_{-\infty}^\infty \nabla \left( \ell_{\tau}(y - \boldsymbol{X}^\top \boldsymbol{\beta}) - \ell_{\tau}(y) \right) f(y \mid \boldsymbol{X}) \, dy \\
&= -2 \boldsymbol{X} \left\{ \tau \int_{\boldsymbol{X}^{\top} \boldsymbol{\beta}}^{+\infty} (y - \boldsymbol{X}^{\top} \boldsymbol{\beta}) f(y \mid \boldsymbol{X}) \, \mathrm{d}y \right. \\
&\quad \left. + (1 - \tau) \int_{-\infty}^{\boldsymbol{X}^{\top} \boldsymbol{\beta}} (y - \boldsymbol{X}^{\top} \boldsymbol{\beta}) f(y \mid \boldsymbol{X}) \, \mathrm{d}y \right\}.
\end{align*}

\end{enumerate}
Thus, the derivative becomes:
\begin{align}
\nabla T_2(\boldsymbol{\beta}) 
&= -2\mathbb{E} \left\{ \boldsymbol{X} \left\{\tau\int_{\boldsymbol{X}^{\top} \boldsymbol{\beta}}^{+\infty} (y-\boldsymbol{X}^{\top} \boldsymbol{\beta}) f(y\vert \boldsymbol{X})\,\mathrm{d}y +(1-\tau)\, \int_{-\infty}^{\boldsymbol{X}^{\top}\boldsymbol{\beta}}\, (y-\boldsymbol{X}^{\top} \boldsymbol{\beta})\, f(y\vert \boldsymbol{X})\,\mathrm{d}y\right\} \right\}.
\end{align}
According to Assumptions \ref{assumption 2} and \ref{assumption 3}, the function $\theta \mapsto \int_{-\infty}^{\theta}(y-\theta) f(y\mid \boldsymbol X) dy,$ is continuously differentiable, since  its derivative, $-F(\theta|\boldsymbol{X})$ is dominated by $1$ almost surly for every $\boldsymbol X=\boldsymbol x$.  
Since all conditions for differentiability are satisfied, the function \(T_2(\boldsymbol{\beta})\) is differentiable with respect to \(\boldsymbol{\beta}\).
 We can then derive inside the expectation and get
 $\nabla^2 T_2(\boldsymbol \beta)= 2\mathbb{E}\left(  \boldsymbol{X}\, \boldsymbol{X}^{\top}\Psi_\tau(Y-\boldsymbol{X}^{\top} \boldsymbol{\beta}) \right)$.
By the previous analysis, we conclude that $T$ is $\mathcal{C}^2$, with $\nabla^2 T=(1-\gamma)\cdot \nabla^2 T_1+\gamma\cdot \nabla^2 T_2$, i.e.,

\begin{eqnarray*}
\nabla^2 T(\boldsymbol \beta)
&=& \mathbb{E}\left(  \boldsymbol{X}\, \boldsymbol{X}^{\top}\Delta_\tau(Y-\boldsymbol{X}^{\top} \boldsymbol{\beta}) \right)+(1-\gamma)\cdot \mathbb{E}(\boldsymbol{X}\boldsymbol{X} ^{\top}f(\boldsymbol{X}^{\top}\boldsymbol{\beta})   )\\
&=& \,\mathbb{E}\left\{  \boldsymbol{X}\boldsymbol{X}^{\top}  \left( \Delta_\tau(Y-\boldsymbol{X}^{\top} \boldsymbol{\beta})   +(1-\gamma)\cdot f(\boldsymbol{X}^{\top}\boldsymbol{\beta}) \right)  \right\}.
\end{eqnarray*}
Let  $\delta= \min \{ \tau,1- \tau \}$, we have the inequality, 
$\Delta_\tau(Y-\boldsymbol{X}^{\top} \boldsymbol{\beta})  +(1-\gamma)\cdot f(\boldsymbol{X}^{\top} \boldsymbol{\beta})   > 2\gamma \delta.$ Set $\kappa =2\gamma \delta$, it follows that $\nabla^2 T(\boldsymbol{\beta}) - \kappa \mathbb{E}(\boldsymbol{X} \boldsymbol{X}^{\top})$ 
is positive semi-definite. 

Under Assumption \ref{assumption 5}, using the same  argument as in the proof of Theorem 3  in  \cite{newey_asymmetric_1987}, i.e., write  a second order mean expansion of 
$T( \boldsymbol{\beta}) $ and we get
\begin{eqnarray} \label{equation 0}
    T(\boldsymbol{\beta})-T(\boldsymbol{\tilde \beta})&=&\left[ \nabla T(\boldsymbol{\beta})\right]^{\top}(\boldsymbol{\beta}-\boldsymbol{\tilde{\beta}} )+(\boldsymbol{\beta}-\boldsymbol{\tilde{\beta}} )^{\top}  \left[ \nabla^2 T(\boldsymbol{\bar \beta})\right]  (\boldsymbol{\beta}-\boldsymbol{\tilde{\beta}} )\notag\\
    &\ge&\left[ \nabla T(\boldsymbol{ \beta})\right]^{\top}  (\boldsymbol{\beta}-\boldsymbol{\tilde{\beta}} )+p\lambda_{x} \kappa\norm{\boldsymbol{\beta}-\boldsymbol{\tilde{\beta}} } ^{2},\label{last inequality}
\end{eqnarray}
where $\boldsymbol{\bar{\beta}}$ is the mean value and $\lambda_{x} $ is the minimum eigenvalue of $\mathbb{E}( \boldsymbol{X}\boldsymbol{X}^{\top}),$ which is positive by Assumption \ref{assumption 5}.\\
Based on the methodology outlined in  \cite{newey_asymmetric_1987} and following the identical procedure as described in their Theorem 2, divide both sides of the  inequality \eqref{last inequality} by  $\norm{\boldsymbol{\beta}-\boldsymbol{\tilde{\beta}} }^{2} $  and fix $\boldsymbol{\Tilde{\beta}}$,  
 then $T ( \boldsymbol{\beta})>T( \boldsymbol{\tilde{\beta}}) $ as  $\norm{ \boldsymbol{\beta}-\boldsymbol{\tilde{\beta}} }\to +\infty $.  
Hence, there exist some $r>0$, such that $T(\boldsymbol{\beta})>T(\boldsymbol{\tilde{\beta}}), \quad \forall \boldsymbol{\beta} \in \overline{\mathbb{B}}(\boldsymbol{\tilde{\beta}}, r),$
with $\overline{\mathbb{B}}(\boldsymbol{\tilde{\beta}},r )$ being the complement of a closed ball centered at $ \boldsymbol{\tilde{\beta}}   $ with radius $r.$ That means that $T( \boldsymbol{\beta})>T ( \boldsymbol{\tilde{\beta}})$ is true outside the closed ball $\mathbb{B}(\boldsymbol{\tilde{\beta}},r )$ centered at $\boldsymbol{\tilde{ \beta}}$. Consider the function $\boldsymbol{\beta} \mapsto T(\boldsymbol{\beta})$ over the closed ball $\mathbb{B}(\boldsymbol{\tilde{\beta}},r ).$ It follows from the continuity of the function $ T(\cdot)$ and the compactness of the closed ball $\mathbb{B}(\boldsymbol{\tilde{\beta}},r ),$ that $ T(\cdot,\tau,\gamma)$  has a minimum $\boldsymbol{\beta}(\tau)$ in $\mathbb{B}(\boldsymbol{\tilde{\beta}},r )$ and since 
$T ( \boldsymbol{\beta}(\tau))  < T ( \boldsymbol{\tilde{\beta}}),\, $ then $\boldsymbol{\beta}(\tau),$ is a global minimum. By differentiability, $\boldsymbol{\beta}(\tau)$ satisfies, $\nabla T(\boldsymbol{\beta})=\boldsymbol{0}$. From   \eqref{equation 0} with $\boldsymbol{\beta}(\tau)=\boldsymbol{\tilde{\beta}}_0(\tau)$, the uniqueness of the solution follows. So, \ref{Q B} holds for $T$.\\
Based on the preceding analysis, Theorem \(\ref{theorem 3}\) follows directly as an application of Lemma \(\ref{lemma hjort 1}\).

\section*{The proof of Theorem $\ref{theorem 4}$}

To prove the theorem, we follow the idea in \cite{pollard1991asymptotics}, \cite{hjort2011asymptotics}, and \cite{knight1998limiting}, who showed that the limiting distribution of \(\sqrt{n}(\hat{\boldsymbol{\beta}}(\tau) - \boldsymbol{\tilde{\beta}}_{0}(\tau))\) follows from considering the objective function defined for any $\boldsymbol{u}\in \mathbb{R}^p$, as follows
\begin{equation}
 \mathcal{L}_n(\boldsymbol{u}) = \sum_{i=1}^n [C_{\tau}^{\,\gamma} \left(\varepsilon_i - \frac{\boldsymbol{x}_i^\top \boldsymbol{u}}{\sqrt{n}}\right) - C_{\tau}^{\,\gamma}(\varepsilon_i)],    
\end{equation}
where \(\varepsilon_i = y_i - \boldsymbol{x}_i^\top \boldsymbol{\tilde {\beta}}_{0}(\tau)\). The function \(\mathcal{L}_n(\cdot)\) is  convex and is minimized by
$\boldsymbol{\hat u}_n = \sqrt{n}(\boldsymbol{\hat \beta}(\tau) - \boldsymbol{\tilde \beta}_{0}(\tau)).$ Denote  $G_i :=\nabla_{s}(\ell_\tau\big(\varepsilon_i - s\big))\mid_{s=0}   =-\ell'_\tau(\varepsilon_i),$ as the first-order derivative of the function \(\ell_\tau\big(\varepsilon_i - s\big)\) at the point \( s = 0 \), $a_\tau(s) = \tau - \mathds{1}\left\{ s < 0\right\}$   and $d(\tau) =  (1-\tau)F_\varepsilon(0) + \tau\big(1-F_\varepsilon(0)\big).$
We may write 
\begin{align}
  \mathcal{L}_n(\boldsymbol{u}) &= \sum_{i=1}^n [C_{\tau}^{\,\gamma} \left(\varepsilon_i - \frac{\boldsymbol{x}_i^\top \boldsymbol{u}}{\sqrt{n}}\right) - C_{\tau}^{\,\gamma}(\varepsilon_i)]\notag\\
&=  (1-\gamma)\sum_{i=1}^n [\rho_\tau\left(\varepsilon_i - \frac{\boldsymbol{x}_i^\top \boldsymbol{u}}{\sqrt{n}}\right) - \rho_\tau(\varepsilon_i)]+\gamma\sum_{i=1}^n [\ell_\tau\left(\varepsilon_i - \frac{\boldsymbol{x}_i^\top \boldsymbol{u}}{\sqrt{n}}\right) - \ell_\tau(\varepsilon_i)]\notag\\
&=(1-\gamma) \sum_{i=1}^n [-a_\tau(\varepsilon_i)\frac{\boldsymbol{x}_i^\top\boldsymbol{u}}{\sqrt{n}}+ \int_0^{\frac{\boldsymbol{x}_i^\top \boldsymbol{u}}{\sqrt{n}}} \big(\mathds{1}\left\{ \varepsilon_i \leq s\right\} - \mathds{1}\left\{ \varepsilon_i \leq 0\right\} \big)  \mathrm{d}s]+\gamma\sum_{i=1}^n [\ell_\tau\left(\varepsilon_i - \frac{\boldsymbol{x}_i^\top \boldsymbol{u}}{\sqrt{n}}\right) - \ell_\tau(\varepsilon_i)]\label{eq knight}\\
&=(1-\gamma) \sum_{i=1}^n [-a_\tau(\varepsilon_i)\frac{\boldsymbol{x}_i^\top\boldsymbol{u}}{\sqrt{n}}+ \int_0^{\frac{\boldsymbol{x}_i^\top \boldsymbol{u}}{\sqrt{n}}} \big(\mathds{1}\left\{ \varepsilon_i \leq s\right\} - \mathds{1}\left\{ \varepsilon_i \leq 0\right\} \big)  \mathrm{d}s]+\gamma\sum_{i=1}^n G_i \frac{\boldsymbol{x}_i^\top \boldsymbol{u}}{\sqrt{n}}\notag\\
&+ \gamma\sum_{i=1}^n [\ell_\tau\left(\varepsilon_i - \frac{\boldsymbol{x}_i^\top \boldsymbol{u}}{\sqrt{n}}\right) - \ell_\tau(\varepsilon_i)-G_i \frac{\boldsymbol{x}_i^\top \boldsymbol{u}}{\sqrt{n}}  ]\\
&=- \frac{1}{\sqrt{n}} \sum_{i=1}^n   [  (1-\gamma) a_\tau(\varepsilon_i)+ \gamma G_i ]\boldsymbol{x}_i^\top \boldsymbol{u}+\int_0^{\frac{\boldsymbol{x}_i^\top \boldsymbol{u}}{\sqrt{n}}} \big(\mathds{1}\left\{ \varepsilon_i \leq s\right\} - \mathds{1}\left\{ \varepsilon_i \leq 0\right\} \big)  \mathrm{d}s]\notag\\
&+ \gamma\sum_{i=1}^n [\ell_\tau\left(\varepsilon_i - \frac{\boldsymbol{x}_i^\top \boldsymbol{u}}{\sqrt{n}}\right) - \ell_\tau(\varepsilon_i)-G_i \frac{\boldsymbol{x}_i^\top \boldsymbol{u}}{\sqrt{n}} \boldsymbol{u} ]\notag\\
&=W_{1n}(\boldsymbol{u})+W_{2n}(\boldsymbol{u})+W_{3n}(\boldsymbol u),\label{notation w}
\end{align}
where equation \eqref{eq knight} follows from the following identity (see  \cite{knight1998limiting}) \[
\rho_\tau(s - t) - \rho_\tau(s) = -t a_\tau(s) + \int_0^t \big(\mathds{1}\left\{s \leq v\right\} - \mathds{1}\left\{ s \leq 0\right\}\big) \, dv,
\]
and the components of \eqref{notation w} are given by 
\begin{align}
    W_{1n}(\boldsymbol u)&=- \frac{1}{\sqrt{n}} \sum_{i=1}^n   [  (1-\gamma) a_\tau(\varepsilon_i)+ \gamma G_i ]\boldsymbol{x}_i^\top \boldsymbol{u},\label{w1}\\
    W_{2n}(\boldsymbol u)&=\int_0^{\frac{\boldsymbol{x}_i^\top \boldsymbol{u}}{\sqrt{n}}} \big(\mathds{1}\left\{ \varepsilon_i \leq s\right\} - \mathds{1}\left\{ \varepsilon_i \leq 0\right\} \big)  \mathrm{d}s],\label{w2}\\
    W_{3n}(\boldsymbol u)&=\gamma\sum_{i=1}^n [\ell_\tau\left(\varepsilon_i - \frac{\boldsymbol{x}_i^\top \boldsymbol{u}}{\sqrt{n}}\right) - \ell_\tau(\varepsilon_i)-G_i \frac{\boldsymbol{x}_i^\top \boldsymbol{u}}{\sqrt{n}}  ]\label{w3}.
\end{align}
First, we examine \( W_{1n} \) in \eqref{w1}. Using Assumptions \ref{assumption 6}-\ref{assumption 6-i}, \ref{assumption 6}-\ref{assumption-a}, and \ref{assumption 6}-\ref{assumption-d}, we apply the Lindeberg–Feller central limit theorem  to establish that
  \begin{align}
W_{1n}(\boldsymbol{u}) \overset{\mathcal{}}{\to} -\boldsymbol{u}^\top \boldsymbol{\tilde V},\; \boldsymbol{\tilde V} \sim \mathcal{N}(\boldsymbol{0}, \varsigma(\tau,\gamma)\cdot\boldsymbol{K}),\label{dealing with w1}  
\end{align}
where $\varsigma(\gamma,\tau)=\mathbb{V}\mathrm{ar}((1-\gamma)\cdot a_\tau(\varepsilon) -\varepsilon\Delta_\tau(\varepsilon)),$   $a_\tau(\varepsilon)=\tau-\mathds{1}\{ \varepsilon<0 \},$ and $\Delta_\tau(\varepsilon)=2\gamma\Psi_\tau(\varepsilon).$
Next, we consider the component $W_{2n}$, the function $W_{2n}(\boldsymbol{u})$ can be written as 
\begin{align}
 W_{2n}(\boldsymbol{u}) &= (1-\gamma) \cdot \sum_{i=1}^n \int_0^{\frac{\boldsymbol{x}_i^\top \boldsymbol{u}}{\sqrt{n}}} \big(\mathds{1}\left\{ \varepsilon_i \leq s\right\} - \mathds{1}\left\{ \varepsilon_i \leq 0\right\} \big) \mathrm{d}s\notag\\ %\notag\label{eq B1 B2}  
 &=\sum_{i=1}^n W_{2ni},\notag
\end{align}
with $W_{2ni}(\boldsymbol u)=(1-\gamma)  \int_0^{\frac{\boldsymbol{x}_i^\top \boldsymbol{u}}{\sqrt{n}}} \big(\mathds{1}\left\{ \varepsilon_i \leq s\right\} - \mathds{1}\left\{ \varepsilon_i \leq 0\right\} \big)\mathrm{d}s.$ It follows that
 
\begin{align}
 W_{2n}(\boldsymbol{u}) = \sum_{i=1}^n\mathbb{E}[W_{2ni}(\boldsymbol{u})] + \sum_{i=1}^n\left\{W_{2ni}(\boldsymbol{u}) - \mathbb{E}[W_{2ni}(\boldsymbol{u})]\right\} \label{B_in}  
\end{align}
and
\begin{align*}
\mathbb{E}[W_{2ni}(\boldsymbol{u})] &= (1-\gamma) \int_0^{\frac{\boldsymbol{x}_i^\top \boldsymbol{u}}{\sqrt{n}}} \Big( F(\boldsymbol{x}_i^\top \boldsymbol{\tilde{\beta}}_0(\tau) + v \mid \boldsymbol{x}_i)  - F(\boldsymbol{x}_i^\top \boldsymbol{\tilde{\beta}}_0(\tau) \mid \boldsymbol{x}_i) \Big) \, \mathrm{d}v \\
&= \frac{1-\gamma}{\sqrt{n}} \int_0^{\boldsymbol{x}_i^\top \boldsymbol{u}} \Big( F\left(\boldsymbol{x}_i^\top \boldsymbol{\tilde{\beta}}_0(\tau) + \frac{s}{\sqrt{n}} \mid \boldsymbol{x}_i \right)  - F(\boldsymbol{x}_i^\top \boldsymbol{\tilde{\beta}}_0(\tau) \mid \boldsymbol{x}_i) \Big) \, \mathrm{d}s\\
&= \frac{1-\gamma}{n} \int_0^{\boldsymbol{x}_i^\top \boldsymbol{u}} \sqrt{n}\Big( F\left(\boldsymbol{x}_i^\top \boldsymbol{\tilde{\beta}}_0(\tau) + \frac{s}{\sqrt{n}} \mid \boldsymbol{x}_i \right)  - F(\boldsymbol{x}_i^\top \boldsymbol{\tilde{\beta}}_0(\tau) \mid \boldsymbol{x}_i) \Big) \, \mathrm{d}s.
\end{align*}
Expanding the integral leads to
\begin{eqnarray*}
\mathbb{E}[W_{2ni}(\boldsymbol{u})] &=& \frac{1-\gamma}{n} \int_0^{\boldsymbol{x}_i^\top \boldsymbol{u}}  f(\boldsymbol{x}^{\top}_i\boldsymbol{\tilde{\beta}}_0(\tau)  \mid \boldsymbol{x}_i) s \, \mathrm{d}s + o(1) \\
&=& \frac{1-\gamma}{2n} f(\boldsymbol{x}^{\top}_i\boldsymbol{\tilde{\beta}}_0(\tau)  \mid \boldsymbol{x}_i) (\boldsymbol{u}^\top \boldsymbol{x}_i)^2 + o(1) \\
&=& \frac{1-\gamma}{2n} \boldsymbol{u}^\top f(\boldsymbol{x}^{\top}_i\boldsymbol{\tilde{\beta}}_0(\tau)  \mid \boldsymbol{x}_i) \boldsymbol{x}_i \boldsymbol{x}_i^\top \boldsymbol{u} + o(1).
\end{eqnarray*}
Summing over \(i=1,\cdots, n\) gives

\begin{align}
   \sum_{i=1}^{n}\mathbb{E}[W_{2ni}(\boldsymbol{u})] \to \frac{1-\gamma}{2} \boldsymbol{u}^\top \boldsymbol{J} \boldsymbol{u}.\label{esperance conv} 
\end{align}
So we get the following bound (see \cite{koenker_quantile_2005})
\[
\mathbb{V}\mathrm{ar}(W_{2n}(\boldsymbol{u}))=(1-\gamma)^2\sum_{i=1}^{n} \mathbb{E}\left(W_{2ni}(\boldsymbol{u})-\mathbb{E}[W_{2ni}(\boldsymbol{u})]\right)^2 \leq \frac{(1-\gamma)^2}{\sqrt{n}} \max \norm{\boldsymbol{x}_i^\top \boldsymbol{u}} \sum_{i=1}^n\mathbb{E}[W_{2ni}(\boldsymbol{u})].
\]
Based on \eqref{B_in}, \eqref{esperance conv} and  Assumption \ref{assumption 6}-\ref{assumption-c}) then implies that

\begin{equation}
 W_{2n}(\boldsymbol{u})\to \frac{1}{2}[(1-\gamma)\boldsymbol{u}^\top \boldsymbol{J} \boldsymbol{u}].\label{dealinf with w2}   
\end{equation}
Now, we consider the component $W_{3n}$ in equation \eqref{w3}. Define
\[
b_i(s) := \mathbb{E}\left[\ell_\tau(\varepsilon_i - s) - \ell_\tau(\varepsilon_i)\right],
\]
 and by a Taylor expansion, one can get
\[
b_i(s) = \mathbb{E}(G_i)+d(\tau)s^2 + o(s^2), 
\]
where 
\[
d(\tau) =  (1-\tau)F_\varepsilon(0) + \tau\big(1-F_\varepsilon(0)\big).
\]
Since $\mathbb{E}(G_i)=0$ based on Assumption \ref{assumption 6}-\ref{assumption 6-i} and     \(\frac{1}{n} \sum_{i=1}^n \boldsymbol{x}_i \boldsymbol{x}_i^\top \to \boldsymbol{K}\)  according to Assumption \ref{assumption 6}-\ref{assumption-a}), with \( \boldsymbol{u} \) fixed, we have

\begin{align*}
\mathbb{E}\left[\sum_{i=1}^n \ell_\tau\left(\varepsilon_i - \frac{\boldsymbol{x}_i^\top \boldsymbol{u}}{\sqrt{n}}\right) - \ell_\tau(\varepsilon_i)\right]
&= \sum_{i=1}^n b_i\left(\frac{\boldsymbol{x}_i^\top \boldsymbol{u}}{\sqrt{n}}\right) \\
&=d(\tau) \boldsymbol{u}^\top \left(\frac{1}{n} \sum_{i=1}^n \boldsymbol{x}_i \boldsymbol{x}_i^\top \right) \boldsymbol{u} + o(1).   
\end{align*}
Then 
\begin{align*}
\mathbb{E}\left[\sum_{i=1}^n \ell_\tau\left(\varepsilon_i - \frac{\boldsymbol{x}_i^\top \boldsymbol{u}}{\sqrt{n}}\right) - \ell_\tau(\varepsilon_i)\right]
 =d(\tau) \boldsymbol{u}^\top \left(\frac{1}{n} \sum_{i=1}^n \boldsymbol{x}_i \boldsymbol{x}_i^\top \right) \boldsymbol{u} +o(1),   
\end{align*}
hence
\begin{align}
\mathbb{E}[W_{3n}(\boldsymbol{u})] =d(\tau) \boldsymbol{u}^\top \left(\frac{1}{n} \sum_{i=1}^n \boldsymbol{x}_i \boldsymbol{x}_i^\top \right) \boldsymbol{u}+ o(1). \label{W_3 petit tau}  
\end{align}
Denote 
\[
L_{i,n}(\boldsymbol{u}) = \ell_\tau\left(\varepsilon_i - \frac{\boldsymbol{x}_i^\top \boldsymbol{u}}{\sqrt{n}}\right) - \ell_\tau(\varepsilon_i) - G_i \frac{\boldsymbol{x}_i^\top \boldsymbol{u}}{\sqrt{n}}.
\]
By Taylor expansion, there exists \(\zeta_i\) between \(\varepsilon_i - \frac{\boldsymbol{x}_i^\top \boldsymbol{u}}{\sqrt{n}}\) and \(\varepsilon_i\) such that
\[
|L_{i,n}(\boldsymbol u)| = \frac{\ell''_\tau(\zeta_i)}{2}\cdot \left(\frac{\boldsymbol{x}_i^\top \boldsymbol{u}}{\sqrt{n}}\right)^2 \leq \max(\tau, 1-\tau) \cdot \left(\frac{\boldsymbol{x}_i^\top \boldsymbol{u}}{\sqrt{n}}\right)^2, 
\]
because
\[
\ell''_\tau(\zeta_i) = 2\left((1-\tau)\mathds{1}\{\zeta_i \leq 0\} + \tau \mathds{1}\{\zeta_i > 0\}\right).
\]
Since 
\[
|L_{i,n}(\boldsymbol u)|^2  \leq \max(\tau, 1-\tau) ^2\cdot \left(\frac{      \boldsymbol{u}^\top  \boldsymbol{x}_i  \boldsymbol{x}_i^\top \boldsymbol{u}}{n}\right)^2 \leq \max(\tau, 1-\tau) ^2\cdot \left(\frac{      \boldsymbol{u}^\top  \boldsymbol{x}_i  \boldsymbol{x}_i^\top \boldsymbol{u}}{n}\right) \left(\frac{      \boldsymbol{u}^\top  \boldsymbol{x}_i  \boldsymbol{x}_i^\top \boldsymbol{u}}{n}\right)   , 
\]
then \[
\sum_{i=1}^n \mathbb{E}[L_{i,n}^2(\boldsymbol u)] \leq \max(\tau, 1-\tau) ^2 \cdot    \sum_{i=1}^n\left( \boldsymbol{u}^\top \frac{\boldsymbol{x}_i \boldsymbol{x}_i^\top}{n} \boldsymbol{u}   \cdot   
  \frac{\norm{\boldsymbol{x}_i}^2  \norm{\boldsymbol{u}}^2 }{\sqrt{n}^2}   \right). 
\]
It follows that  \[
\sum_{i=1}^n \mathbb{E}[L_{i,n}^2(\boldsymbol u)] \leq c \cdot\boldsymbol{u}^\top \left(\sum_{i=1}^n \frac{\boldsymbol{x}_i \boldsymbol{x}_i^\top}{n}\right) \boldsymbol{u} \cdot \max_{1 \leq i \leq n} \left(\frac{\norm{\boldsymbol{x}_i}}{\sqrt{n}}\right)^2 \norm{\boldsymbol{u}}^2, 
\]
where \(c:=\max(\tau,1-\tau)^2\).
Under the Assumptions $\ref{assumption 6}$-$\ref{assumption-a}$) and $\ref{assumption 6}$-$\ref{assumption-c}$), we have \(\frac{1}{n} \sum_{i=1}^n \boldsymbol{u}^\top \boldsymbol{x}_i \boldsymbol{x}_i^\top \boldsymbol{u} \to \boldsymbol{u}^\top \boldsymbol{K} \boldsymbol{u}\) and \(\max_{1 \leq i \leq n} \norm{\boldsymbol{x}_i}/\sqrt{n} \to 0\), respectively.\\
Consequently, we obtain
\begin{align}
  \sum_{i=1}^n \mathbb{E}[L_{i,n}^2(\boldsymbol{u})] \to 0.\label{E l^2 to 0}  
\end{align}
Now, write
\begin{align}\label{the term}
W_{3n}(\boldsymbol{u}) = \mathbb{E}\left[W_{3n}(\boldsymbol{u})\right] 
+ \sum_{i=1}^n \ell_\tau\left(\varepsilon_i - \frac{\boldsymbol{x}_i^\top \boldsymbol{u}}{\sqrt{n}}\right) - \ell_\tau(\varepsilon_i)-G_i\frac{\boldsymbol{x}_i^\top}{\sqrt{n}} - \mathbb{E}\left[\sum_{i=1}^n \ell_\tau\left(\varepsilon_i - \frac{\boldsymbol{x}_i^\top \boldsymbol{u}}{\sqrt{n}}\right) - \ell_\tau(\varepsilon_i)-G_i\frac{\boldsymbol{x}_i^\top}{\sqrt{n}}\right].  
\end{align}
Based on \eqref{W_3 petit tau}, the term in \eqref{the term} simplifies to
\[
W_{3n}(\boldsymbol{u}) =d(\tau) \boldsymbol{u}^\top \left(\frac{1}{n} \sum_{i=1}^n \boldsymbol{x}_i \boldsymbol{x}_i^\top \right) \boldsymbol{u}+  \sum_{i=1}^n [L_{i,n}(\boldsymbol{u}) - \mathbb{E}(L_{i,n}(\boldsymbol{u}))] + o(1), 
\]
The residual term satisfies
\[
\mathbb{E}\left[\left(\sum_{i=1}^n (L_{i,n}(\boldsymbol{u}) - \mathbb{E}[L_{i,n}(\boldsymbol{u})])\right)^2\right] \leq \sum_{i=1}^n \mathbb{E}[L_{i,n}^2(\boldsymbol{u})] 
\]
because 
\begin{align*}
\mathbb{E}\left[\left(\sum_{i=1}^n (L_{i,n}(\boldsymbol{u}) - \mathbb{E}[L_{i,n}(\boldsymbol{u})])\right)^2\right] 
&= \mathbb{E}\left[ \sum_{i=1}^n (L_{i,n}(\boldsymbol{u}) - \mathbb{E}[L_{i,n}(\boldsymbol{u})])^2 \right. \\
&\quad +  2 \sum_{1 \leq i < j \leq n} \left\{L_{i,n}(\boldsymbol{u}) - \mathbb{E}[L_{i,n}(\boldsymbol{u})]\right\} \left\{L_{j,n}(\boldsymbol{u}) - \mathbb{E}[L_{j,n}(\boldsymbol{u})]\right\}.
\end{align*}
This leads to 
\begin{eqnarray*}
\mathbb{E}\left[\left(\sum_{i=1}^n (L_{i,n}(\boldsymbol{u}) - \mathbb{E}[L_{i,n}(\boldsymbol{u})])\right)^2\right] 
& = & \sum_{i=1}^n \mathbb{E}\left[  (L_{i,n}(\boldsymbol{u}) - \mathbb{E}[L_{i,n}(\boldsymbol{u})])^2 \right] \\
& & + 2 \sum_{1\leq i<j\leq n} \left\{ \mathbb{E}[L_{i,n}(\boldsymbol{u})] - \mathbb{E}[L_{i,n}(\boldsymbol{u})] \right\}  \left\{ \mathbb{E}[L_{j,n}(\boldsymbol{u})] - \mathbb{E}[L_{j,n}(\boldsymbol{u})] \right\}.
\end{eqnarray*}
Hence, one has
\begin{align}
 \mathbb{E}\left[\left(\sum_{i=1}^n (L_{i,n}(\boldsymbol{u}) - \mathbb{E}[L_{i,n}(\boldsymbol{u})])\right)^2\right] &= \sum_{i=1}^n  \mathbb{V}\mathrm{ar}[  (L_{i,n}(\boldsymbol{u}) )]\notag \\
 &= \sum_{i=1}^n  \mathbb{E}[  (L_{i,n}(\boldsymbol{u}) )^2]-  (\mathbb{E}[  L_{i,n}(\boldsymbol{u} )])^2\notag \\
 &\leq \sum_{i=1}^n  \mathbb{E}[  (L_{i,n}(\boldsymbol{u}) )^2],   \label{inequalty}
\end{align}
then $\mathbb{E}\left[\left(\sum_{i=1}^n (L_{i,n}(\boldsymbol{u}) - \mathbb{E}[L_{i,n}(\boldsymbol{u})])\right)^2\right]\to 0$  according to  \eqref{E l^2 to 0} and \eqref{inequalty}.\\
Therefore, we have
\begin{align}
W_{3n}(\boldsymbol{u}) = d(\tau) \boldsymbol{u}^\top \left(\frac{1}{n} \sum_{i=1}^n \boldsymbol{x}_i \boldsymbol{x}_i^\top \right) \boldsymbol{u}+ o_p(1). \label{dealing with W_3}
\end{align}
Based on \eqref{dealing with w1}, \eqref{dealinf with w2} and \eqref{dealing with W_3}, one has

\[
\mathcal{L}_n(\boldsymbol{u}) \overset{\mathcal D}{\to} \mathcal{L}_\infty(\boldsymbol{u}) := -\boldsymbol{u}^\top \boldsymbol{\tilde V} + \frac{1}{2} \boldsymbol{u}^\top \boldsymbol{\tilde {J}} \boldsymbol{u},
\]
with $\boldsymbol{\tilde J}=\frac{1}{2}[(1-\gamma)\boldsymbol{J}+2\gamma d(\tau)\boldsymbol{K}]$ and $\boldsymbol{\tilde V} \xrightarrow[]{\mathcal D } \mathcal{N}(\boldsymbol{0}, \varsigma(\tau,\gamma)\cdot\boldsymbol{K})$.\\
The convexity of the limiting objective function \(\mathcal{L}_\infty(\cdot)\) ensures the uniqueness of the minimizer (see, e.g., \cite{knight1998limiting} and \cite{pollard1991asymptotics}) and consequently the convergence of the sequence  of minimizers $\boldsymbol{u}_n$ of the corresponding sequence of objective functions $\mathcal{L}_n(\cdot)$, namely
\[
\hat{\boldsymbol{u}}_n = \arg\min \mathcal{L}_n(\boldsymbol{u}) \overset{\mathcal{D} }{\to} \boldsymbol{u}_\infty = \arg\min \mathcal{L}_\infty(\boldsymbol{u}).
\]
To find the minimizer, take the gradient of $\mathcal{L}_\infty(\boldsymbol{u})$ with respect to $\boldsymbol{u}$ and set it to zero.
\[
\nabla \mathcal{L}_\infty(\boldsymbol{u}) = -\boldsymbol{\tilde{V}} + \boldsymbol{\tilde{J}} \boldsymbol{u} = 0.
\]
Solving for $\boldsymbol{u}$, we get the candidate minimizer 
\[
\boldsymbol{u}_\infty = (\boldsymbol{\tilde{J}})^{-1} \boldsymbol{\tilde{V}}.
\]
To make sure that $\boldsymbol{u}_\infty$ is the unique minimizer, we compute the second derivative  of $\mathcal{L}_\infty(\boldsymbol{u})$
\[
\nabla^2 \mathcal{L}_\infty(\boldsymbol{u}) = \boldsymbol{\tilde{J}}.
\]
Since $\boldsymbol{\tilde{J}}$ is positive definite (since a linear combination of positive definite matrices with strictly positive weights is always positive definite), $\mathcal{L}_\infty(\boldsymbol{u})$ is strictly convex, ensuring that $\boldsymbol{u}_\infty$ is the unique minimizer.
Finally, we obtain
\[
\boldsymbol{u}_\infty = (\boldsymbol{\tilde J})^{-1} \boldsymbol{\tilde{V} }.
\]
Given that $\boldsymbol{\tilde{V}} \sim \mathcal{N}(\boldsymbol{0}, \varsigma(\tau,\gamma) \cdot \boldsymbol{K})$, then the asymptotic distribution of $\boldsymbol{\hat u}_n$ is 
\[
\boldsymbol{\hat u}_n\xrightarrow[]{\mathcal{D}} \mathcal{N}\left(\boldsymbol{0}, \varsigma(\tau, \gamma) \cdot (\boldsymbol{\tilde{J}})^{-1} \boldsymbol{K} (\boldsymbol{\tilde{J}})^{-1}\right).
\]
Thus, the asymptotic normality  in Theorem \ref{theorem 4} is established.

\bibliographystyle{apalike}
\bibliography{main}

\end{document}